\documentclass[12pt]{article}
\usepackage{graphicx}
\usepackage{amsfonts}
\usepackage{amssymb,amsmath}
\usepackage{color}
\usepackage[colorlinks=true,linkcolor=blue,citecolor=blue]{hyperref}

\newcommand{\blue}{\color{blue}}

\begin{document}

\begin{titlepage}

\begin{flushright}
ICRR-Report-609-2011-26\\
IPMU 12-0024 \\
UT-12-02\\
\end{flushright}

\begin{center}

{\Large \bf  
Non-Gaussian isocurvature perturbations\\ in dark radiation}

\vskip .45in

{
Etsuko Kawakami$^{a}$, 
Masahiro Kawasaki$^{a,b}$,
Koichi Miyamoto$^a$,\\
Kazunori Nakayama$^{b,c}$
and
Toyokazu Sekiguchi$^d$}

\vskip .45in

{\em
$^a$Institute for Cosmic Ray Research,
University of Tokyo, Kashiwa 277-8582, Japan \vspace{0.2cm}\\
$^b$Institute for the Physics and Mathematics of the Universe,
University of Tokyo, Kashiwa 277-8568, Japan \vspace{0.2cm}\\
$^c$Department of Physics, University of Tokyo, Bunkyo-ku, Tokyo 113-0033, Japan \vspace{0.2cm}\\
$^d$Department of Physics and Astrophysics, Nagoya University, Nagoya 464-8602, Japan\\
}

\end{center}

\vskip .4in

\begin{abstract}

We study non-Gaussian properties of the isocurvature perturbations in the dark radiation, which
consists of the active neutrinos and extra light species, if exist. 
We first derive expressions for the bispectra of primordial perturbations
which are mixtures of curvature and dark radiation isocurvature perturbations.
We also discuss CMB bispectra produced in our model
and forecast CMB constraints on the non-linearity parameters based on the Fisher matrix analysis. 
Some concrete particle physics motivated models are presented
in which large isocurvature perturbations in extra light species and/or the 
neutrino density isocurvature perturbations as well as their non-Gaussianities may be generated.
Thus detections of non-Gaussianity in the dark radiation isocurvature perturbation
will give us an opportunity to identify the origin of extra light species and lepton asymmetry.

\end{abstract}

\end{titlepage}

\setcounter{page}{1}

\section{Introduction} 
\label{sec:introduction}

Recently, several cosmological observations
independently suggest that the effective number  
of neutrino species in the Universe is 
larger than the standard value, i.e. $\Delta N_{\rm eff}\equiv N_{\rm eff}-3.04\simeq 1$. 
According to recent observations of the primordial abundances of light 
elements, it is constrained as $N_{\rm eff}=3.68^{+0.80}_{-0.70}$
at 2$\sigma$ level (with slight dependence on the center value on the 
measured neutron lifetime)~\cite{Izotov:2010ca}.
On the other hand, recent observations of the cosmic microwave background (CMB) anisotropy
at small scales in combination with WMAP~\cite{Komatsu:2010fb} and standard distance rulers
\cite{Percival:2009xn,Riess:2011yx,Riess:2009pu}, 
give $N_{\rm eff}=4.56\pm0.75$~\cite{Dunkley:2010ge} 
and $N_{\rm eff}=3.86\pm0.42$~\cite{Keisler:2011aw} at 1$\sigma$ level.
These results may be evidences for the existence of 
extra radiation component, other than the three species of active neutrinos, in the Universe.
See also Refs.~\cite{Hamann:2010bk,Hamann:2011ge} for limits on the mass of extra radiation component.
Motivated by these observations, models for explaining $\Delta N_{\rm eff}\simeq1$ were proposed
\cite{Ichikawa:2007jv,Jaeckel:2008fi,Nakayama:2010vs,Fischler:2010xz,Kawasaki:2011ym,Hall:2011zq,Hasenkamp:2011em,Kawasaki:2011rc,Marsh:2011bf,Menestrina:2011mz,Kobayashi:2011hp,Hooper:2011aj,Jeong:2012hp}.
The Planck and other projected CMB observations
will improve constraints on $N_{\rm eff}$ by an order of magnitude 
(see e.g. Refs.~\cite{Bashinsky:2003tk,Ichikawa:2008pz}), 
and $\Delta N_{\rm eff}\simeq1$ can be clearly tested in the near future.

Once it will be proven that extra radiation indeed exists, 
it is important to understand the origin of extra radiation in the early Universe.
In the previous work~\cite{Kawasaki:2011rc}, it is argued that 
one way to probe this is to see how they fluctuate at the cosmological scales.
Observationally, the extra radiation and neutrinos are not discriminable
and we call the mixed fluid of them as dark radiation (DR).
The DR can have isocurvature perturbations, depending on how they are 
produced in the inflationary Universe.
It was shown that the DR isocurvature perturbations
affect the CMB anisotropy and constraints on
the amplitude of isocurvature perturbations were derived using
recent CMB and other cosmological observations.
Future forecasts on the constraint were discussed in Ref.~\cite{DiValentino:2011sv}.

While perturbations are assumed to be Gaussian in the most part
of Ref.~\cite{Kawasaki:2011rc}, the possibility of large
non-Gaussianities in the DR isocurvature perturbations was also briefly pointed out. 
In this paper, we present detailed analysis on these non-Gaussianities.
Non-Gaussianities in the DR isocurvature modes would have rich information on the properties of DR.
We note that there are several studies on 
non-Gaussianities in the cold dark matter (CDM) and baryon isocurvature perturbations 
\cite{Kawasaki:2008sn,Kawasaki:2008jy,Langlois:2008vk,Kawasaki:2008pa,
Hikage:2008sk,Kawakami:2009iu,Hikage:2009rt,
Langlois:2010fe,Langlois:2011zz,Langlois:2011hn}.
However, this is the first paper that studies non-Gaussianities in the 
DR isocurvature perturbations, including
those in the neutrino density isocurvature perturbations.
In this paper we focus on the local type non-Gaussianities at bispectrum level. 

The paper is organized as follows: 
We first present bispectrum generated from mixtures of 
primordial DR isocurvature and adiabatic perturbations
in Section~\ref{sec:NG}. In Section~\ref{sec:CMB}, we apply these results 
to the CMB angular bispectrum and discuss how non-Gaussianities in 
DR isocurvature perturbations manifest in the CMB 
anisotropy. Then we discuss constraints on these non-Gaussianities
from CMB observations in Section~\ref{sec:Fisher}.
Here we forecast constraints from the Planck satellite and an ideal survey 
limited by the cosmic variance, based on the Fisher matrix analysis.
We mention some particle physics models which may lead to 
large isocurvature perturbations in the extra radiation and neutrinos
as well as non-Gaussianities in them in Section~\ref{sec:model}.
The final section is devoted to summary.





\section{Non-Gaussian curvature and isocurvature perturbations} 
\label{sec:NG}

In this section we derive formulae for the (non-Gaussian) curvature/isocurvature perturbations
based on the $\delta N$-formalism~\cite{Sasaki:1995aw,Lyth:2004gb}.
We mostly follow formalism in Ref.~\cite{Kawasaki:2011rc}.
Various modes of primordial perturbations, including the adiabatic mode 
$\zeta$ and some kind of isocurvature mode $S$, 
can be generated from fluctuations in scalar fields and given as
\begin{eqnarray}
\zeta&=&N_{\phi_i}\delta\phi_i+\frac{1}{2}N_{\phi_i\phi_j}\delta\phi_i\delta\phi_j+\dots,
\label{eq:primordial}\\
S&=&S_{\phi_i}\delta\phi_i+\frac{1}{2}S_{\phi_i\phi_j}\delta\phi_i\delta\phi_j+\dots\notag.
\end{eqnarray}
Here, $\delta \phi_i$ is quantum fluctuation of a scalar field $\phi_i$, 
whose mass is smaller than the Hubble parameter during inflation, $H_{\rm inf}$.
Hereafter, we concentrate on the isocurvature perturbation 
in the dark radiation (DR) denoted by $S_{\rm DR}$. 
We here again emphasize that the DR consists of both active neutrinos and extra light particle species.

We can express the power spectra
of the auto- and cross- correlation functions of 
$\zeta$ and $S_{\rm DR}$ as follows,
\begin{equation}
\begin{split}
	\langle \zeta(\vec k_1)\zeta(\vec k_2) \rangle 
		& \equiv (2\pi)^3\delta(\vec k_1+\vec k_2) P^{\zeta\zeta}(k_1), \\
	\langle \zeta(\vec k_1) S_{\rm DR}(\vec k_2) \rangle 
		& \equiv (2\pi)^3\delta(\vec k_1+\vec k_2) P^{\zeta S_{\rm DR}}(k_1), \\
	\langle  S_{\rm DR}(\vec k_1) S_{\rm DR}(\vec k_2) \rangle 
		& \equiv (2\pi)^3\delta(\vec k_1+\vec k_2) P^{S_{\rm DR}S_{\rm DR}}(k_1).
	\label{Pzeta}
\end{split}
\end{equation}
where
\begin{equation}
\begin{split}
	P^{\zeta\zeta} (k) &= N_{\phi_i}^2P_{\delta \phi}(k), \\
	P^{\zeta S_{\rm DR}} (k) &=  N_{\phi_i} S_{\phi_i} P_{\delta \phi}(k), \\
	P^{S_{\rm DR}S_{\rm DR}} (k) &= S_{\phi_i}^2 P_{\delta \phi}(k),     \label{power}
\end{split}
\end{equation}
Here, we have neglected higher order terms and $P_{\delta \phi}(k)$ is the power spectrum of the fluctuations of the scalar fields,
\begin{gather}
	\langle  \delta \phi_i(\vec k_1) \delta \phi_j(\vec k_2) \rangle 
	\equiv (2\pi)^3\delta(\vec k_1+\vec k_2) P_{\delta \phi}(k_1)\delta_{ij}, \\
	P_{\delta \phi}(k) = \frac{H_{\rm inf}^2}{2k^3}\left( \frac{k}{k_0} \right)^{n_s-1},
\end{gather}
where $n_s$ is the scalar spectral index\footnote{
	The scalar spectral indices for $\phi$ and $\sigma$ do not coincide in general.
	In the following we assume they are the same just for simplicity.
} 
and $k_0$ is the pivot scale chosen as $k_0=0.002{\rm Mpc}^{-1}$.
Note that the above power spectra and correlation have same spectral shape up to this order.
We therefore adopt $P_{\zeta}\equiv P^{\zeta\zeta}$ as the normalization of power spectra
and other power spectra can be expressed in the form $P_{\zeta}$ times some constants 
hereafter.

The bispectra of $\zeta$ and $S_{\rm DR}$ are defined by the following equations,
\begin{equation}
\begin{split}
	\langle \zeta(\vec k_1)\zeta(\vec k_2)\zeta(\vec k_3) \rangle 
		& \equiv (2\pi)^3\delta(\vec k_1+\vec k_2 + \vec k_3) B_{\zeta\zeta\zeta}(k_1,k_2,k_3), \\
	\langle \zeta(\vec k_1)\zeta(\vec k_2) S_{\rm DR}(\vec k_3) \rangle
		& \equiv (2\pi)^3\delta(\vec k_1+\vec k_2 + \vec k_3) B_{\zeta\zeta S}(k_1,k_2,k_3), \\
	\langle \zeta(\vec k_1) S_{\rm DR}(\vec k_2) S_{\rm DR}(\vec k_3) \rangle
		& \equiv (2\pi)^3\delta(\vec k_1+\vec k_2 + \vec k_3) B_{\zeta SS}(k_1,k_2,k_3), \\
	\langle  S_{\rm DR}(\vec k_1) S_{\rm DR}(\vec k_2) S_{\rm DR}(\vec k_3) \rangle 
		& \equiv (2\pi)^3\delta(\vec k_1+\vec k_2 + \vec k_3) B_{SSS}(k_1,k_2,k_3), 
\end{split}
\end{equation}
so that the primordial bispectrum can be written in the form of 
\begin{eqnarray}
B^{A_1A_2A_3}(k_1,k_2,k_3)=
f_{\rm NL}^{A_1,A_2A_3}(k_1,k_2,k_3)P_\zeta(k_2)P_\zeta(k_3)+
\mbox{(2 cyclics  of \{123\})},
\label{eq:localB}
\end{eqnarray}
where the each of the subscript $A_i$ ($i=1,\,2,\,3$) is either $\zeta$ or $S_{\rm DR}$. 
The coefficients $f_{\rm NL}^{A_1,A_2A_3}$ represent magnitudes of non-Gaussianities and in the following we call them non-Gaussianity parameters.
Note that our definition of $f^{A_1,A_2A_3}_{\rm NL}$ 
is consistent with Ref. \cite{Langlois:2011hn} besides difference in types
of isocurvature perturbations considered.
We also note that if there is only a single scalar field
which sources primordial perturbations
and there are only adiabatic perturbations, 
$f^{\zeta,\zeta\zeta}_{\rm NL}$ is related to
the ordinary non-Gaussianity parameter
$f_{\rm NL}$ via
\begin{equation}
f_{\rm NL}^{\zeta, \zeta\zeta}=\frac{6}{5}f_{\rm NL}.
\end{equation}
By using the expansion (\ref{eq:primordial}), we can explicitly write the non-Gaussianity parameters as
\begin{equation}
\begin{split}
	f_{\rm NL}^{\zeta,\zeta\zeta} &= \frac{N_{\phi_i}N_{\phi_j}N_{\phi_i\phi_j}}{(N_{\phi_i}^2)^2}
	+\frac{N_{\phi_i \phi_j}N_{\phi_j \phi_k}N_{\phi_k\phi_i} }{(N_{\phi_i}^2)^3}\Delta_{\zeta}^2 \ln(k_b L),\\
	f_{\rm NL}^{S_{\rm DR},\zeta\zeta} &= \frac{S_{\phi_i}N_{\phi_j}N_{\phi_i\phi_j}}{(N_{\phi_i}^2)^2}
	+\frac{S_{\phi_i \phi_j}N_{\phi_j \phi_k}N_{\phi_k\phi_i} }{(N_{\phi_i}^2)^3}\Delta_{\zeta}^2 \ln(k_b L),\\
	f_{\rm NL}^{\zeta,S_{\rm DR}\zeta} =f_{\rm NL}^{\zeta,\zeta S_{\rm DR}}
	&= \frac{N_{\phi_i}N_{\phi_j}S_{\phi_i\phi_j}}{(N_{\phi_i}^2)^2}
	+\frac{N_{\phi_i \phi_j}N_{\phi_j \phi_k}S_{\phi_k\phi_i} }{(N_{\phi_i}^2)^3}\Delta_{\zeta}^2 \ln(k_b L),\\
	f_{\rm NL}^{\zeta,S_{\rm DR}S_{\rm DR}} &= \frac{N_{\phi_i}S_{\phi_j}S_{\phi_i\phi_j}}{(N_{\phi_i}^2)^2}
	+\frac{N_{\phi_i \phi_j}S_{\phi_j \phi_k}S_{\phi_k\phi_i} }{(N_{\phi_i}^2)^3}\Delta_{\zeta}^2 \ln(k_b L),\\
	f_{\rm NL}^{S_{\rm DR},\zeta S_{\rm DR}} =f_{\rm NL}^{S_{\rm DR},S_{\rm DR}\zeta}
	&= \frac{S_{\phi_i}S_{\phi_j}N_{\phi_i\phi_j}}{(N_{\phi_i}^2)^2}
	+\frac{S_{\phi_i \phi_j}S_{\phi_j \phi_k}N_{\phi_k\phi_i} }{(N_{\phi_i}^2)^3}\Delta_{\zeta}^2 \ln(k_b L),\\
	f_{\rm NL}^{S_{\rm DR},S_{\rm DR}S_{\rm DR}} &= \frac{S_{\phi_i}S_{\phi_j}S_{\phi_i\phi_j}}{(N_{\phi_i}^2)^2}
	+\frac{S_{\phi_i \phi_j}S_{\phi_j \phi_k}S_{\phi_k\phi_i} }{(N_{\phi_i}^2)^3}\Delta_{\zeta}^2 \ln(k_b L),
	\label{fNLs}
\end{split}
\end{equation}
where $\Delta_\zeta^2 \equiv (k^3/2\pi^2) P_\zeta(k)$ is the dimensionless power spectrum of the
curvature perturbation, $k_b\equiv {\rm min} \{ k_1,k_2,k_3 \}$ and $L$ is the infrared cutoff scale~\cite{Boubekeur:2005fj,Lyth:2006gd},
which should be set to be a scale comparable to the present horizon scale.
\footnote{
The second term in the RHS of each equation of (\ref{fNLs}) arises from the product of three quadratic terms of $\delta \phi_i$ in $\zeta$ or $S_{\rm DR}$.
We refer to \cite{Hikage:2008sk} for the detail of the derivation of it.
}

\section{CMB bispectrum} 
\label{sec:CMB}

CMB bispectrum from non-Gaussian curvature and 
extra radiation-isocurvature perturbations are to be 
discussed. For CDM isocurvature perturbation, 
similar analysis is done in 
Refs.~\cite{Kawasaki:2008sn,Kawasaki:2008pa,
Hikage:2008sk,Hikage:2009rt,Langlois:2011hn},
which extend the analysis of Ref.~\cite{Komatsu:2001rj} to isocurvature perturbations. 
We here consider the isocurvature perturbations in extra or dark radiation.
We include both the temperature and E-polarization CMB anisotropies.

First, we denote primordial perturbations by $X^A$, 
where the subscript $A$ is either $\zeta$ or $S_{\rm DR}$. 
CMB anisotropy is given by
\begin{equation}
a^P_{lm}=
4\pi(-i)^l \sum_A\int \frac{d^3k}{(2\pi)^3}
g^{AP}_l(k)Y^*_{lm}(\hat k)X^A_{\vec k},
\label{eq:alm}
\end{equation}
where the subscript $P$ represents the type of 
CMB anisotropy and should be either T or E, and
$g^{AP}_l(k)$ is the transfer function at linear order.

First the power spectrum of the CMB anisotropy is expressed as
\begin{equation}
	C_l ^{P_1 P_2} \delta_{l l'}\delta_{mm'} \equiv \langle a_{lm}^{P_1} a_{l' m'}^{* P_2} \rangle.
\end{equation}
It is given in terms of the primordial perturbations as
\begin{equation}
	C_{l}^{P_1 P_2} = \frac{2}{\pi}\sum_{A_1 A_2}\int k^2 dk g_l^{A_1 P_1}(k) g_{l'}^{*A_2 P_2}(k) P^{A_1 A_2}(k).
\end{equation}
where $P^{A_1A_2}(k)$ is the
power spectrum of $X^A$ in the wave number space defined in Eq.~(\ref{Pzeta}),
which is conveniently written as
\begin{equation}
\langle
X^{A_1}(\vec k_1)
X^{* A_2}(\vec k_2)
\rangle
\equiv
P^{A_1A_2}(k_1)(2\pi)^3
\delta^{(3)}(\vec k_1-\vec k_2).
\end{equation}

Let us now consider the bisepctrum of 
CMB anisotropy in the harmonic space, 
\begin{equation}
B^{P_1P_2P_3}_{l_1m_1l_2m_2l_3m_3}
\equiv \langle a^{P_1}_{l_1m_1}
a^{P_2}_{l_2m_2}
a^{P_3}_{l_3m_3}
\rangle.
\label{eq:3point}
\end{equation}
Using Eq.~\eqref{eq:alm}, 
$B^{P_1P_2P_3}_{l_1m_1l_2m_2l_3m_3}$
can be written as
\begin{eqnarray}
B^{P_1P_2P_3}_{l_1m_1l_2m_2l_3m_3}
&=&\sum_{A_1A_2A_3}
\prod^3_{i=1}\left[
4\pi(-i)^{l_i}\int\frac{d^3k_i}{(2\pi)^3}
g^{A_iP_i}_{l_i}(k_i)Y^*_{l_im_i}(\hat k_i)
\right] \notag \\
&&\quad\times
B^{A_1A_2A_3}(k_1,k_2,k_3)(2\pi)^3
\delta^{(3)}(\vec k_1+\vec k_2+\vec k_3), 
\label{eq:Blll}
\end{eqnarray}
where $B^{A_1A_2A_3}(k_1,k_2,k_3)$ is the
bispectrum of $X^A$ in the wave number space defined in Eq.~(\ref{eq:localB}),
which are conveniently written as
\begin{equation}
\langle
X^{A_1}(\vec k_1)
X^{A_2}(\vec k_2)
X^{A_3}(\vec k_3)
\rangle
\equiv
B^{A_1A_2A_3}(k_1,k_2,k_3)(2\pi)^3
\delta^{(3)}(\vec k_1+\vec k_2+\vec k_3).
\end{equation}
Due to the statistical isotropy assumed here,
$B^{A_1A_2A_3}$ is independent of 
$\hat k_i$.

Using the following formulae
\begin{eqnarray}
(2\pi)^3\delta^{(3)}(\vec k_1+\vec k_2+\vec k_3)
&=&\int d^3r e^{i(\vec k_1+\vec k_2+\vec k_3)\cdot \vec r},\\
\int d\hat k Y^*_{lm}(\hat k)
e^{i\vec k\cdot \vec r}
&=&4\pi i^l j_l(kr)Y^*_{lm}(\hat r),
\end{eqnarray}
Eqs.~\eqref{eq:Blll} can be reduced into
\begin{eqnarray}
B^{P_1P_2P_3}_{l_1m_1l_2m_2l_3m_3}
&=&\sum_{A_1A_2A_3}
\int r^2dr
\prod_{i=1}^3\left[\frac{2}{\pi}\int k_i^2dk_i
\,g^{A_iP_i}_{l_i}(k_i)j_{l_i}(k_ir)
\right] \notag\\
&&\quad\times
B^{A_1A_2A_3}(k_1,k_2,k_3)
\int d\hat r\prod_{i=1}^3\left[Y_{l_im_i}^*(\hat r)\right].
\label{eq:Blll2}
\end{eqnarray}

In Eq.~\eqref{eq:Blll2}, we can factor out
the Gaunt integral, 
\begin{equation}
\mathcal G^{m_1m_2m_3}_{l_1l_2l_3}=
\int d\hat r
\prod^3_{i=1}\left[Y^*_{l_im_i}(\hat r)\right]
\end{equation}
which manifests the statistical isotropy. 
In terms of the Wigner-3j symbol, 
$\mathcal G^{m_1m_2m_3}_{l_1l_2l_3}$
can be rewritten as 
\begin{equation}
\mathcal G^{m_1m_2m_3}_{l_1l_2l_3}
=\sqrt{\frac{(2l_1+1)(2l_2+1)(2l_3+1)}{4\pi}}
\begin{pmatrix}
l_1 & l_2 & l_3 \\
0 & 0 & 0 
\end{pmatrix}
\begin{pmatrix}
l_1 & l_2 & l_3 \\
m_1 & m_2 & m_3 \\
\end{pmatrix}.
\end{equation}
Then we obtain
\begin{equation}
B^{P_1P_2P_3}_{l_1m_1,l_2m_2,l_3m_3}
=\mathcal G^{m_1m_2m_3}_{l_1l_2l_3}
b^{P_1P_2P_3}_{l_1l_2l_3},
\end{equation}
where $b^{P_1P_2P_3}_{l_1l_2l_3}$
is the reduced bispectrum given by
\begin{eqnarray}
b^{P_1P_2P_3}_{l_1l_2l_3}
=\sum_{A_1A_2A_3}
\int r^2dr
\prod_{i=1}^3\left[\frac{2}{\pi}\int k_i^2dk_i
\,g^{A_iP_i}_{l_i}(k_i)j_{l_i}(k_ir)
\right]B^{A_1A_2A_3}(k_1,k_2,k_3)
\label{eq:redb}
\end{eqnarray}
This is the most general expression of
CMB bispectrum, in the presence of non-adiabatic 
primordial scalar perturbations. This is applicable to 
any types of primordial non-Gaussianities.

Now we focus on primordial perturbations with local-type non-Gaussianity, 
which can be written in the form of Eq.~\eqref{eq:primordial}.

Given the primordial bispectrum of Eq.~\eqref{eq:localB},
the reduced CMB bispectrum
$b^{P_1P_2P_3}_{l_1l_2l_3}$
can be written as
\begin{equation}
b^{P_1P_2P_3}_{l_1l_2l_3}
=\sum_{A_1A_2A_3}\left[f^{A_1,A_2A_3}_{\rm NL}
b^{A_1P_1, A_2P_2A_3P_3}_{l_1l_2l_3}
+\mbox{(2 cyclics  of \{123\})}\right].
\label{eq:redb2}
\end{equation}
Here, $b^{A_1P_1, A_2P_2A_3P_3}_{l_1l_2l_3}$
is defined as 
\begin{equation}
b^{A_1P_1, A_2P_2A_3P_3}_{l_1l_2l_3}
\equiv
\int r^2dr
\alpha^{A_1P_1}_{l_1}(r)
\beta^{A_2P_2}_{l_2}(r)
\beta^{A_3P_3}_{l_3}(r)
\end{equation}
where
\begin{eqnarray}
\alpha^{AP}_l(r)&\equiv&
\frac{2}{\pi}\int k^2dk g^{AP}_l(k)j_l(kr), \\
\beta^{AP}_l(r)&\equiv&
\frac{2}{\pi}\int k^2dk P_\zeta(k)g^{AP}_l(k)j_l(kr).   \label{betal}
\end{eqnarray}

As long as only two kinds of initial perturbations $\zeta$ and $S_\mathrm{DR}$ 
are included, there are only six independent non-Gaussian parameters \cite{Langlois:2011hn}.
For latter convenience, we denote them by $f_{\rm NL}^{(i)}$ ($a=1,\dots,6$), which is defined as
\begin{eqnarray}
f^{(1)}_{\rm NL}\equiv f^{\zeta,\zeta\zeta}_{\rm NL}, &
f^{(2)}_{\rm NL}\equiv f^{S_\mathrm{DR},\zeta\zeta}_{\rm NL}, &
f^{(3)}_{\rm NL}\equiv f^{\zeta,\zeta S_\mathrm{DR}}_{\rm NL}=
f^{\zeta,S_\mathrm{DR}\zeta}_{\rm NL}, \\
f^{(4)}_{\rm NL}\equiv f^{\zeta,S_\mathrm{DR}S_\mathrm{DR}}_{\rm NL}, &
f^{(5)}_{\rm NL}\equiv f^{S_\mathrm{DR},\zeta S_\mathrm{DR}}_{\rm NL}=
f^{S_\mathrm{DR},S_\mathrm{DR}\zeta }_{\rm NL}, &
f^{(6)}_{\rm NL}\equiv f^{S_\mathrm{DR},S_\mathrm{DR}S_\mathrm{DR}}_{\rm NL}. 
\end{eqnarray}
We also pile up $b^{A_1P_1,A_2P_2A_3P_3}_{l_1l_2l_3}$ into
six types of reduced bispectra $b^{(a)~P_1P_2P_3}_{l_1l_2l_3}$: 
\begin{eqnarray}
b^{(1)~P_1P_2P_3}_{l_1l_2l_3}&\equiv&
b^{\zeta P_1,\zeta P_2\zeta P_3}_{l_1l_2l_3}
+\mbox{(2 cyclics  of \{123\})}, \\
b^{(2)~P_1P_2P_3}_{l_1l_2l_3}&\equiv&
b^{\bar S_\mathrm{DR} P_1,\zeta P_2\zeta P_3}_{l_1l_2l_3}
+\mbox{(2 cyclics  of \{123\})}, \\
b^{(3)~P_1P_2P_3}_{l_1l_2l_3}&\equiv&
\left[b^{\bar\zeta P_1,S_\mathrm{DR} P_2\zeta P_3}_{l_1l_2l_3}+
b^{\bar\zeta P_1,\zeta P_2S_\mathrm{DR} P_3}_{l_1l_2l_3}\right]
+\mbox{(2 cyclics  of \{123\})}, \\
b^{(4)~P_1P_2P_3}_{l_1l_2l_3}&\equiv&
b^{\zeta P_1,S_\mathrm{DR} P_2S_\mathrm{DR} P_3}_{l_1l_2l_3}+
\mbox{(2 cyclics  of \{123\})}, \\
b^{(5)~P_1P_2P_3}_{l_1l_2l_3}&\equiv&
\left[b^{S_\mathrm{DR}P_1,\zeta P_2S_\mathrm{DR} P_3}_{l_1l_2l_3}
+b^{S_\mathrm{DR} P_1,\bar S_\mathrm{DR} P_2\zeta P_3}_{l_1l_2l_3}\right]
+\mbox{(2 cyclics  of \{123\})}, \\
b^{(6)~P_1P_2P_3}_{l_1l_2l_3}&\equiv&
b^{\bar S_\mathrm{DR} P_1S_\mathrm{DR} P_2S_\mathrm{DR} P_3}_{l_1l_2l_3}
+\mbox{(2 cyclics  of \{123\})}
\end{eqnarray}
Then the total CMB bispectrum in Eq.~\eqref{eq:redb2} 
can be rewritten as
\begin{equation}
b^{P_1P_2P_3}_{l_1l_2l_3}=
\sum^{6}_{a=1}
f^{(a)}_{\rm NL}b^{(a)~P_1P_2P_3}_{l_1l_2l_3}.
\end{equation}

\begin{figure}
  \begin{center}
    \begin{tabular}{ccc}
      \hspace{-10mm}
      \resizebox{60mm}{!}{\includegraphics{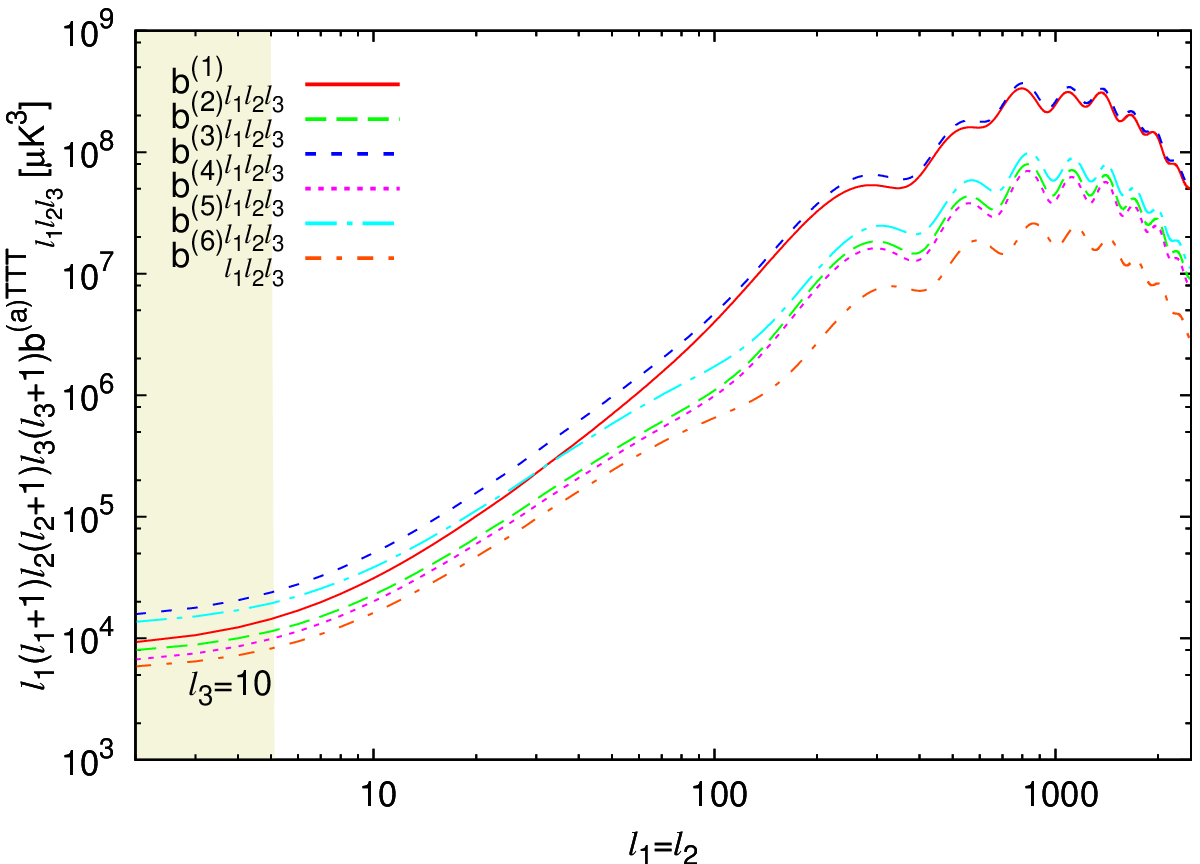}} &
      \hspace{-5mm}
      \resizebox{60mm}{!}{\includegraphics{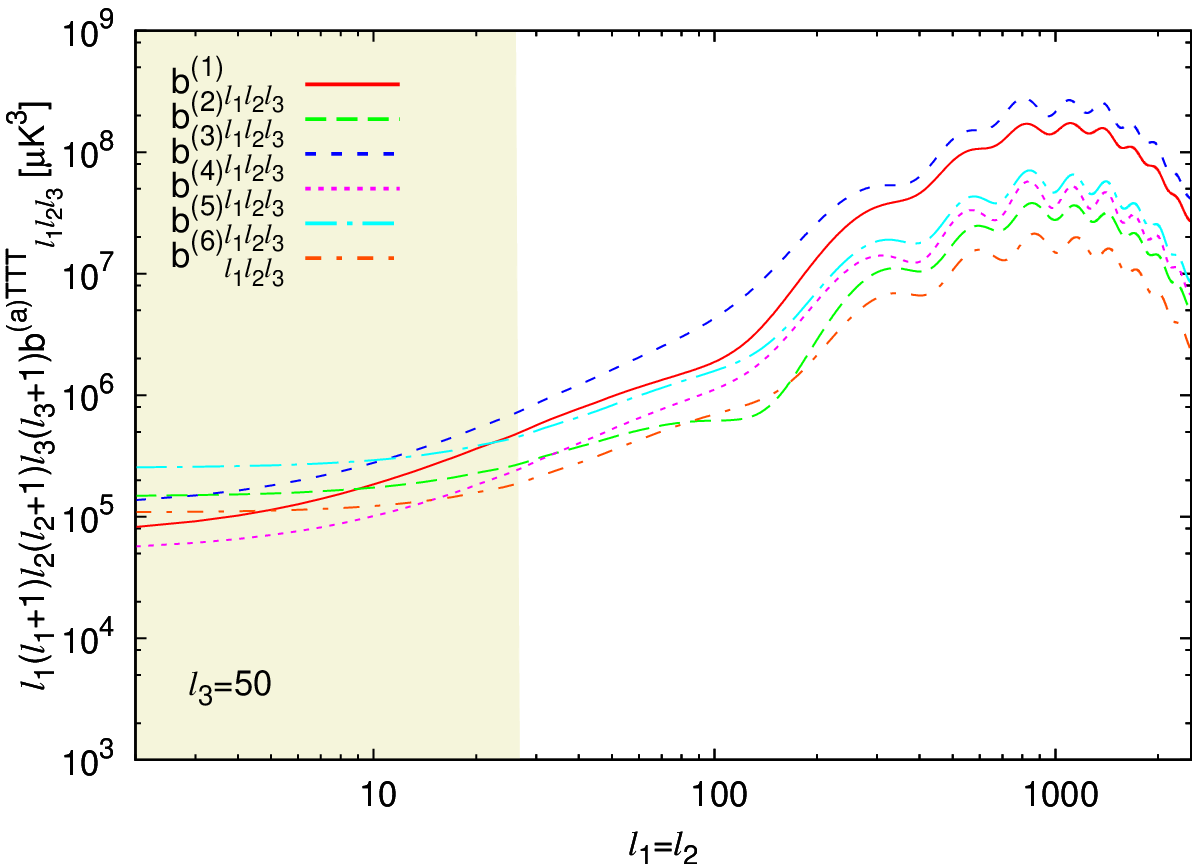}} &
      \hspace{-5mm}
      \resizebox{60mm}{!}{\includegraphics{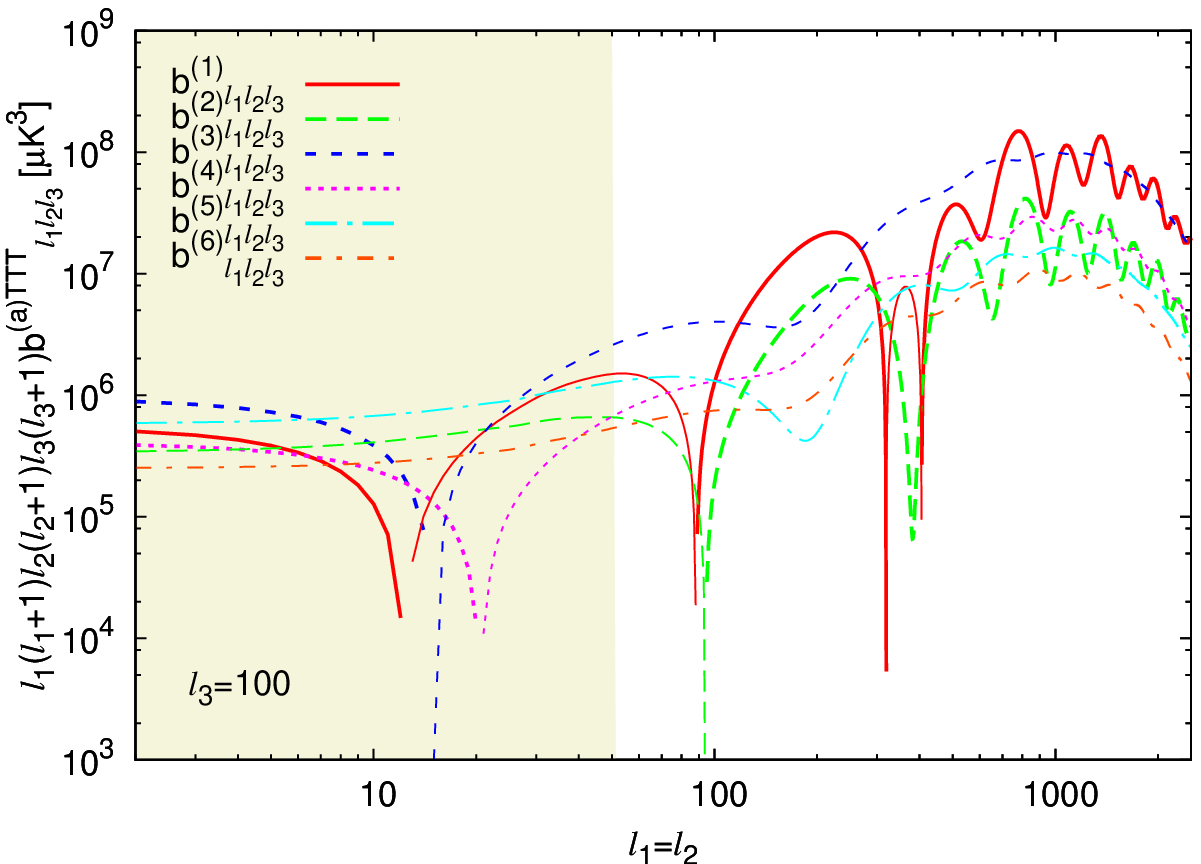}} \\
      \hspace{-10mm}
      \resizebox{60mm}{!}{\includegraphics{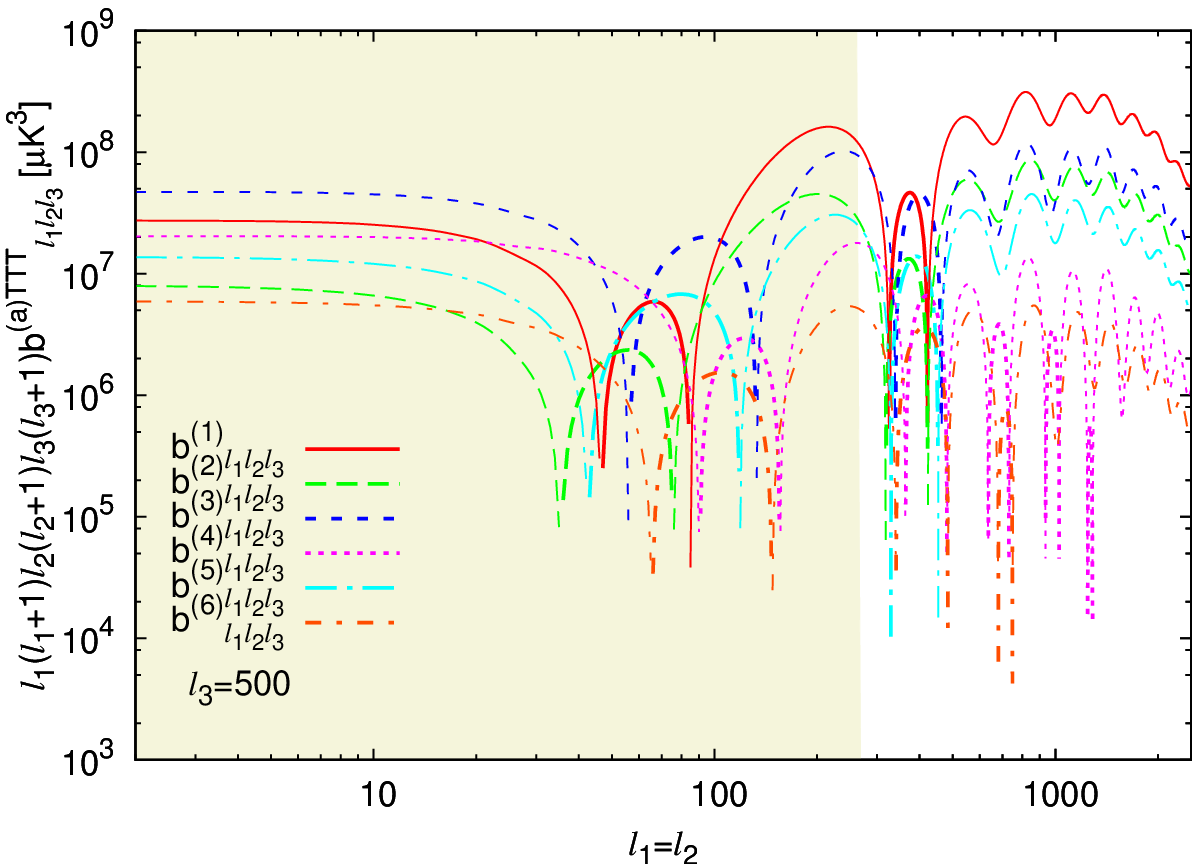}} &
      \hspace{-5mm}
      \resizebox{60mm}{!}{\includegraphics{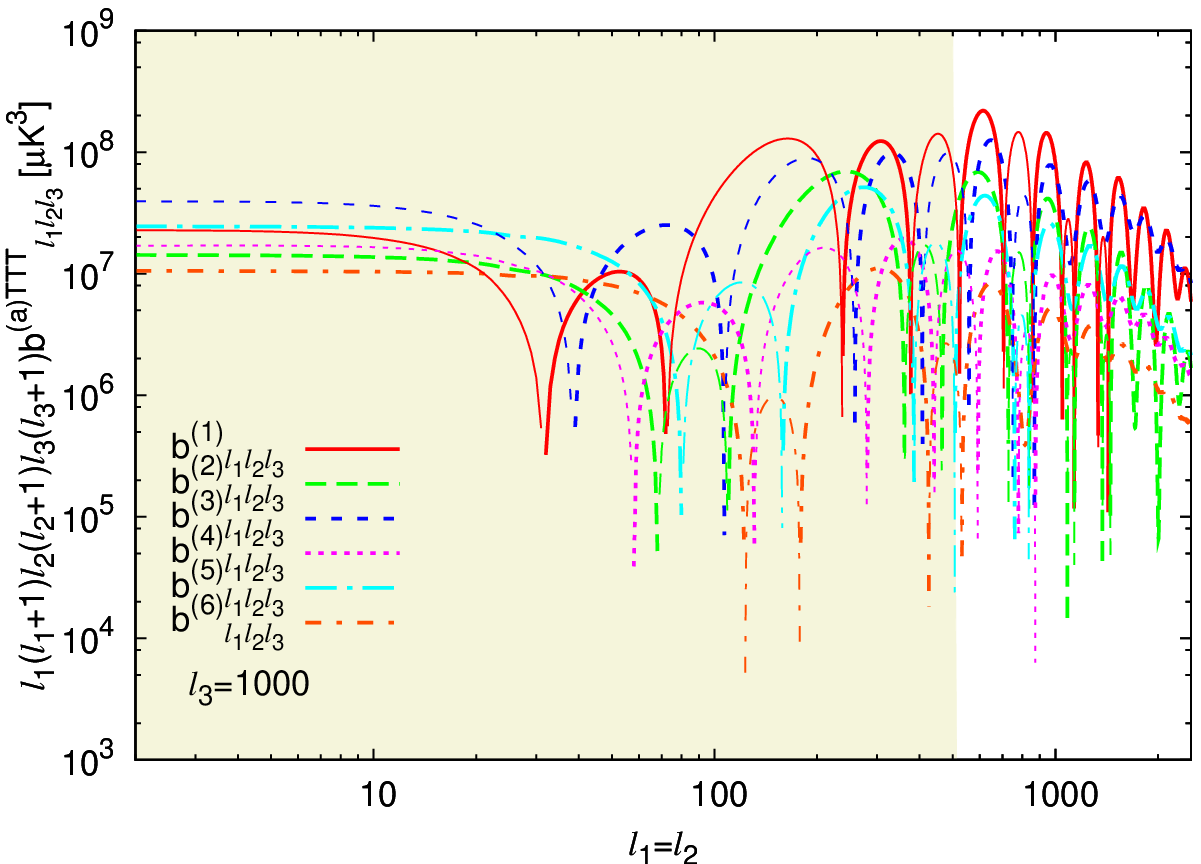}} &
      \hspace{-5mm}
      \resizebox{60mm}{!}{\includegraphics{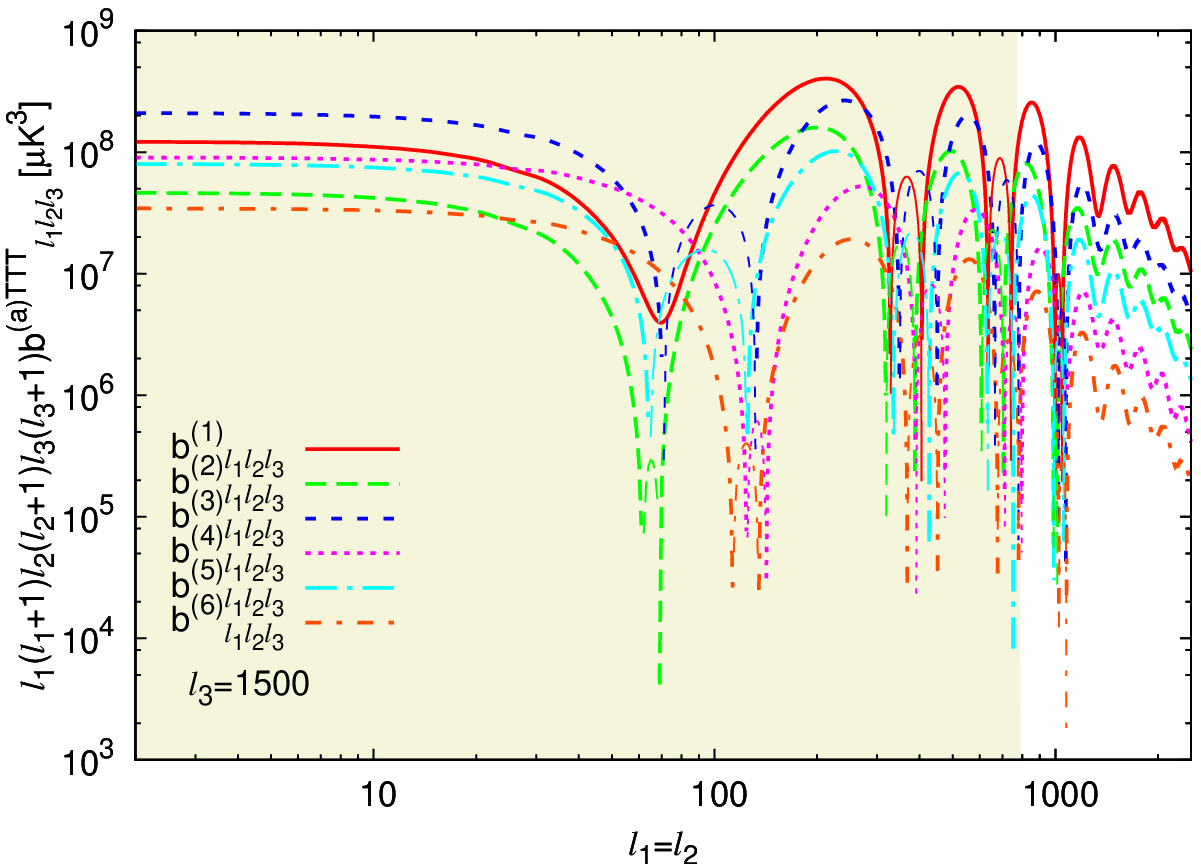}} \\
    \end{tabular}
  \end{center}
  \caption{Shown are the temperature bispectra $b^{(a)TTT}_{l_1l_2l_3}$.
  $b^{(1)TTT}_{l_1l_2l_3}$ (solid red), $b^{(2)TTT}_{l_1l_2l_3}$ (short-dashed green), 
  $b^{(3)TTT}_{l_1l_2l_3}$ (dotted blue), $b^{(4)TTT}_{l_1l_2l_3}$ (dot-dashed), 
  $b^{(5)TTT}_{l_1l_2l_3}$ (long-dashed), $b^{(6)TTT}_{l_1l_2l_3}$ (dot-dot-dashed) 
  in isosceles triangular configurations with $l_1=l_2$ are plotted as function 
  of $l_1$ with fixed $l_3$. $l_3$ is set to 10 (top left), 50 (top middle), 
  100 (top right),  500 (bottom left),  1000 (bottom middle),  1500 (bottom right).
  In each panel, the shaded region at low multipoles shows configurations which
  fail to satisfy the triangular condition i.e. $|l_1-l_2|\ge l_3\ge l_1+l_2$.
  }
  \label{fig:TTT}
\end{figure}

\begin{figure}
  \begin{center}
    \begin{tabular}{ccc}
      \hspace{-10mm}
      \resizebox{60mm}{!}{\includegraphics{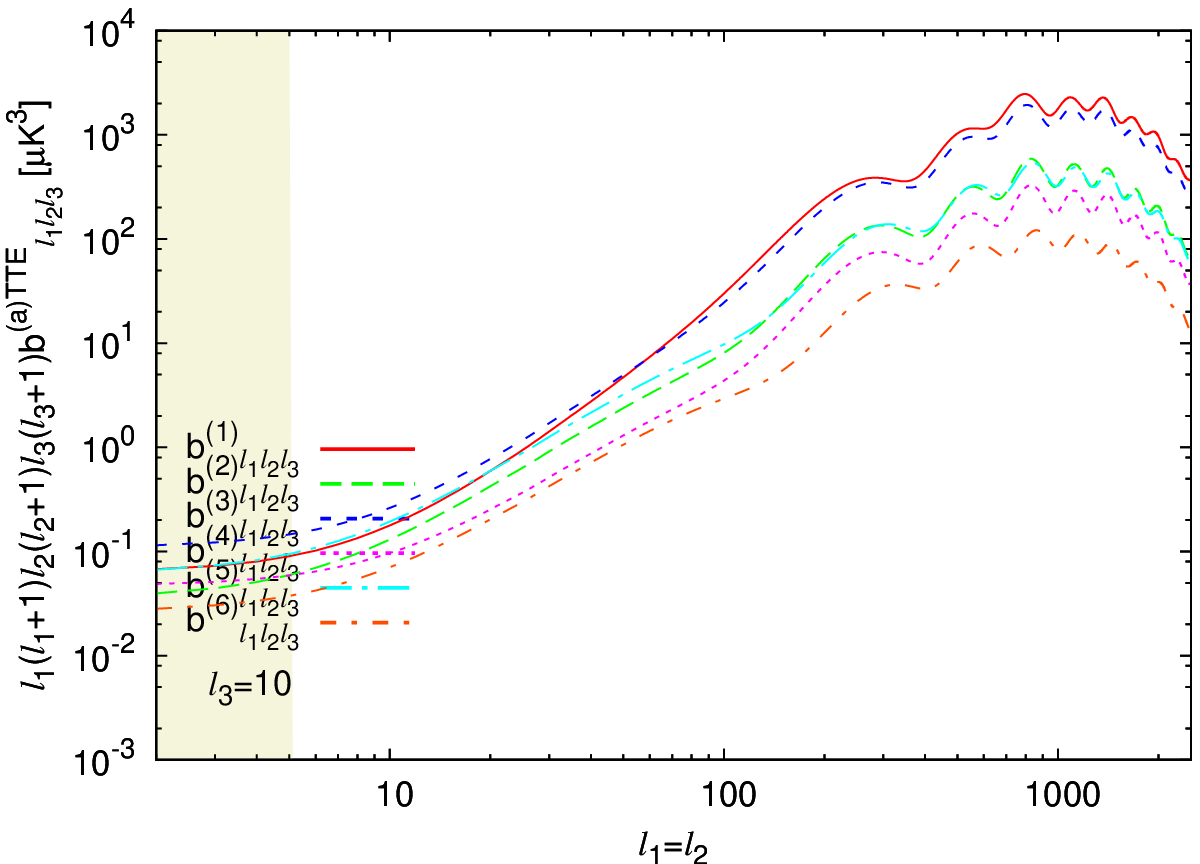}} &
      \hspace{-5mm}
      \resizebox{60mm}{!}{\includegraphics{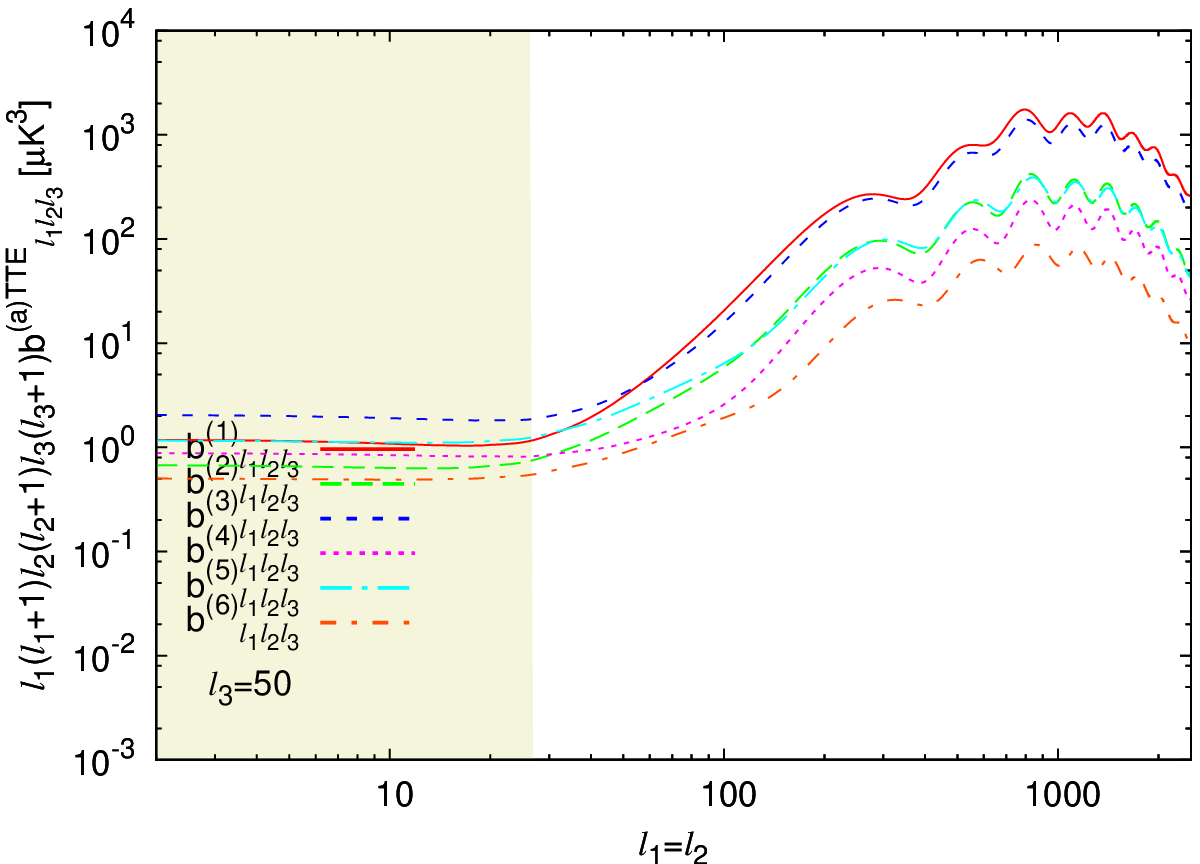}} &
      \hspace{-5mm}
      \resizebox{60mm}{!}{\includegraphics{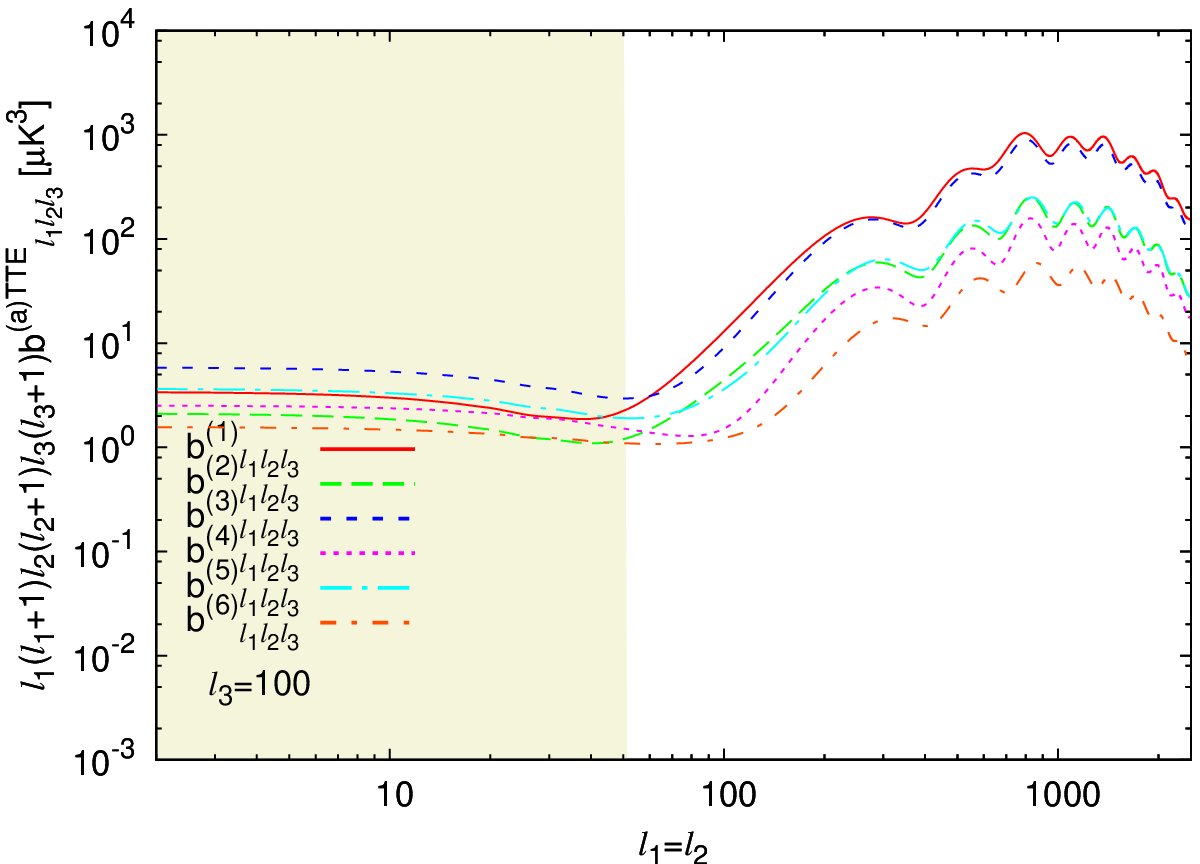}} \\
      \hspace{-10mm}
      \resizebox{60mm}{!}{\includegraphics{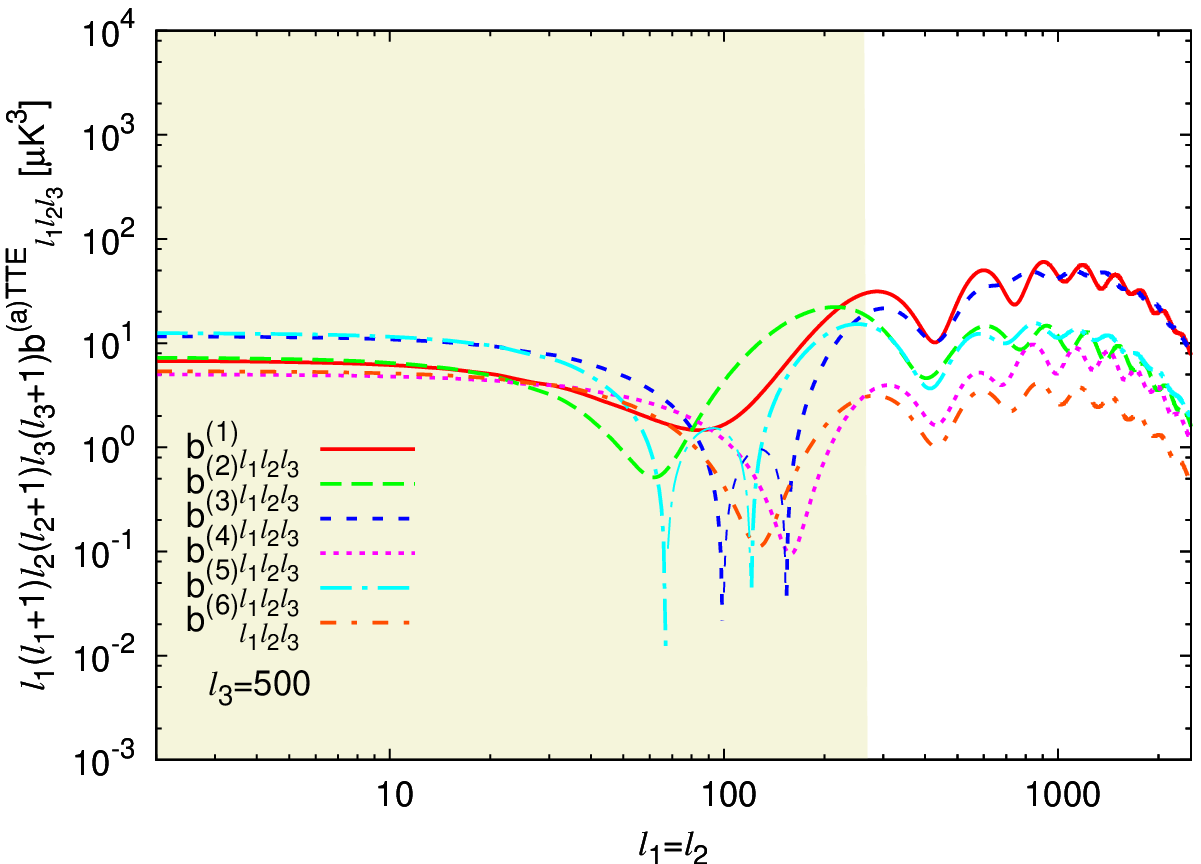}} &
      \hspace{-5mm}
      \resizebox{60mm}{!}{\includegraphics{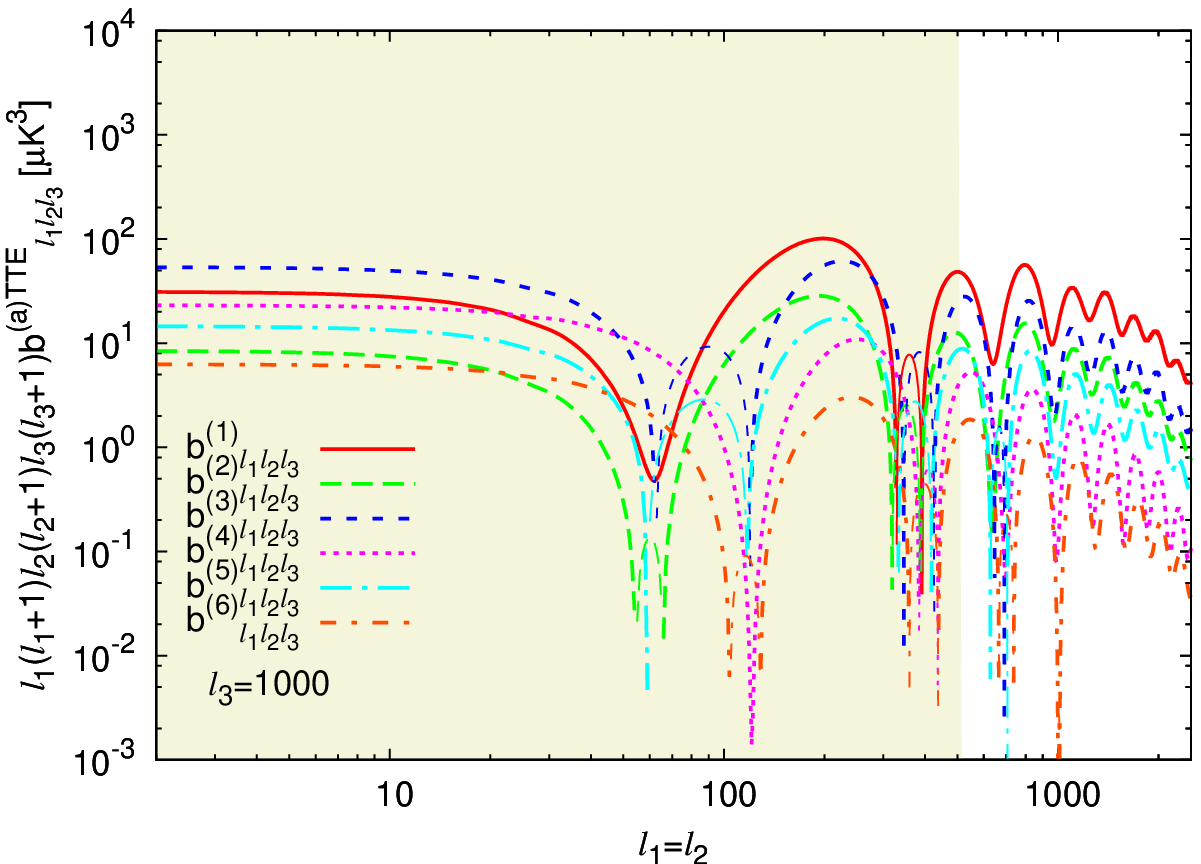}} &
      \hspace{-5mm}
      \resizebox{60mm}{!}{\includegraphics{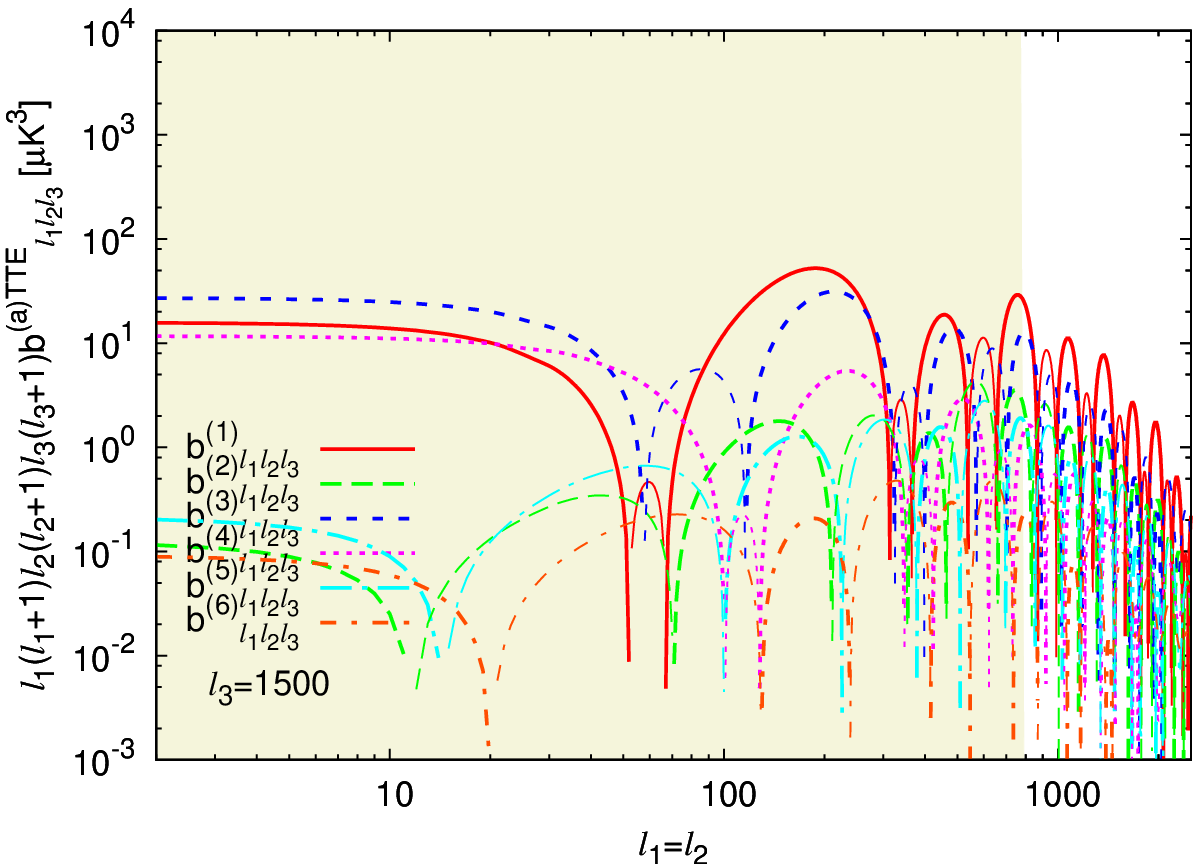}} \\
    \end{tabular}
  \end{center}
  \caption{Same figure as in Fig. \ref{fig:TTT} but for $b^{(a)TTE}_{l_1l_2l_3}$.}
  \label{fig:TTE}
\end{figure}

\begin{figure}
  \begin{center}
    \begin{tabular}{ccc}
      \hspace{-10mm}
      \resizebox{60mm}{!}{\includegraphics{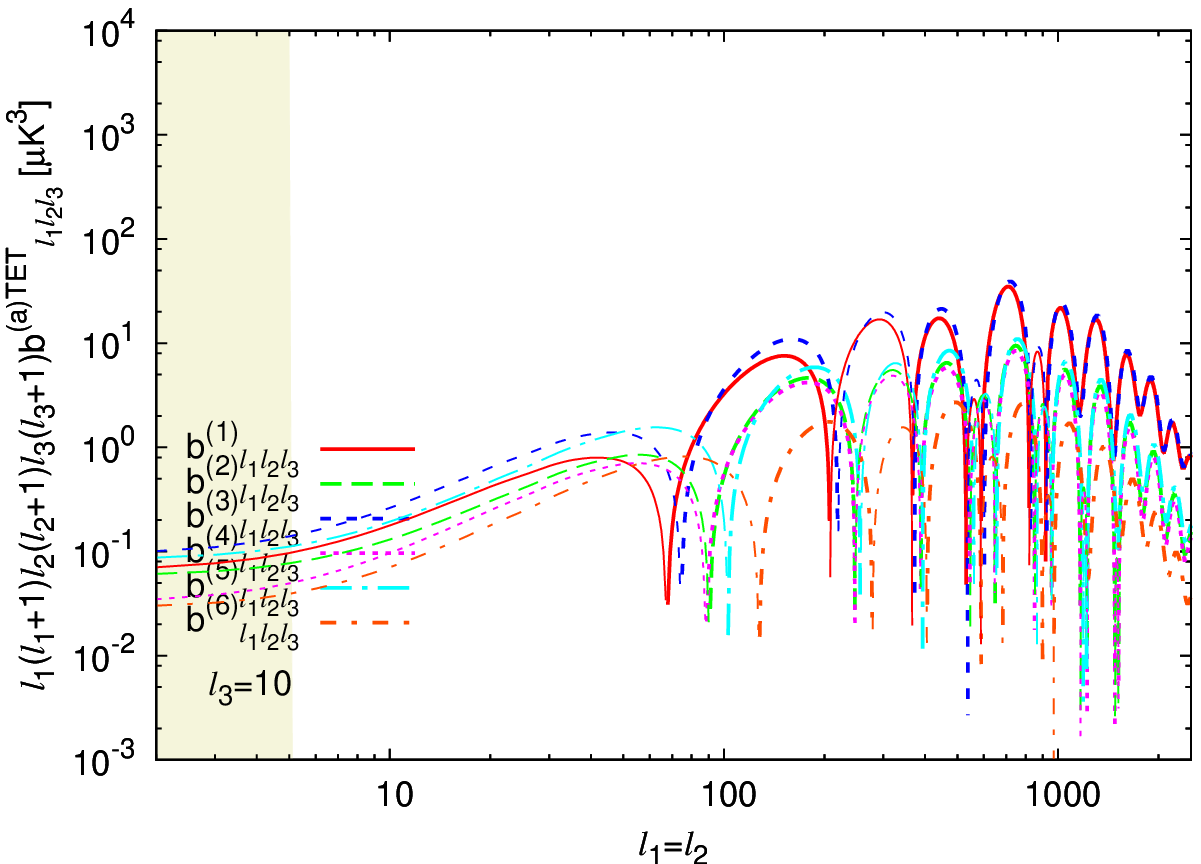}} &
      \hspace{-5mm}
      \resizebox{60mm}{!}{\includegraphics{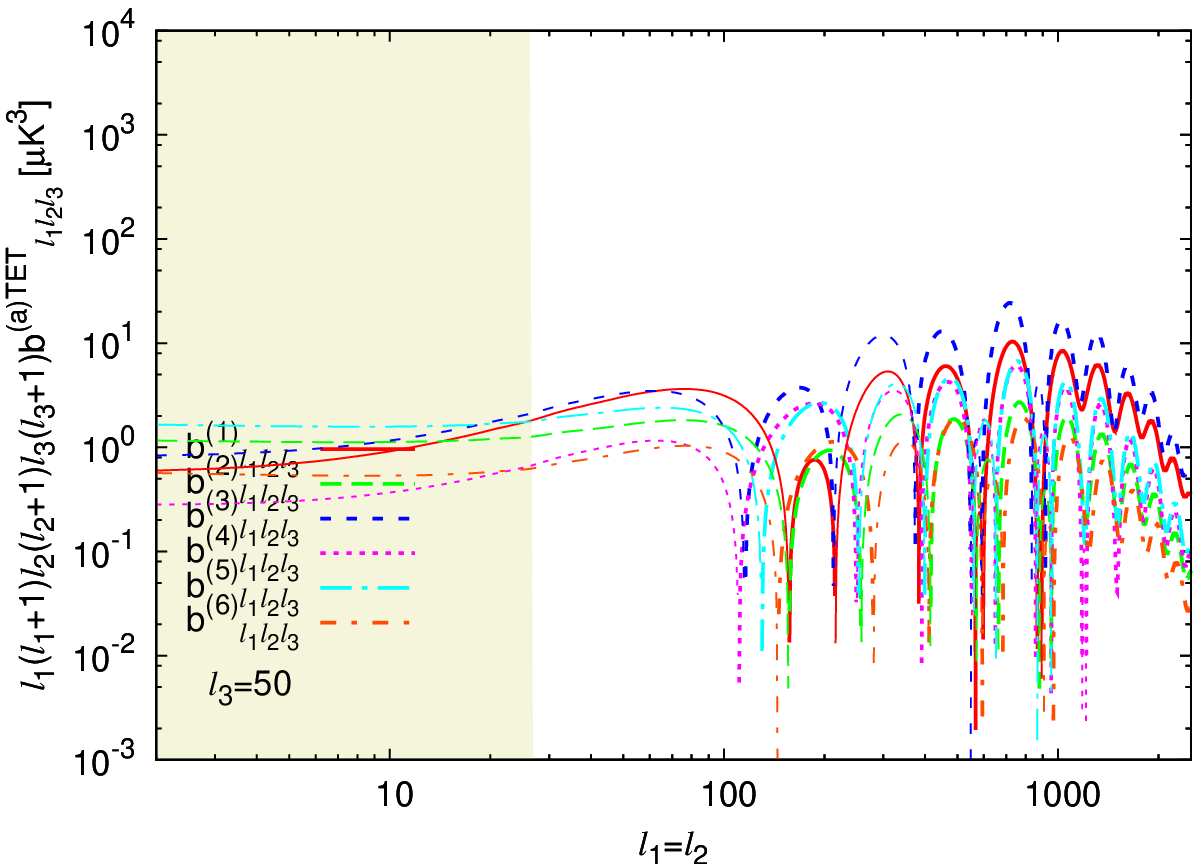}} &
      \hspace{-5mm}
      \resizebox{60mm}{!}{\includegraphics{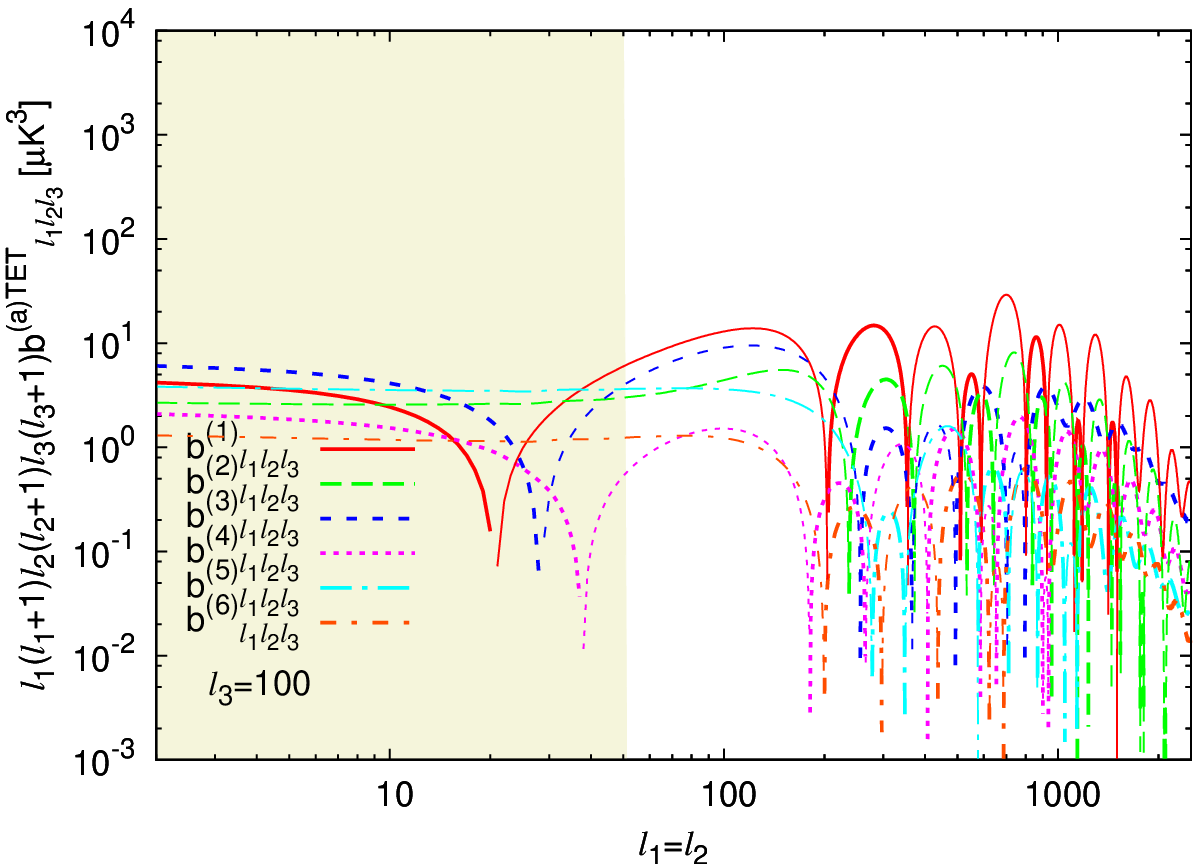}} \\
      \hspace{-10mm}
      \resizebox{60mm}{!}{\includegraphics{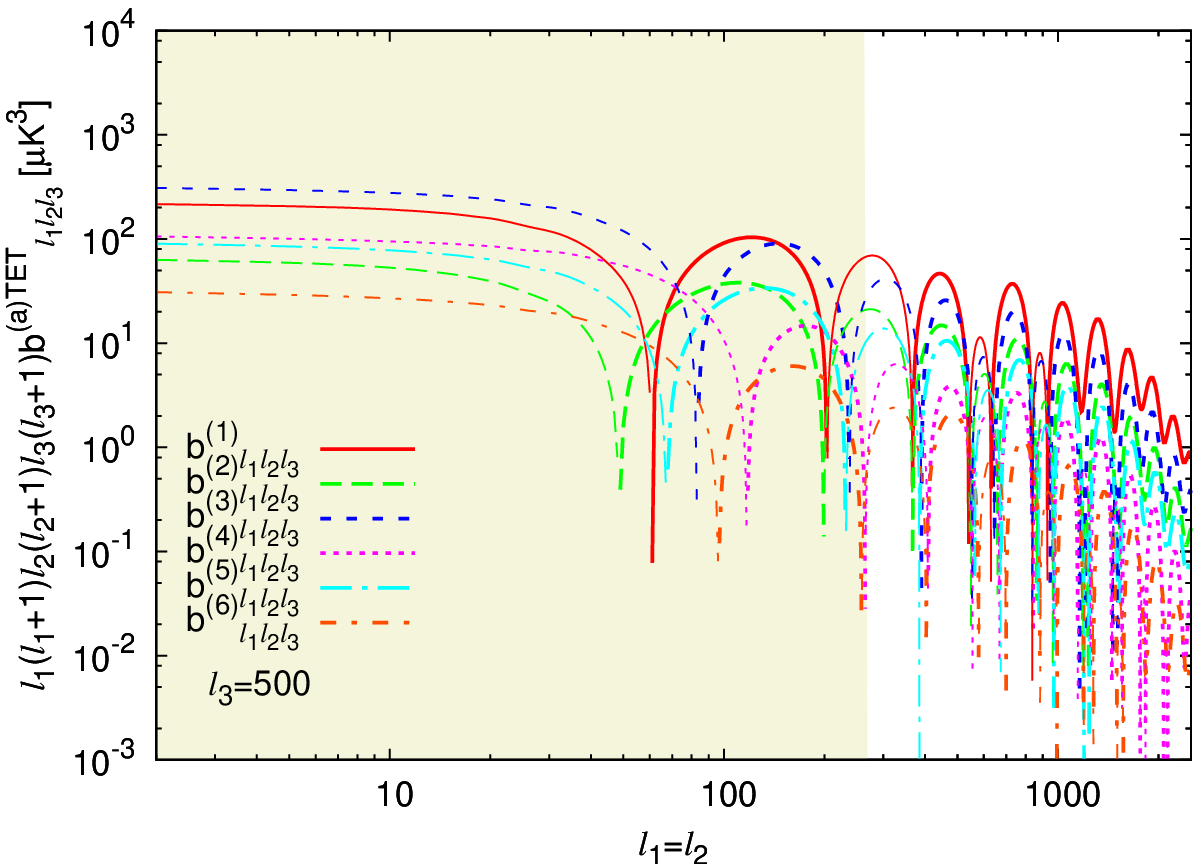}} &
      \hspace{-5mm}
      \resizebox{60mm}{!}{\includegraphics{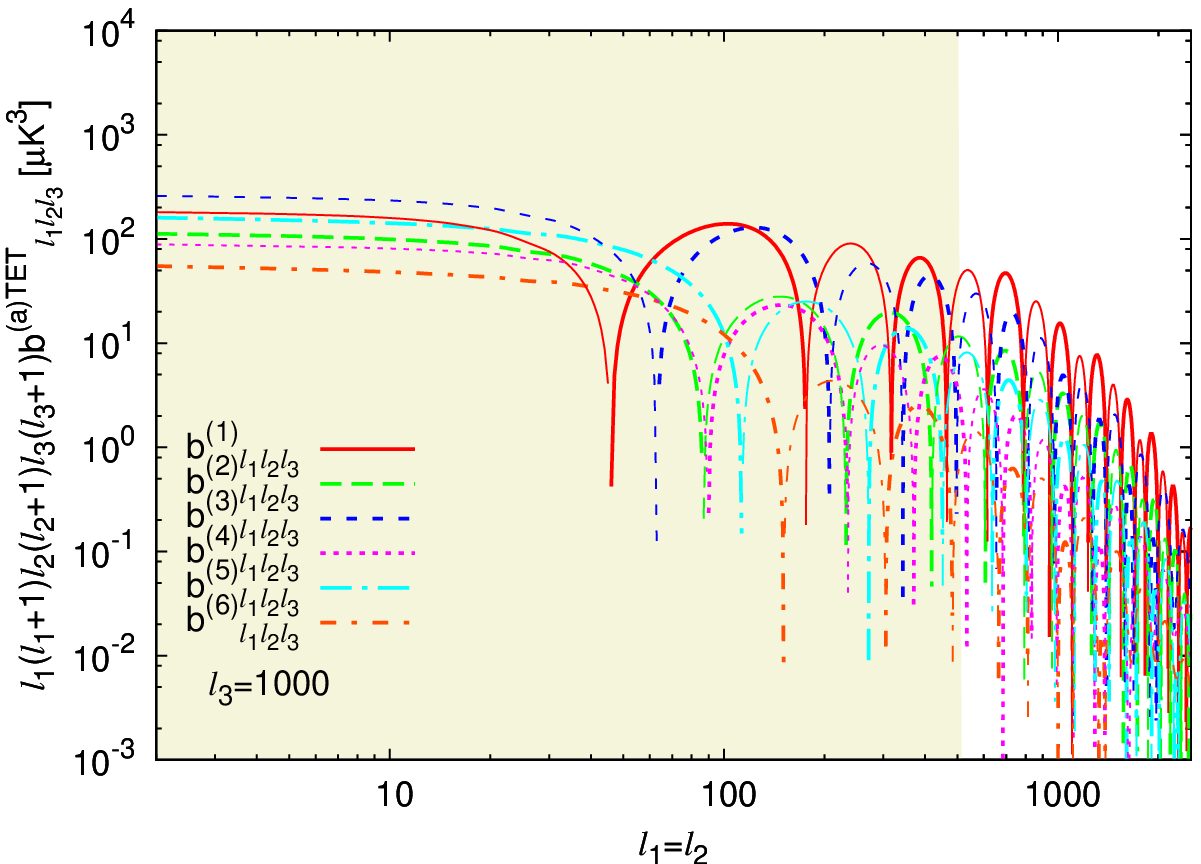}} &
      \hspace{-5mm}
      \resizebox{60mm}{!}{\includegraphics{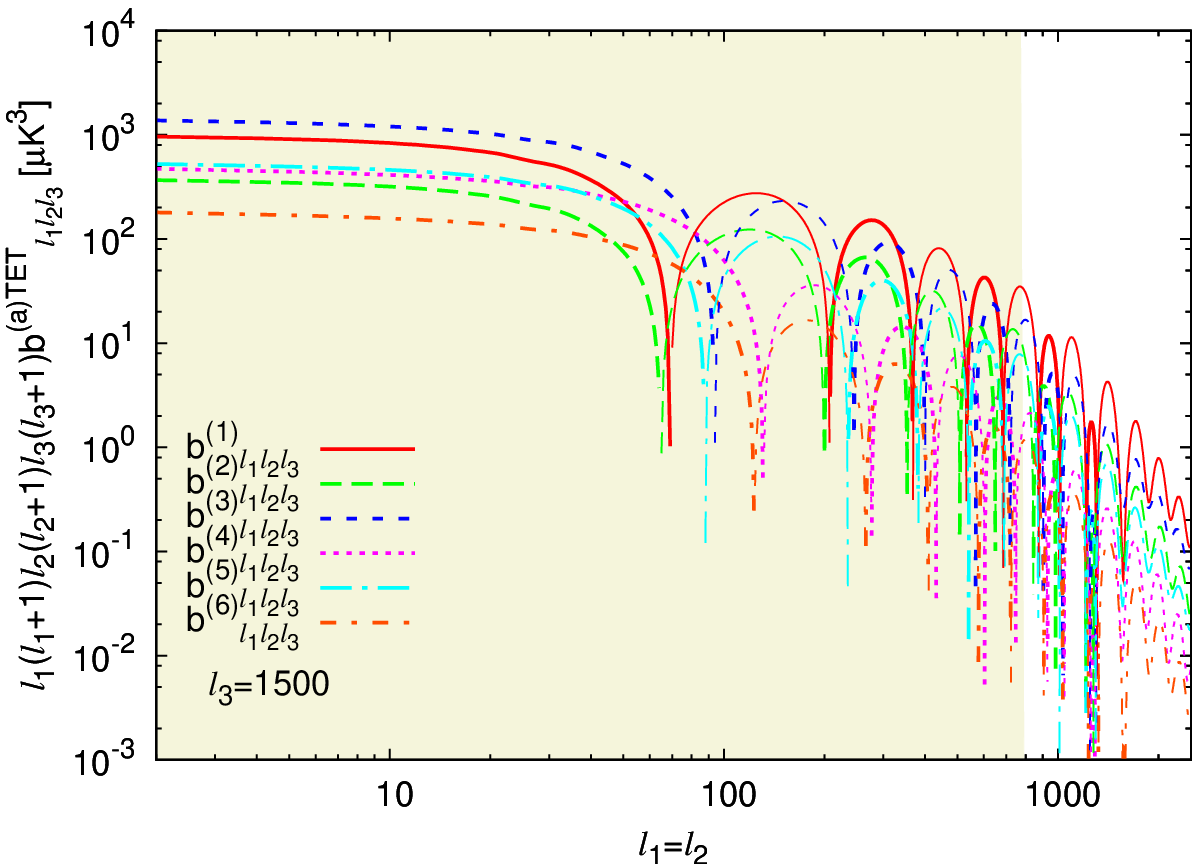}} \\
    \end{tabular}
  \end{center}
  \caption{Same figure as in Fig. \ref{fig:TTT} but for $b^{(a)TET}_{l_1l_2l_3}$.}
  \label{fig:TET}
\end{figure}

Fig. \ref{fig:TTT} shows the temperature bispectra $b^{(a)TTT}_{l_1l_2l_3}$
in isosceles triangular configurations with $l_1=l_2$. 
Cosmological parameters adopted here are the mean parameters 
for the flat power-law $\Lambda$CDM model  
from the WMAP 7-year result \cite{Komatsu:2010fb}.
In numerical calculation, the transfer functions $g^{AP}_l(k)$ 
are computed using the CAMB code \cite{Lewis:1999bs}.
Among six bispectra  $b^{(a)}_{l_1l_2l_3}$, 
$b^{(1)}_{l_1l_2l_3}$ and $b^{(3)}_{l_1l_2l_3}$ tend to be larger 
than others in most configurations, 
although configurations shown in the figure are limited.
On the other hand, $b^{(6)}_{l_1l_2l_3}$ is in general
the smallest. 
In addition, we can see there are peaks and troughs 
in the bispctra, which originate from the acoustic oscillation
of the photon-baryon fluid prior to the recombination.
As is discussed in Ref. \cite{Kawasaki:2011rc}, 
there is slight difference in the phase of the
acoustic oscillation between the adiabatic and neutrino 
isocurvature density modes. This is due from the fact that 
while the acoustic oscillation in the adiabatic mode
is dominantly sourced by the metric perturbations, 
that in the neutrino isocurvature density mode is dominated 
by the initial amplitudes (See Ref. \cite{Kawasaki:2011rc} for more details).
This makes the positions of acoustic peaks and troughs 
differ among bispectra. This can be prominently seen
by comparing $b^{(1)}_{l_1l_2l_3}$ and $b^{(6)}_{l_1l_2l_3}$, 
which respectively originates purely from 
adiabatic and neutrino isocurvature density perturbations
in Fig. \ref{fig:TTT}. However, 
since the phase difference in acoustic oscillation
is not as large as one between the adiabatic and
matter isocurvature modes, the difference in
positions of acoustic peaks is not prominent compared with
the bispectra from mixture of adiabatic and matter 
isocurvature perturbations (See e.g. Ref. \cite{Kawasaki:2008pa}).

We also plotted the bispectra arising from the correlation
of two temperature and one polarization anisotropies, $b^{(a)TTE}_{l_1l_2l_3}$
and $b^{(a)TET}_{l_1l_2l_3}$ in Figs. \ref{fig:TTE} and \ref{fig:TET}, respectively.
From these figures, we can see that above 
discussions on spectral shape of the bispectra are still
qualitatively true when polarization is included.

\section{Forecast for a CMB constraint} 
\label{sec:Fisher}

As we have seen, CMB bispectrum arising from
the non-Gaussian isocurvature perturbations
in extra radiation is distinct from 
the usual one from the non-Gaussian curvature 
perturbations.
Therefore we can discriminate different 
non-Gaussinities in primordial perturbations
from the observation of CMB anisotropy.
To discuss this issue in a quantitative manner, 
we perform a Fisher matrix analysis.

In the limit of weak non-Gaussinity, 
the Fisher matrix for the non-Gaussianity parameters
$f^{(a)}_{\rm NL}$ is given by 
\cite{Komatsu:2001rj,Babich:2004yc,Yadav:2007rk}
\begin{eqnarray}
F_{ab}&=&
\sum_{l_1\le l_2\le l_3}
\frac{(2l_1+1)(2l_2+1)(2l_3+1)}{4\pi}
\begin{pmatrix}
l_1 & l_2 & l_3\\
0 & 0 & 0 
\end{pmatrix}^2\\
&\times&\sum_{P_1P_2P_3}
\sum_{Q_1Q_2Q_3}
b^{(a)~P_1P_2P_3}_{l_1l_2l_3}
[\mathbf{Cov}^{-1}]^{P_1P_2P_3|Q_1Q_2Q_3}_{l_1l_2l_3}
b^{(b)~Q_1Q_2Q_3}_{l_1l_2l_3},
\end{eqnarray}
where 
$[\mathbf{Cov}^{-1}]^{P_1P_2P_3|Q_1Q_2Q_3}_{l_1l_2l_3}$
is the inverse covariance matrix.
Assuming that the observed sky coverage is
unity and the instrumental noise is isotropic, the covariance matrix
$[\mathbf{Cov}]^{P_1P_2P_3|Q_1Q_2Q_3}_{l_1l_2l_3}$
can be given as
\begin{equation}
[\mathbf{Cov}]^{P_1P_2P_3|Q_1Q_2Q_3}_{l_1l_2l_3}
=\Delta_{l_1l_2l_3}\mathcal C^{P_1Q_1}_{l_1}
\mathcal C^{P_2Q_2}_{l_2}\mathcal C^{P_3Q_3}_{l_3},
\end{equation}
where $\mathcal C^{PQ}_l=C^{PQ}_l+N^{PQ}_l$
is the total angular power spectrum, which
is the sum of ones from the CMB $C^{PQ}_l$ 
and instrumental noise $N^{PQ}_l$.
$\Delta_{l_1l_2l_3}$ takes values 6, 2, 1 for the
cases that all $l$'s are the same, only two of them are the same
and otherwise, respectively.

Following \cite{Knox:1995dq}, the
noise power spectrum $N^{PQ}_l$
can be approximated as
\begin{equation}
N^{PQ}_l=\delta_{PQ}
\theta_\mathrm{FWHM}^2\sigma_P^2
\exp\left[l(l+1)
\frac{\theta_\mathrm{FWHM}^2}{8\ln2}
\right],
\end{equation}
where $\theta_\mathrm{FWHM}$ is
the full width at half maximum of the Gaussian
beam, and $\sigma_P$ is the root mean square 
of the instrumental noise par pixel.
For cases of multi-frequency observations, 
$N^{PQ}_l$ is given via the quadrature sum
over all the frequency bands.
Here we study expected constraints from
two survey. One is the on-going Planck survey~\cite{Planck:2006aa} whose 
survey parameters are given in Table~\ref{table:planck}.
Another is a hypothetical survey (hereafter CVL survey)
whose sensitivity is limited by the cosmic variance, i.e. $N^{PQ}_l=0$.
In both cases, we omit the effect of the sky cut 
and the sky coverage is assumed to be unity.

In the analysis, we consider only CMB bispectra 
from primordial non-Gaussianities, assuming contaminations 
from other sources are negligible.
Since many of these sources including point sources and the thermal
Sunyaev-Zel'dovich (SZ) effect have frequency spectra
different from the black body, 
above assumption can be to some extent achieved by exploiting
observations at multi-frequency bands.
Other contaminations such as the lensing of CMB, 
the kinetic SZ effect, and the patchy reionization would
not affect our results significantly.

As our models have six 
non-Gaussianity parameters
$f^{(a)}_{\rm NL}$, the full Fisher 
matrix $F_{ab}$ is a $6\times6$ matrix.
However, it may sometimes occur that 
some of the parameters
are not of primary interest and we want them to be
marginalized over. 
In such the case, the Fisher matrix for the
remaining non-Gaussianity parameters 
can be given as the inverse
of the principal sub-matrix of the inverted
full Fisher matrix \cite{Albrecht:2006um}. 


\begin{table}[htb]
  \begin{center}
  \begin{tabular}{l|c|c|c}
  \hline
  \hline
  bands [GHz] & $\theta_{\rm FWHM}$ [arcmin] & $\sigma_T$ [$\mu$K] & $\sigma_P$ [$\mu$K] \\
  \hline
      $30$ & $33.0$ & $2.0$ & $2.8$ \\ 
      $44$ & $24.0$ & $2.7$ & $3.9$ \\ 
      $70$ & $14.0$ & $4.7$ & $6.7$ \\
      $100$ & $10.0$ & $2.5$ & $4.0$ \\
      $143$ & $7.1$ & $2.2$ & $4.2$ \\
      $217$ & $5.0$ & $4.8$ & $9.8$ \\
      $353$ & $5.0$ & $14.7$ & $29.8$ \\
  \hline
  \hline 
\end{tabular}
  \caption{Survey parameters adopted in our analysis for Planck. 
  $\theta_{\rm FWHM}$ is Gaussian beam width at FWHM, 
  $\sigma_T$ and $\sigma_P$ are temperature and polarization noise, respectively. 
  We assume 1-year duration of observation.
  }
  \label{table:planck}
\end{center}
\end{table}

\begin{figure}
  \begin{center}
    \begin{tabular}{cccccc}
      \hspace{-10mm}
      \resizebox{40mm}{!}{\includegraphics{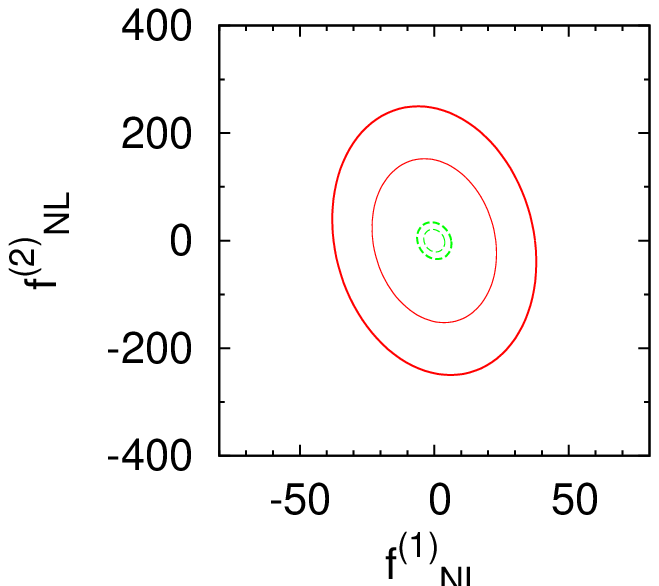}} \\
      \hspace{-10mm}
      \resizebox{40mm}{!}{\includegraphics{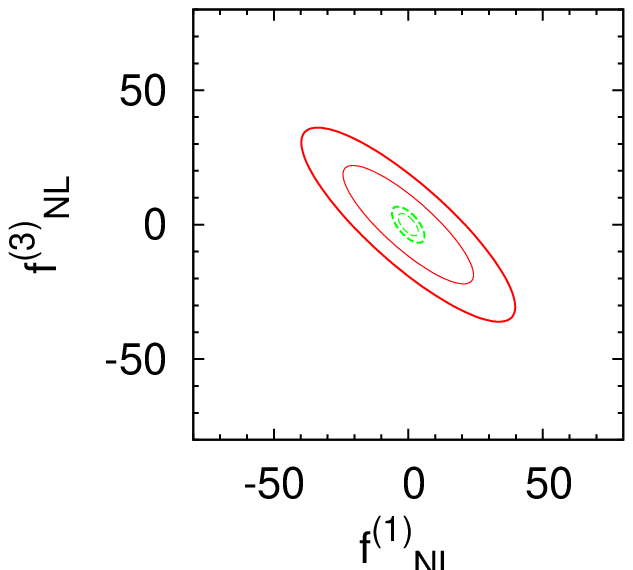}} &
      \hspace{-15mm}
      \resizebox{40mm}{!}{\includegraphics{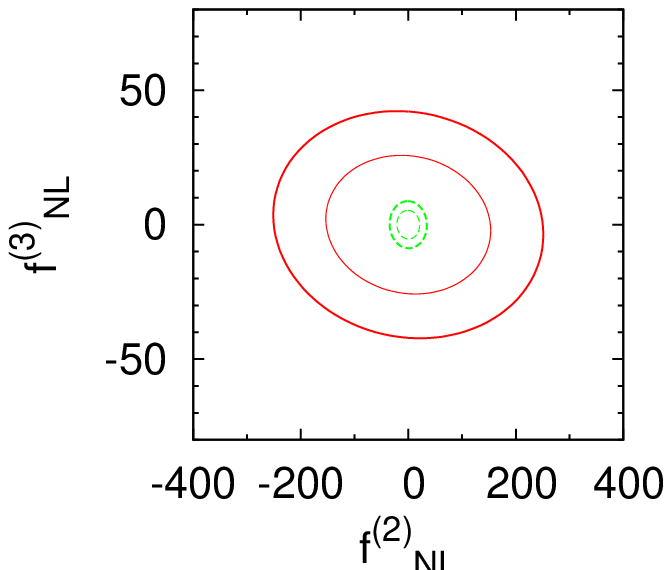}} \\
      \hspace{-10mm}
      \resizebox{40mm}{!}{\includegraphics{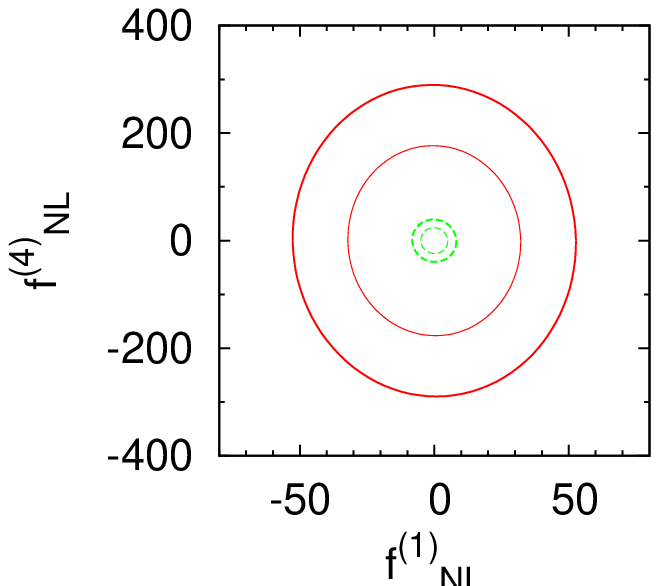}} &
      \hspace{-15mm}
      \resizebox{40mm}{!}{\includegraphics{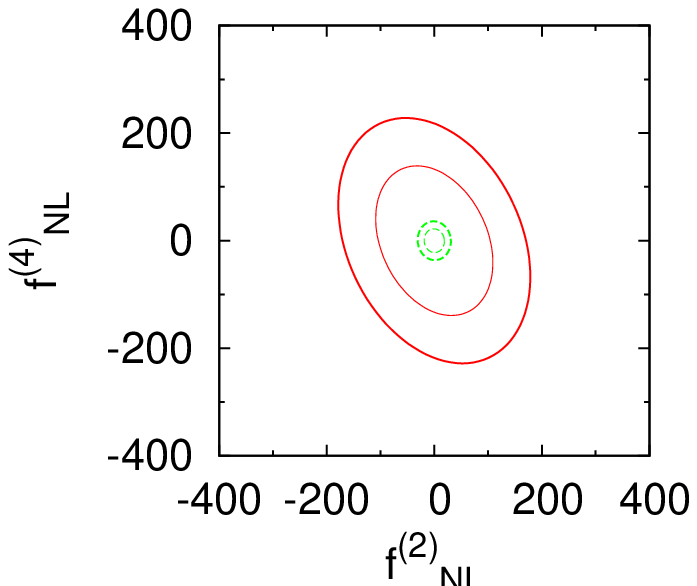}} &
      \hspace{-15mm}
      \resizebox{40mm}{!}{\includegraphics{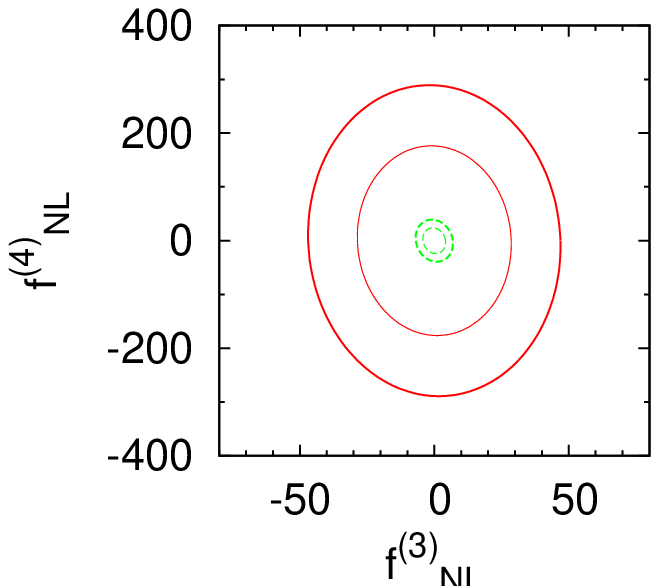}} \\
      \hspace{-10mm}
      \resizebox{40mm}{!}{\includegraphics{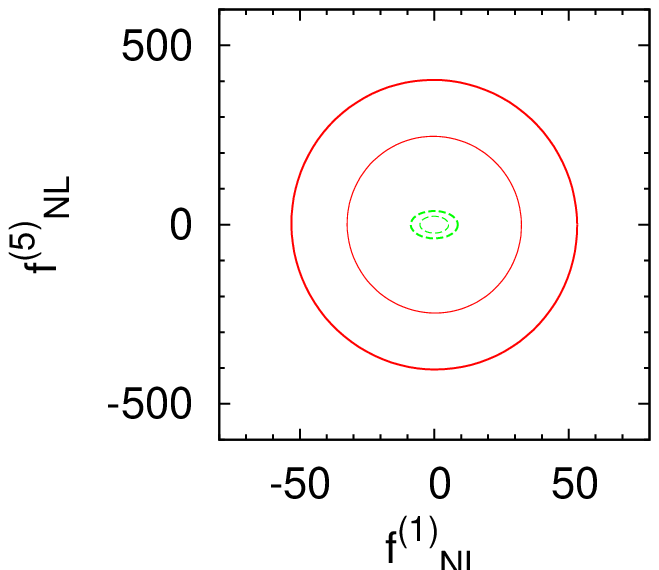}} &
      \hspace{-15mm}
      \resizebox{40mm}{!}{\includegraphics{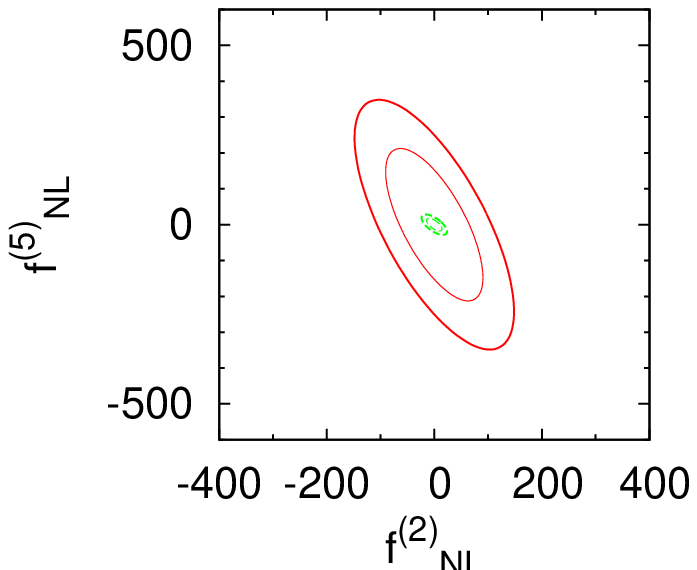}} &
      \hspace{-15mm}
      \resizebox{40mm}{!}{\includegraphics{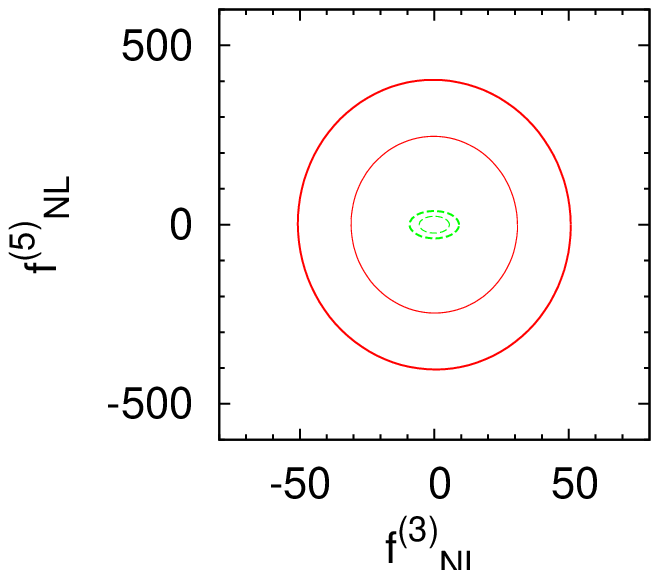}} &
      \hspace{-15mm}
      \resizebox{40mm}{!}{\includegraphics{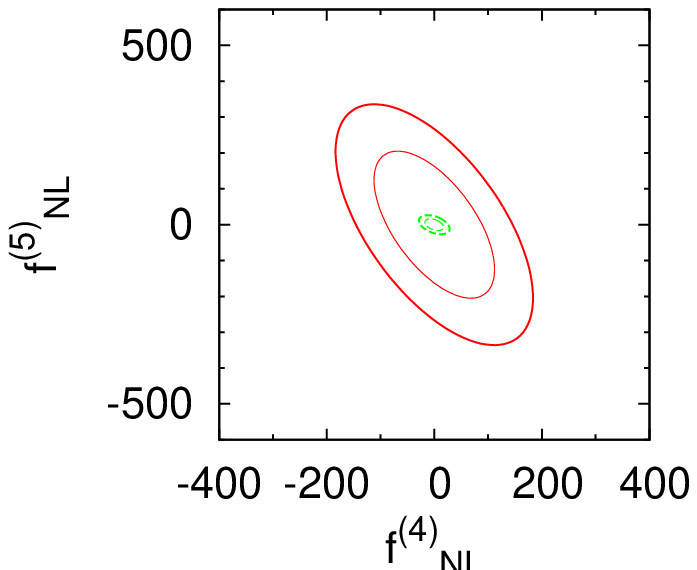}} \\
      \hspace{-10mm}
      \resizebox{40mm}{!}{\includegraphics{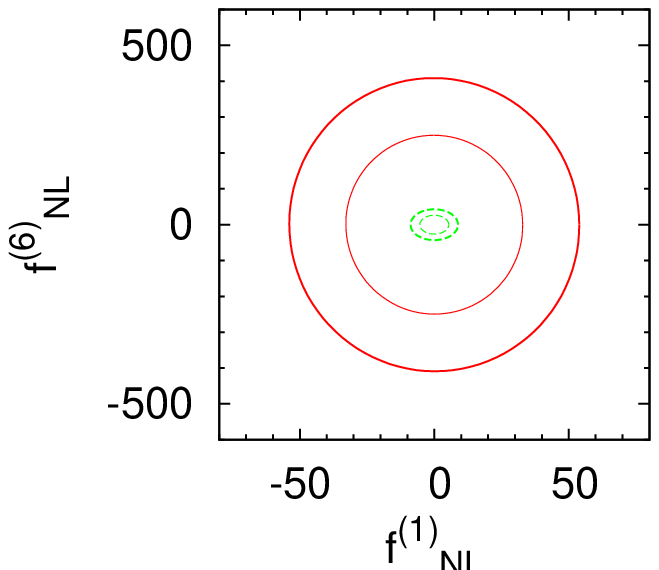}} &
      \hspace{-15mm}
      \resizebox{40mm}{!}{\includegraphics{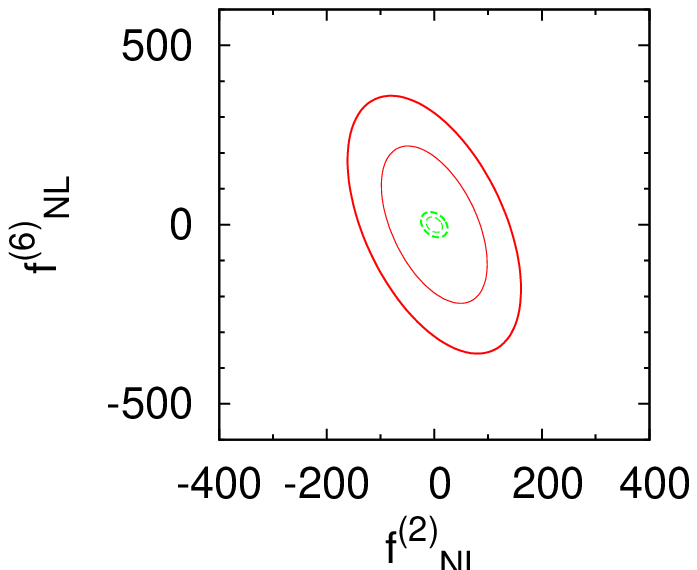}} &
      \hspace{-15mm}
      \resizebox{40mm}{!}{\includegraphics{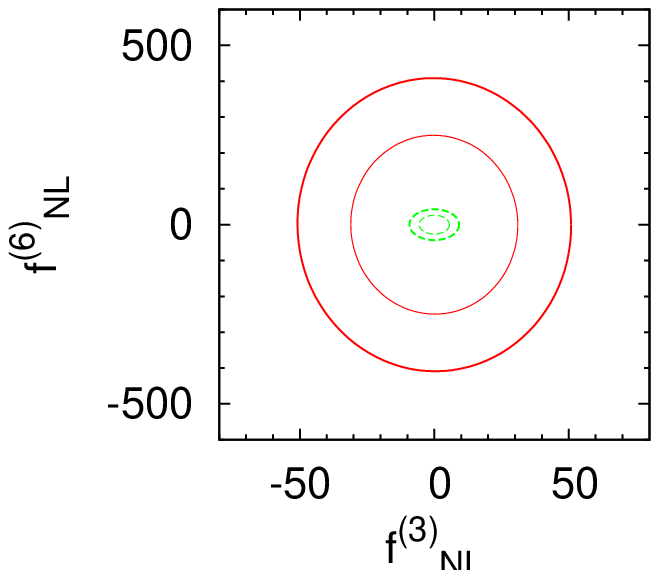}} &
      \hspace{-15mm}
      \resizebox{40mm}{!}{\includegraphics{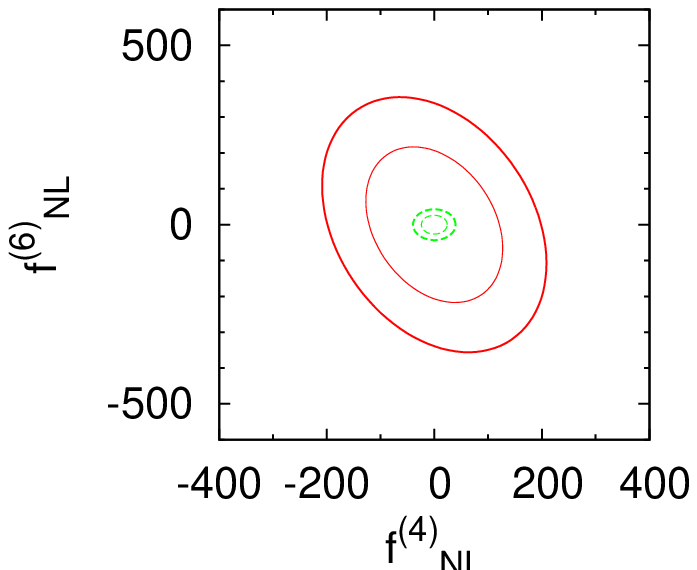}} &
      \hspace{-15mm}
      \resizebox{40mm}{!}{\includegraphics{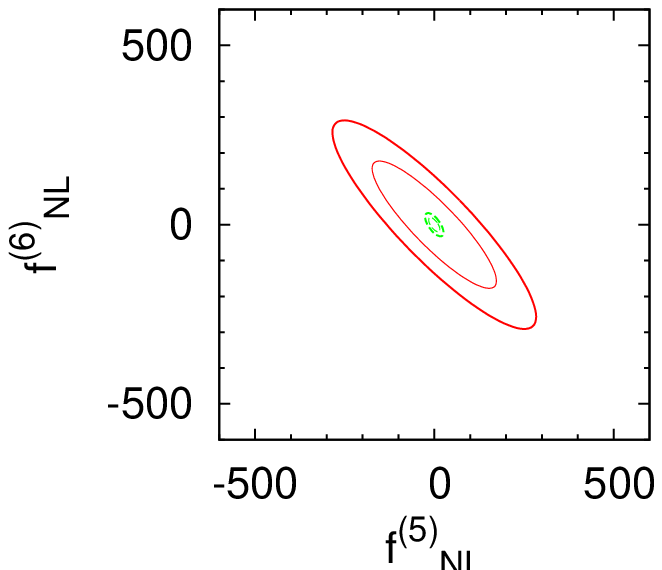}} 
    \end{tabular}
  \end{center}
  \caption{2d marginalized constraints on the non-Gaussianity parameters $f^{(a)}_{\rm NL}$
  expected for Planck (solid red) and CVL (dashed green) surveys.
  $N_\nu$ is fixed to $4$.
  Inner and outer contours correspond to constraints at 1 and 2 $\sigma$ levels.}
  \label{fig:Nnu4_margeFij}
\end{figure}

\begin{figure}
  \begin{center}
    \begin{tabular}{cccccc}
      \hspace{-10mm}
      \resizebox{40mm}{!}{\includegraphics{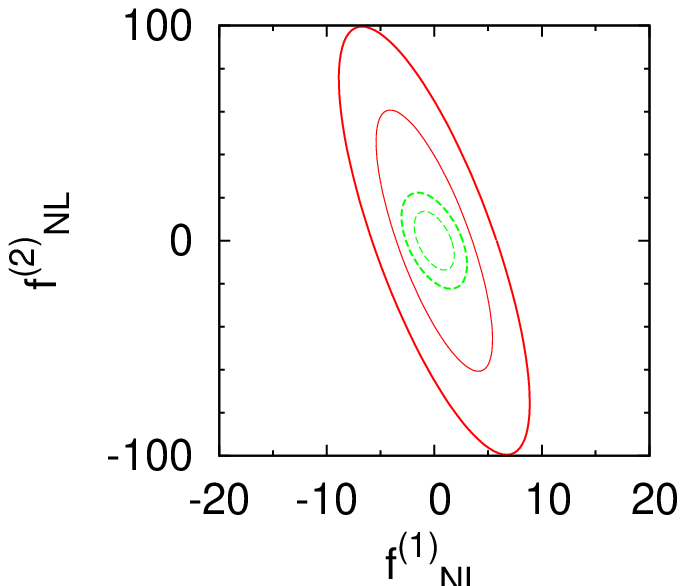}} \\
      \hspace{-10mm}
      \resizebox{40mm}{!}{\includegraphics{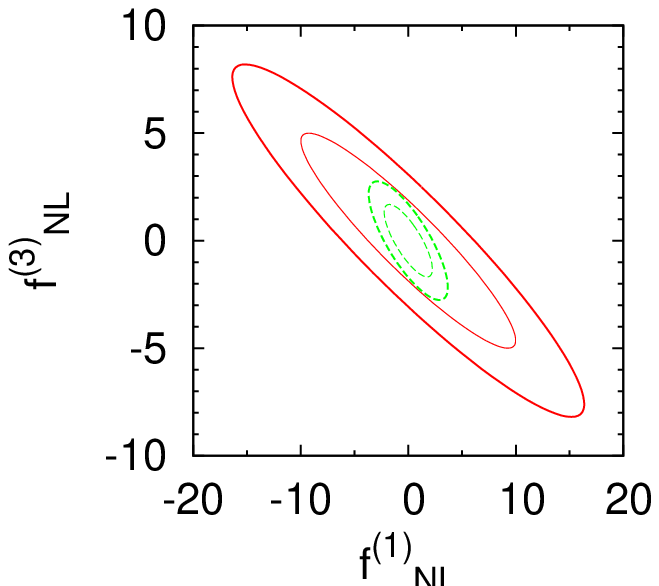}} &
      \hspace{-15mm}
      \resizebox{40mm}{!}{\includegraphics{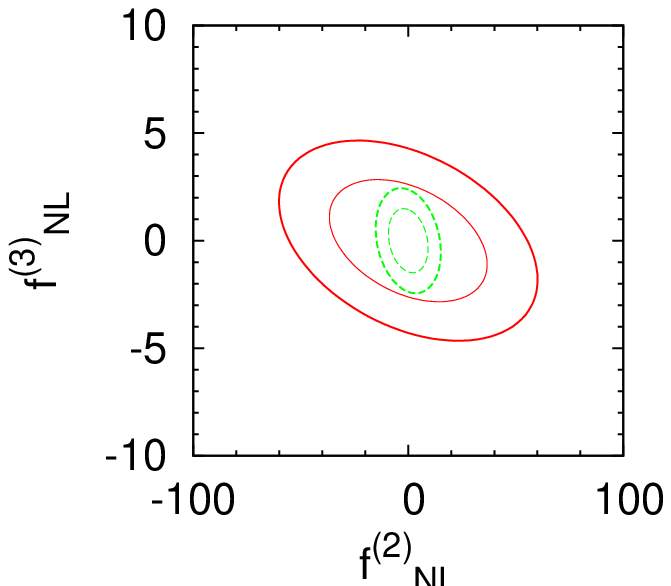}} \\
      \hspace{-10mm}
      \resizebox{40mm}{!}{\includegraphics{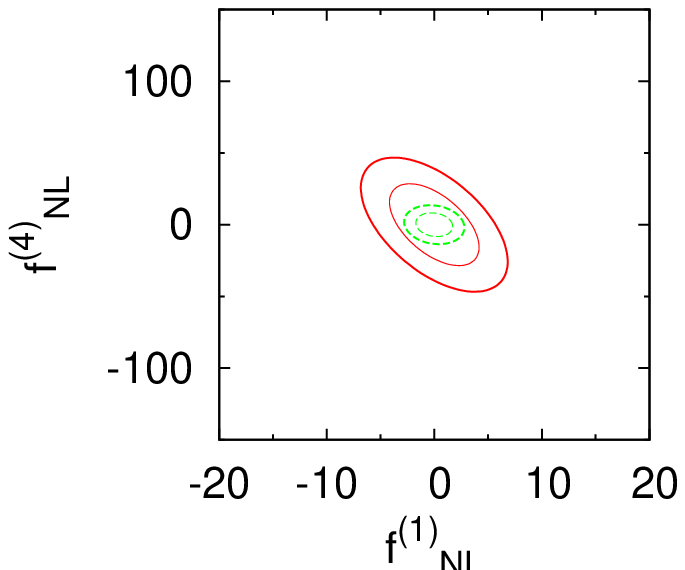}} &
      \hspace{-15mm}
      \resizebox{40mm}{!}{\includegraphics{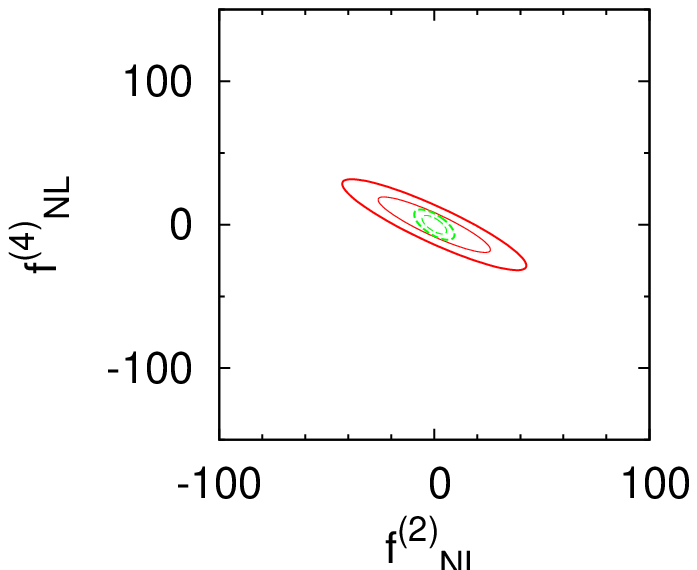}} &
      \hspace{-15mm}
      \resizebox{40mm}{!}{\includegraphics{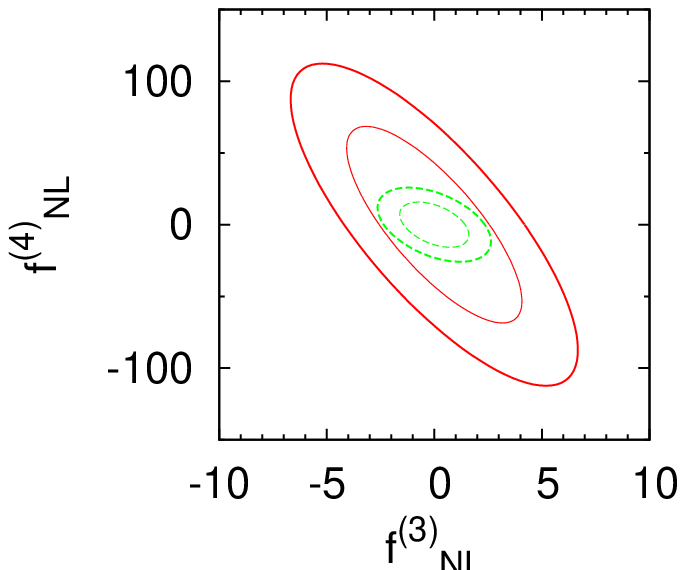}} \\
      \hspace{-10mm}
      \resizebox{40mm}{!}{\includegraphics{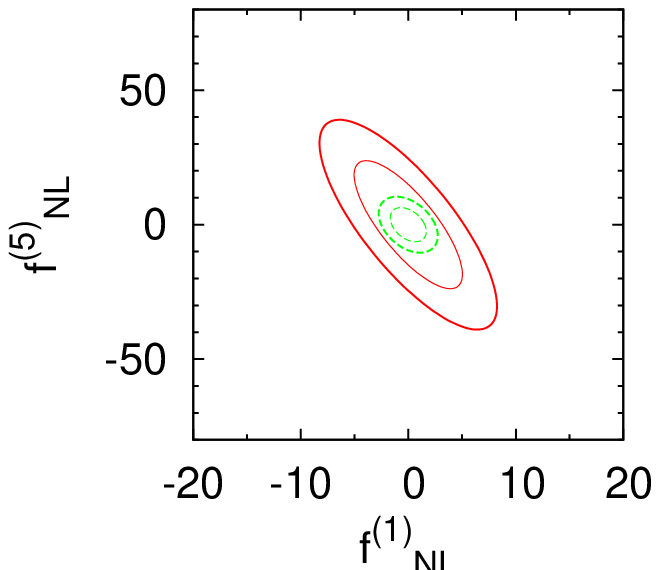}} &
      \hspace{-15mm}
      \resizebox{40mm}{!}{\includegraphics{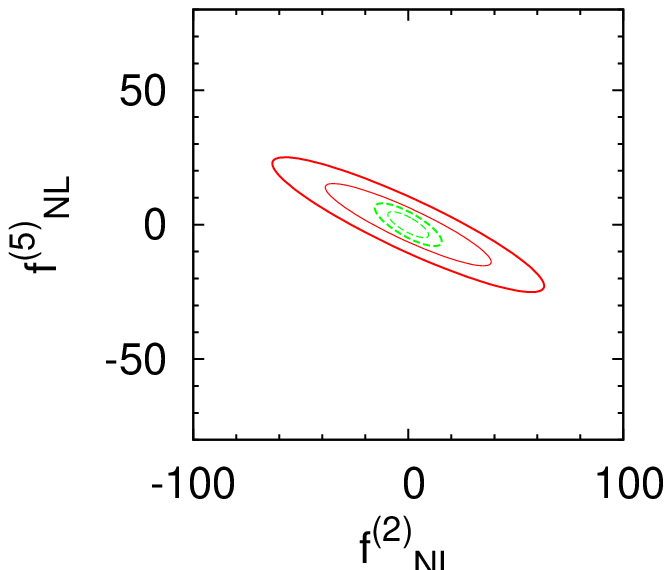}} &
      \hspace{-15mm}
      \resizebox{40mm}{!}{\includegraphics{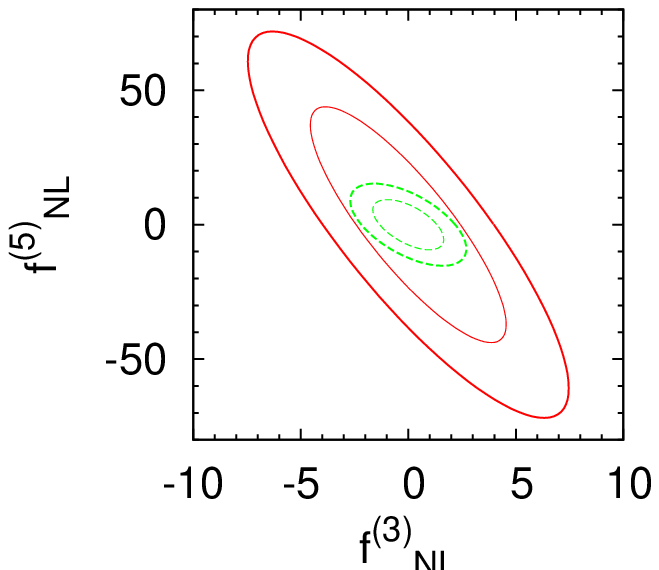}} &
      \hspace{-15mm}
      \resizebox{40mm}{!}{\includegraphics{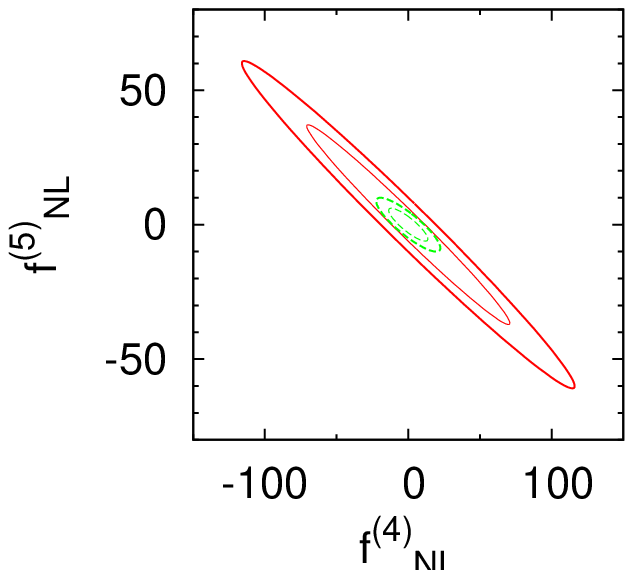}} \\
      \hspace{-10mm}
      \resizebox{40mm}{!}{\includegraphics{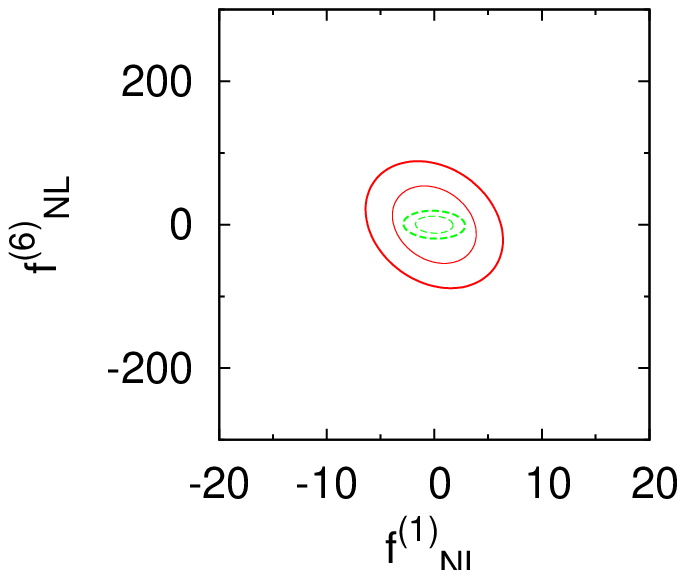}} &
      \hspace{-15mm}
      \resizebox{40mm}{!}{\includegraphics{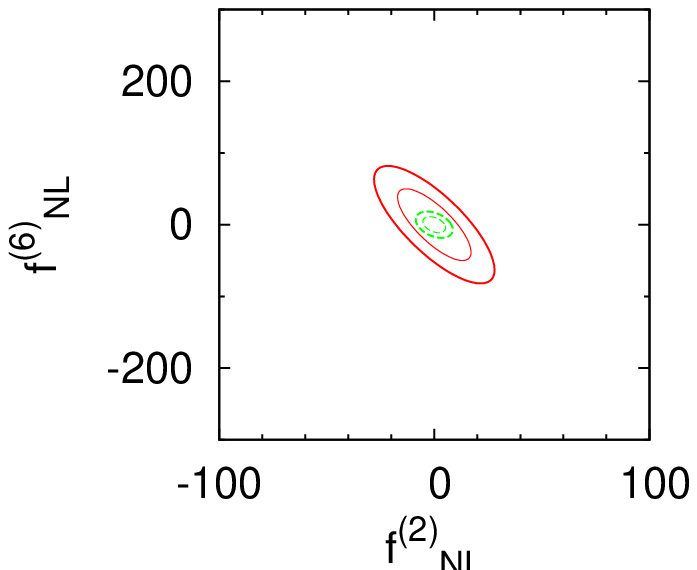}} &
      \hspace{-15mm}
      \resizebox{40mm}{!}{\includegraphics{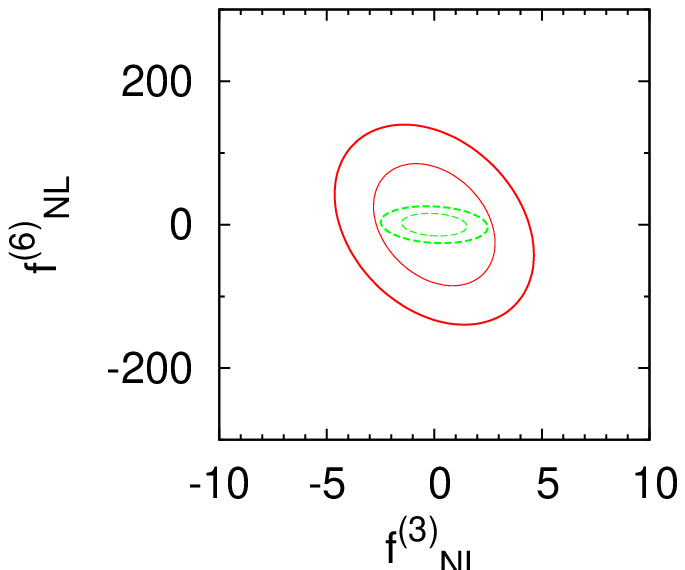}} &
      \hspace{-15mm}
      \resizebox{40mm}{!}{\includegraphics{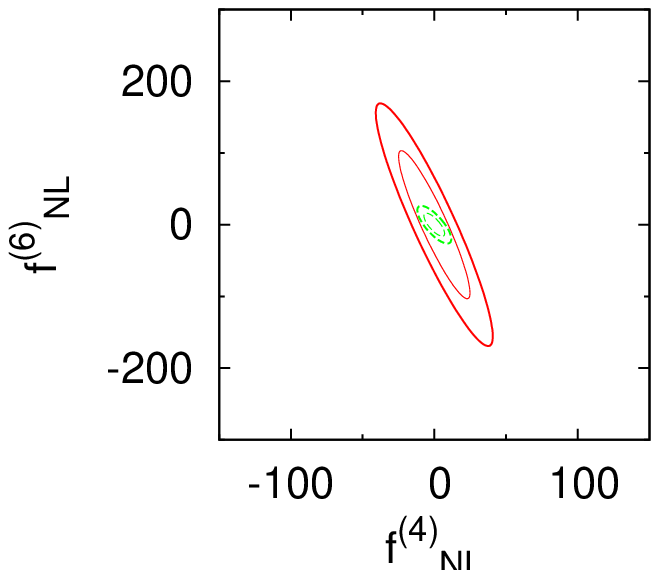}} &
      \hspace{-15mm}
      \resizebox{40mm}{!}{\includegraphics{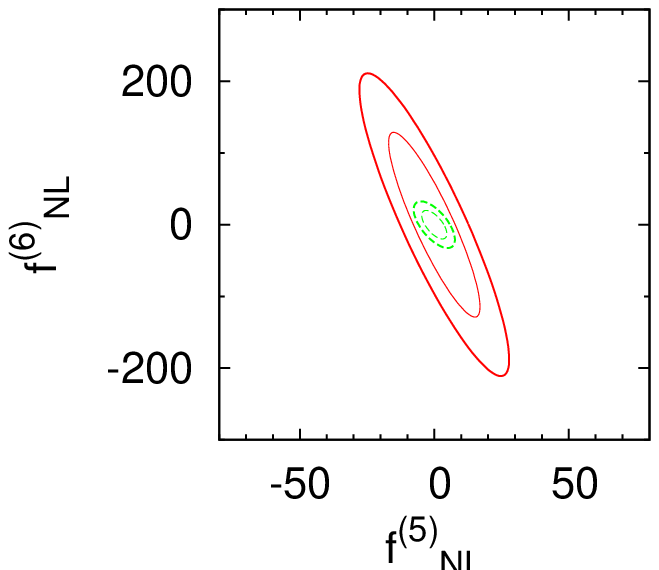}} 
    \end{tabular}
  \end{center}
  \caption{Same figure as in Fig.~\ref{fig:Nnu4_margeFij}, but 
  the non-marginalized constraints are shown here.}
  \label{fig:Nnu4_Fij}
\end{figure}

In Figs.~\ref{fig:Nnu4_margeFij} and  \ref{fig:Nnu4_Fij},
shown are 2-dimensional constraints 
on the non-Gaussianity parameters
expected for Planck and CVL surveys.
Here we fixed $N_\nu$ to 4, which is suggested by recent
observations we mentioned in Introduction.
On each panel, constraints on a pair of $f^{(a)}_{\rm NL}$ are shown; 
other four non-Gaussianity parameters are marginalized over
in Fig.~\ref{fig:Nnu4_margeFij} while they are fixed to zero in
Fig.~\ref{fig:Nnu4_Fij}. 
Hereafter we will refer to constraints shown in 
Fig.~\ref{fig:Nnu4_margeFij} and \ref{fig:Nnu4_Fij}
as marginalized and non-marginalized constraints, respectively.

\begin{table}
\begin{center}\caption{Expected uncertainties for non-Gaussianity parameters
$\Delta f_{\rm NL}^{(a)}$ for the case with $N_{\rm eff}=4$.}
\label{tbl:deltaf_4}
\begin{tabular}{c|rrrrrr}
    \hline\hline
    survey & $f_{\rm NL}^{(1)}$ & $f_{\rm NL}^{(2)}$ & $f_{\rm NL}^{(3)}$
    & $f_{\rm NL}^{(4)}$ & $f_{\rm NL}^{(5)}$ & $f_{\rm NL}^{(6)}$ \\
    \hline\hline
    Planck & 22 & 101 & 21 & 116 & 163 & 164 \\
    \hline
    CVL & 3.5 & 14.0 & 3.7 & 15.9 & 15.4 & 17.3 \\
    \hline\hline
\end{tabular}
\end{center}
\end{table}

In Table \ref{tbl:deltaf_4}, we listed
expected uncertainties in the non-Gaussian parameters 
$\Delta f_{\rm NL}^{(a)}$, which are defined by
\begin{equation}
\Delta f_{\rm NL}^{(a)}\equiv F^{-1}_{aa}.
\end{equation}
From the table as well as figures, we can see that
among the six non-Gaussian parameters $f_{\rm NL}^{(a)}$,
$f_{\rm NL}^{(1)}$ and $f_{\rm NL}^{(3)}$ can be constrained tighter than others.
Planck (a CVL survey) can constrain $f_{\rm NL}^{(1)}$ and 
$f_{\rm NL}^{(3)}$ to about 20 (4). On the other hand, 
expected constraints on other $f_{\rm NL}^{(a)}$
are weaker with factor from five or eight. 
This result is consistent with our discussion in the previous section, where
we showed that in the squeezed configurations, 
$b^{(1)P_1P_2P_3}_{l_1l_2l_3}$ and $b^{(3)P_1P_2P_3}_{l_1l_2l_3}$
are in general larger than other bispecra. 
We can also see that a CVL survey can significantly improve the 
constrains on all non-Gaussianity parameters from Planck 
by an order of magnitude.
On the other hand, in the case of the matter 
isocurvature mode, constraints on some of non-Gaussianity parameters
improve little as we measure higher and higher multipoles as shown 
in Ref. \cite{Langlois:2011hn}.
This difference reflects that CMB anisotropies at high multipoles are damped 
in the case of the matter isocurvature mode, while 
they are comparable in amplitude with the adiabatic mode
in the extra radiation isocurvature mode.

\begin{figure}
  \begin{center}
    \begin{tabular}{cccccc}
      \hspace{-10mm}
      \resizebox{40mm}{!}{\includegraphics{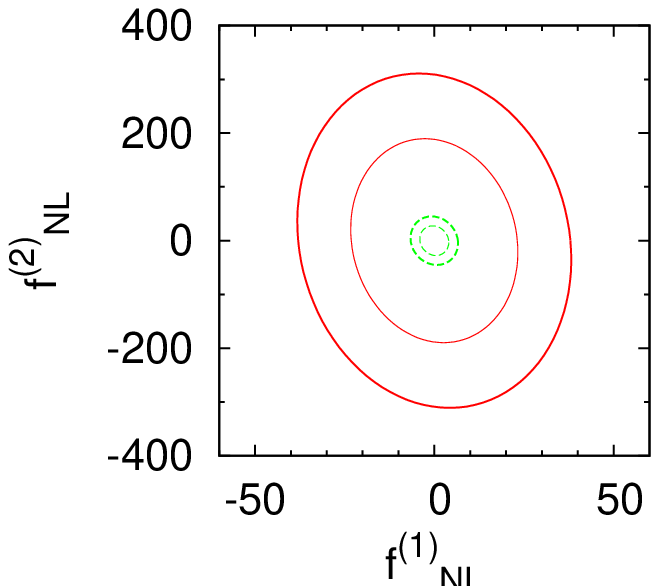}} \\
      \hspace{-10mm}
      \resizebox{40mm}{!}{\includegraphics{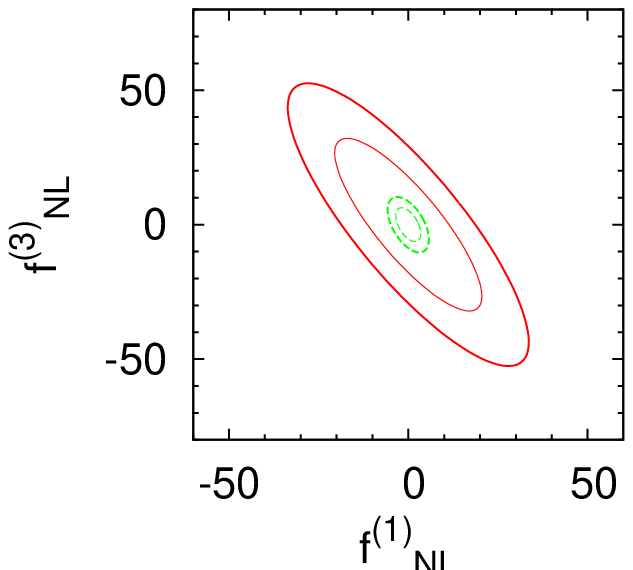}} &
      \hspace{-15mm}
      \resizebox{40mm}{!}{\includegraphics{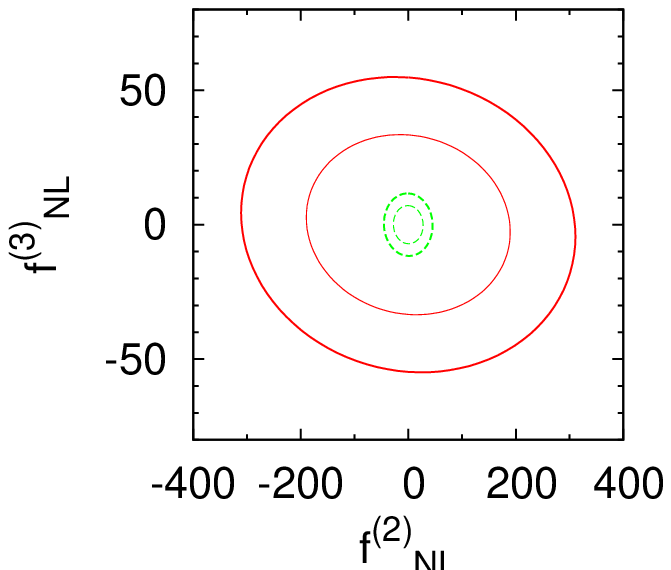}} \\
      \hspace{-10mm}
      \resizebox{40mm}{!}{\includegraphics{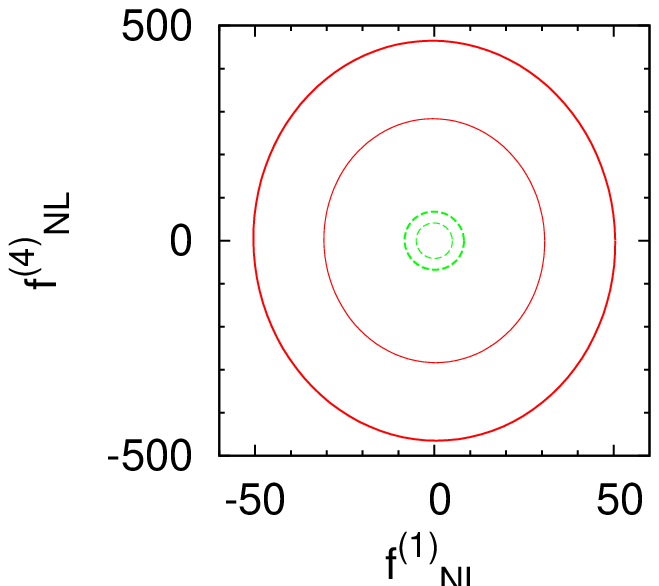}} &
      \hspace{-15mm}
      \resizebox{40mm}{!}{\includegraphics{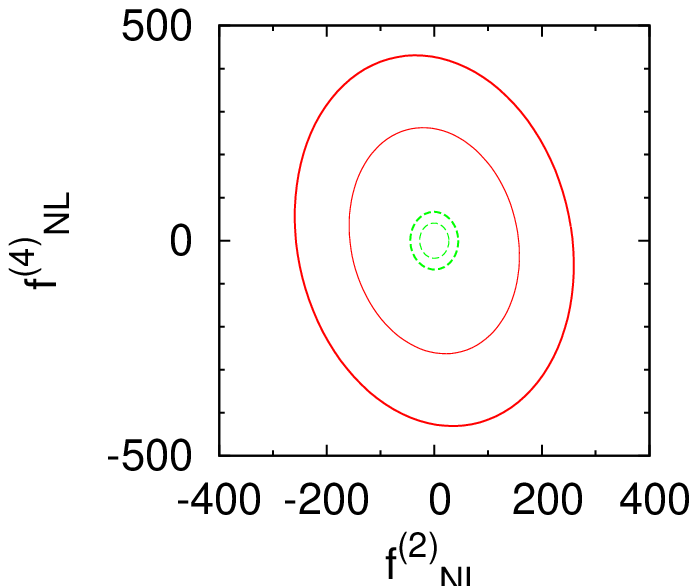}} &
      \hspace{-15mm}
      \resizebox{40mm}{!}{\includegraphics{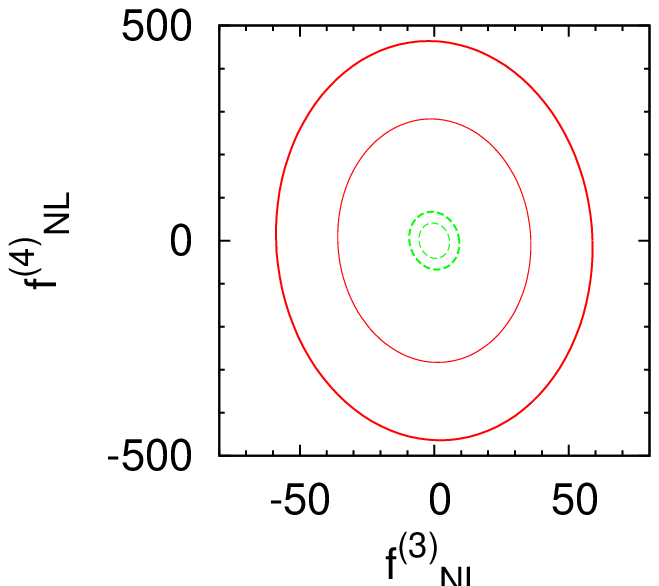}} \\
      \hspace{-10mm}
      \resizebox{40mm}{!}{\includegraphics{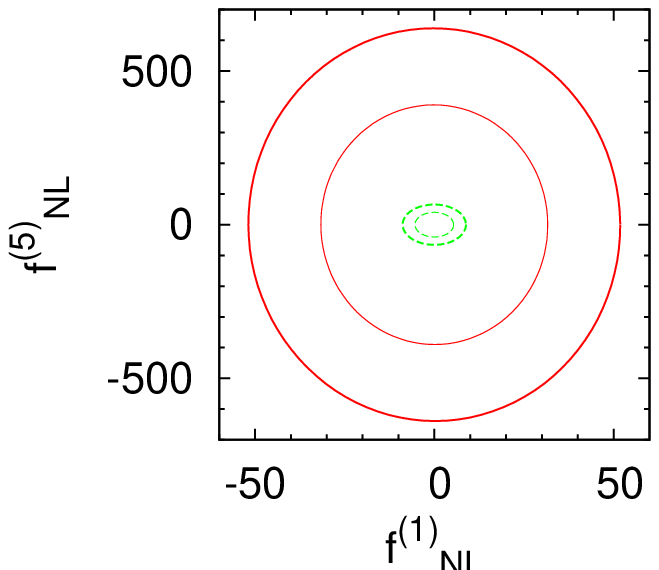}} &
      \hspace{-15mm}
      \resizebox{40mm}{!}{\includegraphics{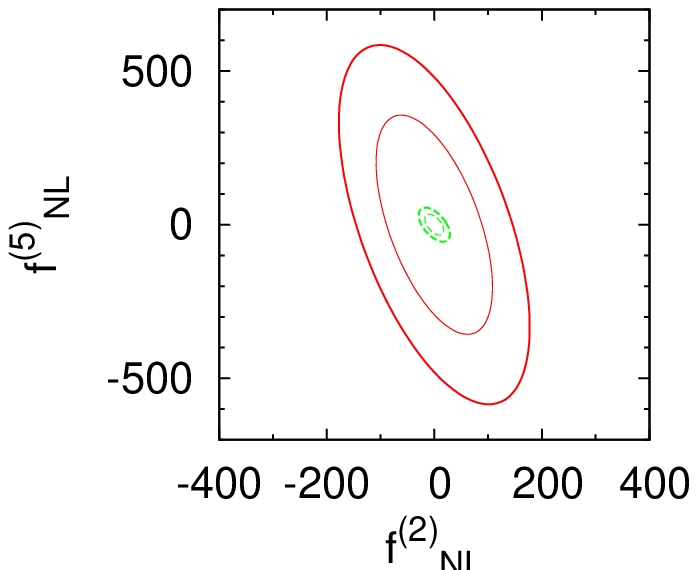}} &
      \hspace{-15mm}
      \resizebox{40mm}{!}{\includegraphics{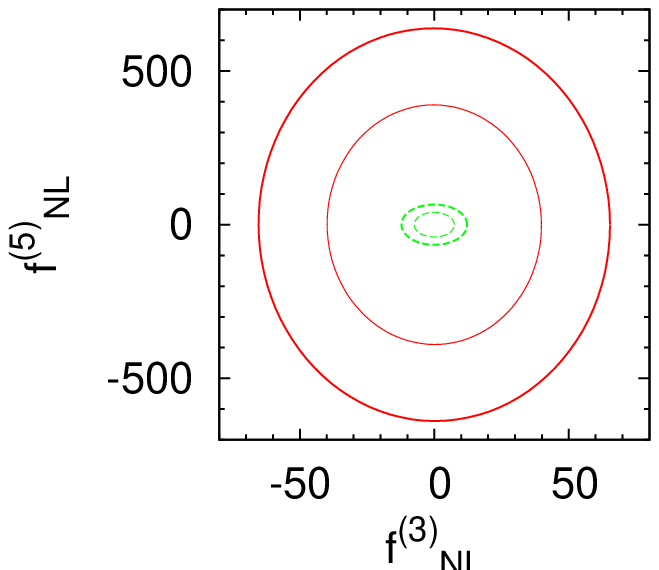}} &
      \hspace{-15mm}
      \resizebox{40mm}{!}{\includegraphics{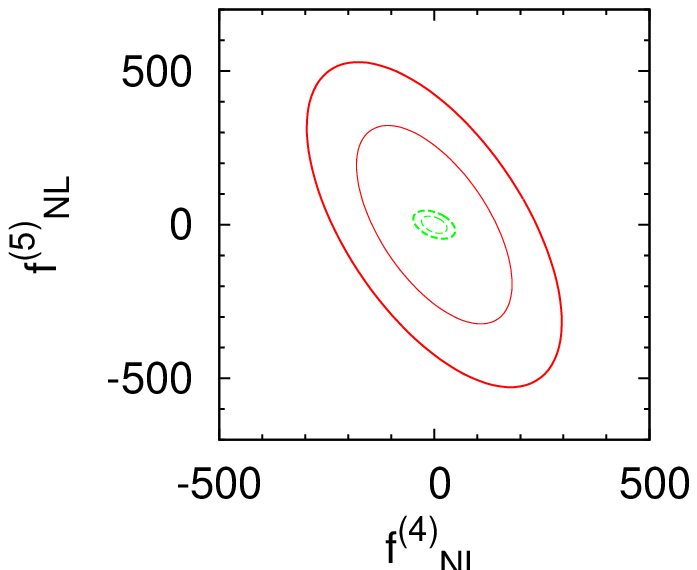}} \\
      \hspace{-10mm}
      \resizebox{40mm}{!}{\includegraphics{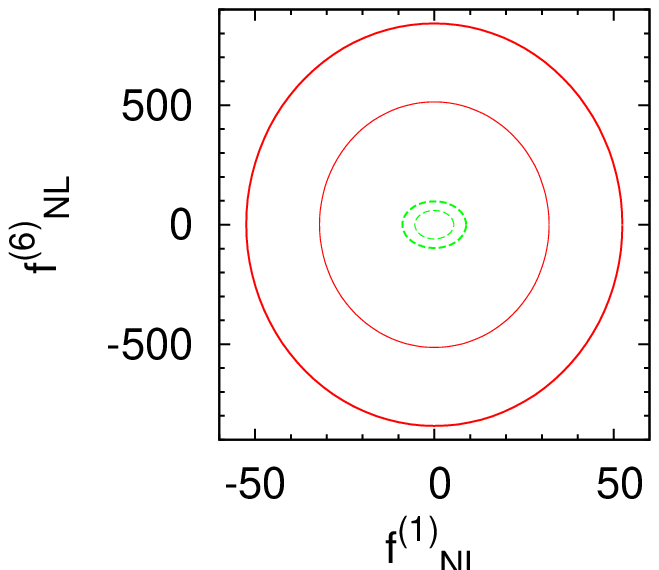}} &
      \hspace{-15mm}
      \resizebox{40mm}{!}{\includegraphics{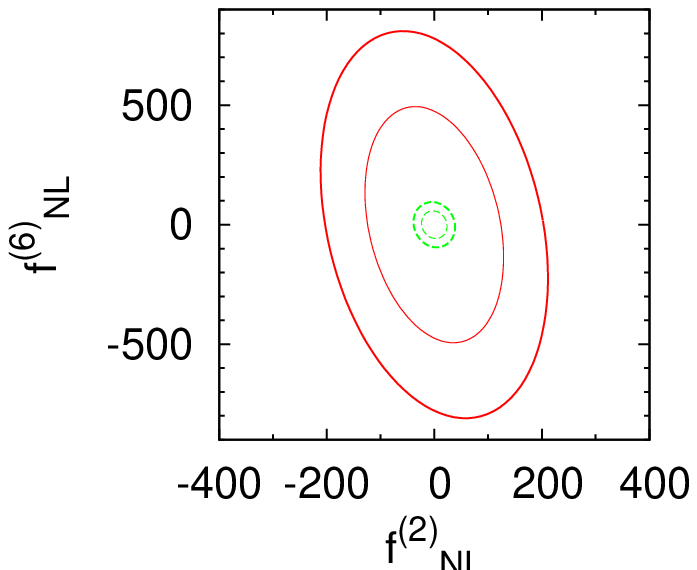}} &
      \hspace{-15mm}
      \resizebox{40mm}{!}{\includegraphics{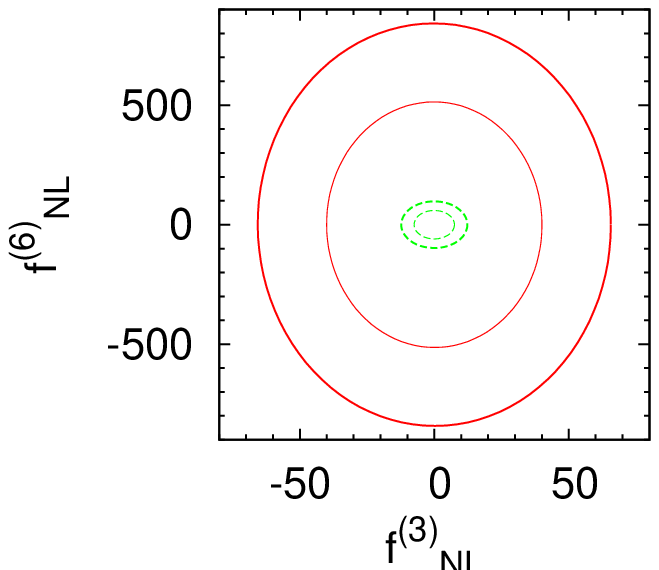}} &
      \hspace{-15mm}
      \resizebox{40mm}{!}{\includegraphics{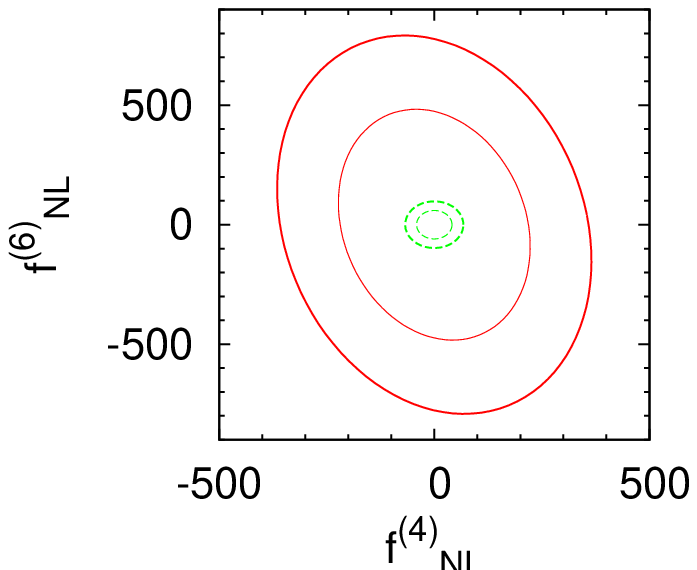}} &
      \hspace{-15mm}
      \resizebox{40mm}{!}{\includegraphics{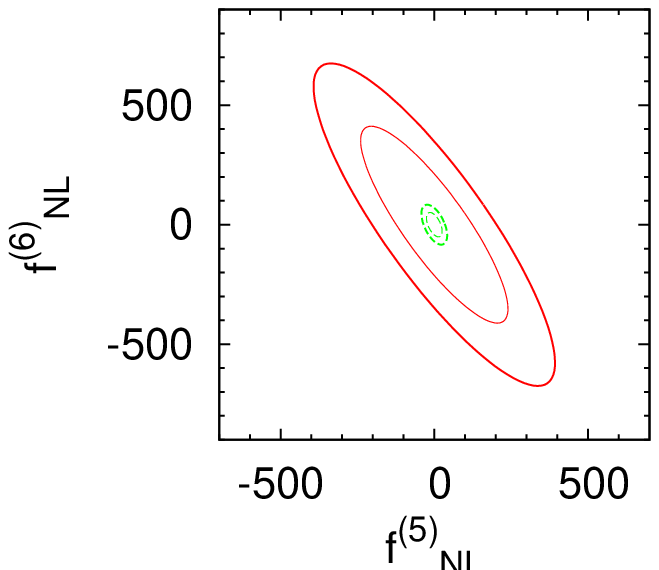}} 
    \end{tabular}
  \end{center}
  \caption{2d marginalized constraints on non-Gaussianity parameters for $N_\nu=3.04$.
  }
  \label{fig:Nnu3_margeFij}
\end{figure}

\begin{figure}
  \begin{center}
    \begin{tabular}{cccccc}
      \hspace{-10mm}
      \resizebox{40mm}{!}{\includegraphics{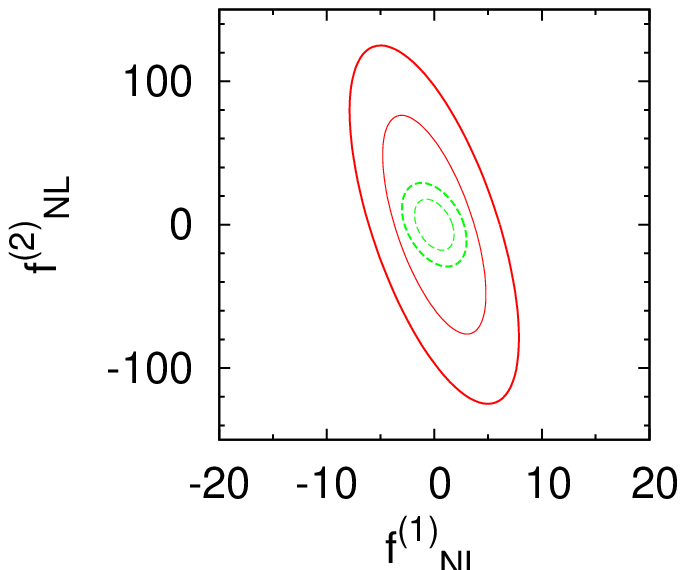}} \\
      \hspace{-10mm}
      \resizebox{40mm}{!}{\includegraphics{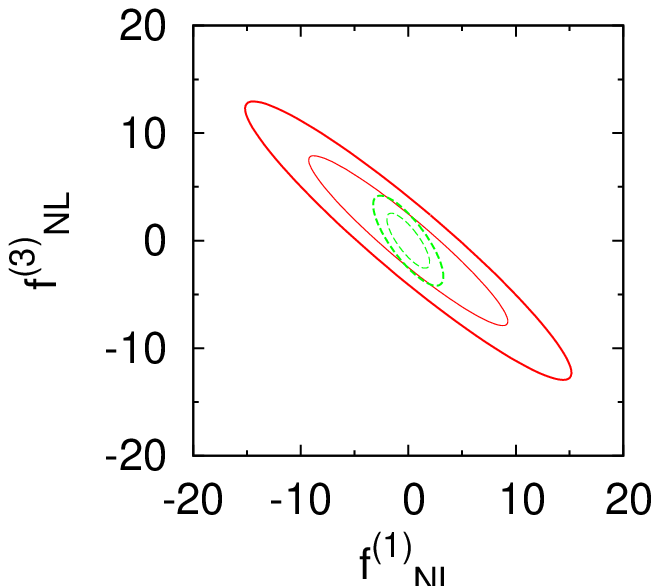}} &
      \hspace{-15mm}
      \resizebox{40mm}{!}{\includegraphics{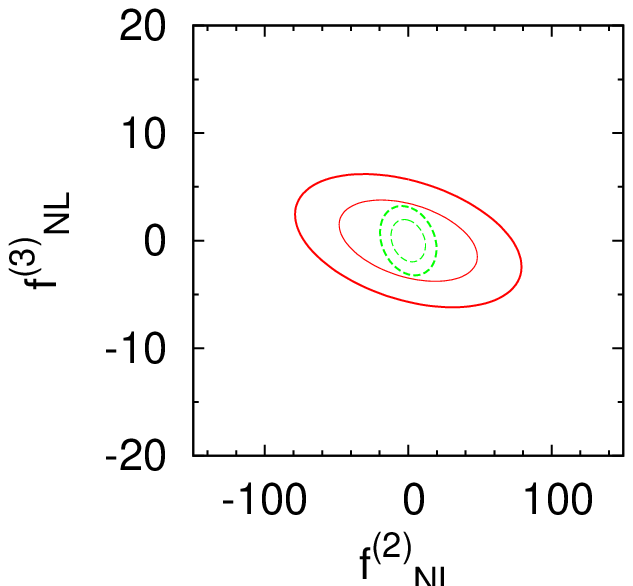}} \\
      \hspace{-10mm}
      \resizebox{40mm}{!}{\includegraphics{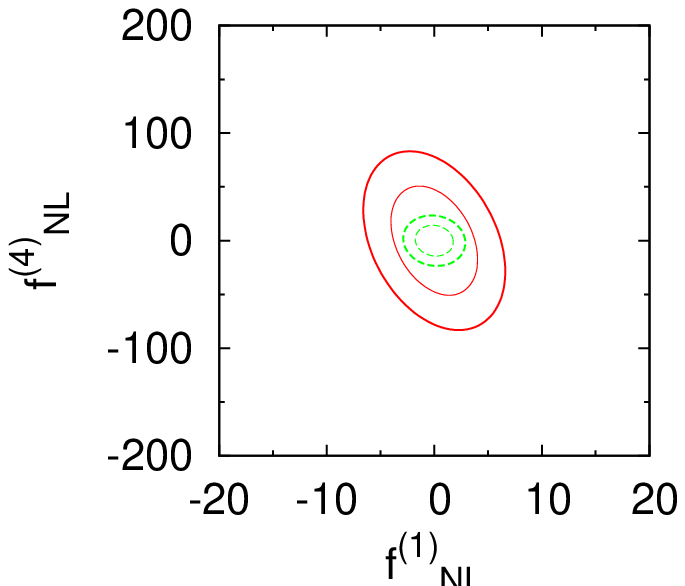}} &
      \hspace{-15mm}
      \resizebox{40mm}{!}{\includegraphics{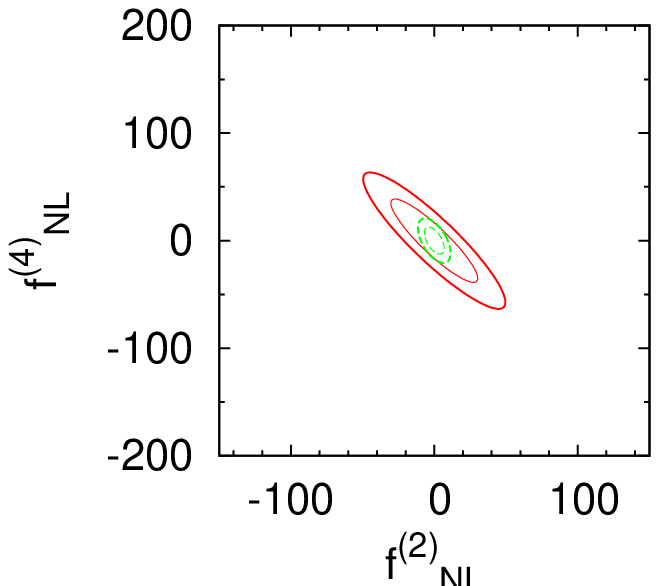}} &
      \hspace{-15mm}
      \resizebox{40mm}{!}{\includegraphics{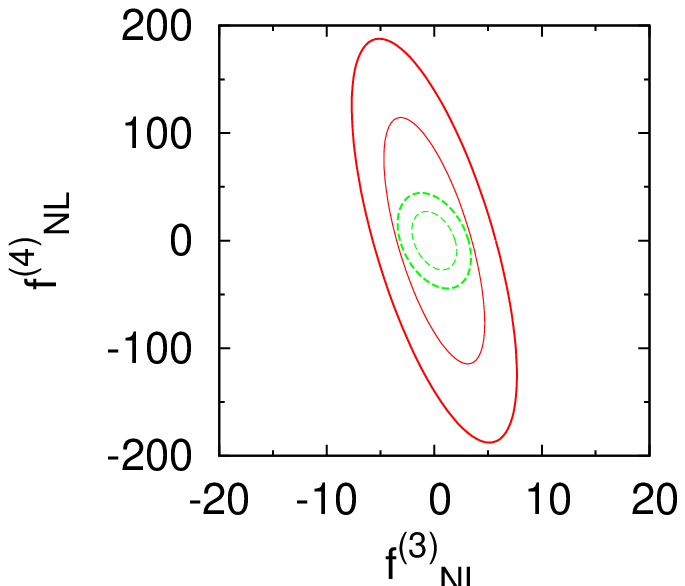}} \\
      \hspace{-10mm}
      \resizebox{40mm}{!}{\includegraphics{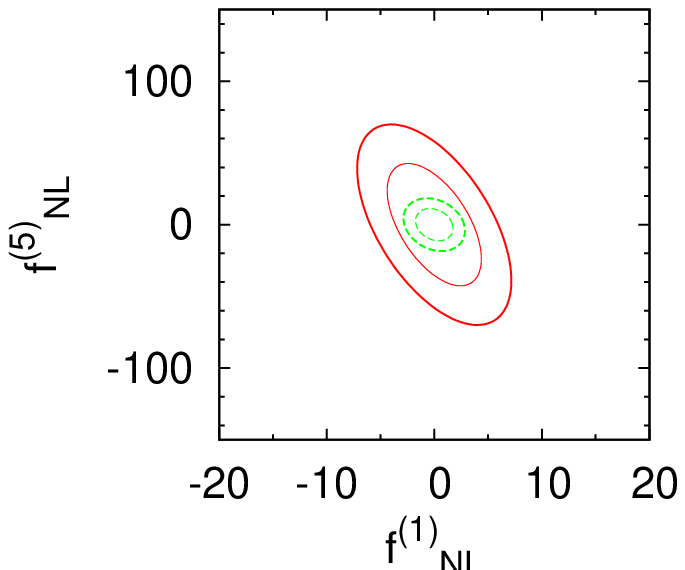}} &
      \hspace{-15mm}
      \resizebox{40mm}{!}{\includegraphics{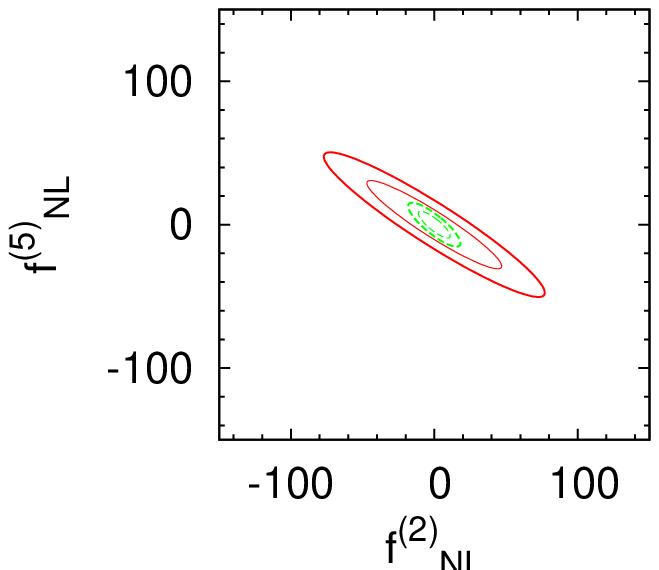}} &
      \hspace{-15mm}
      \resizebox{40mm}{!}{\includegraphics{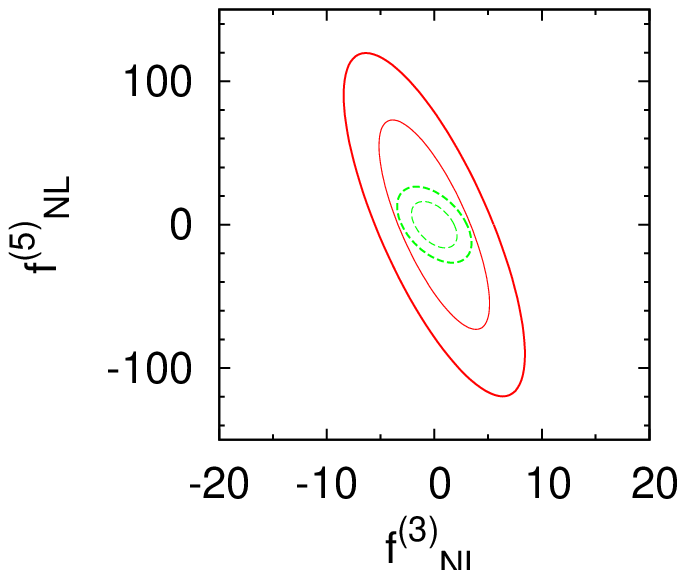}} &
      \hspace{-15mm}
      \resizebox{40mm}{!}{\includegraphics{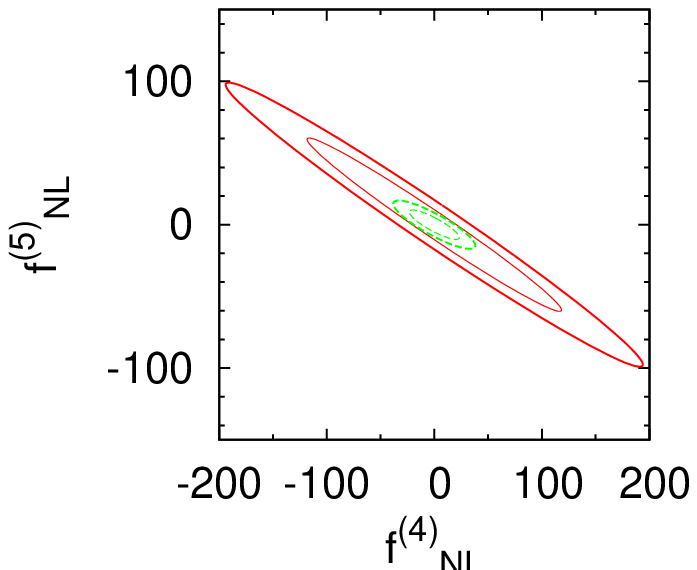}} \\
      \hspace{-10mm}
      \resizebox{40mm}{!}{\includegraphics{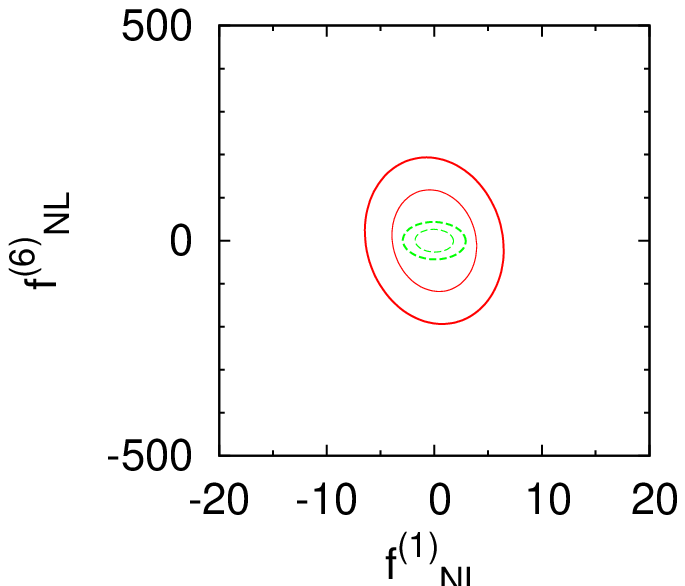}} &
      \hspace{-15mm}
      \resizebox{40mm}{!}{\includegraphics{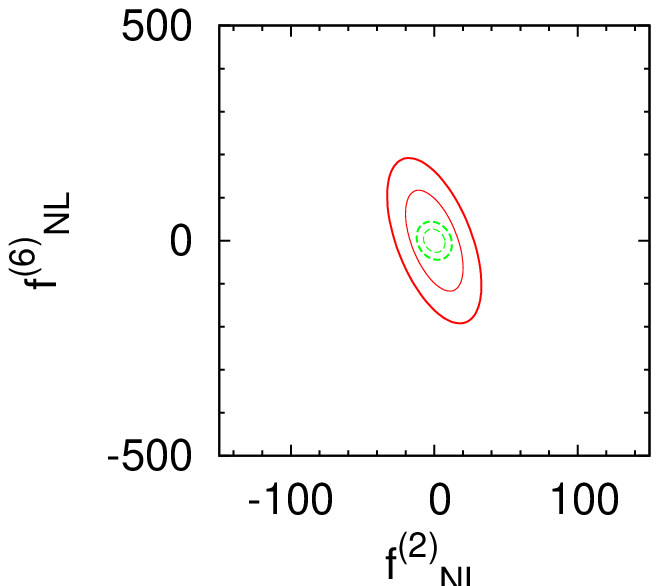}} &
      \hspace{-15mm}
      \resizebox{40mm}{!}{\includegraphics{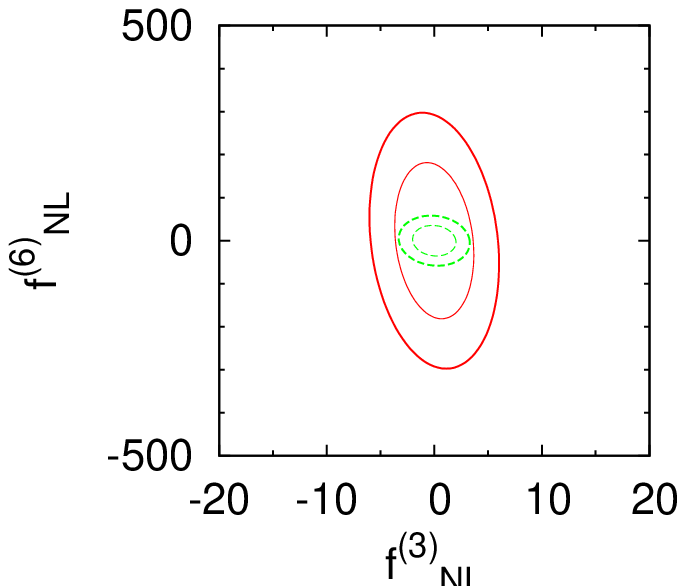}} &
      \hspace{-15mm}
      \resizebox{40mm}{!}{\includegraphics{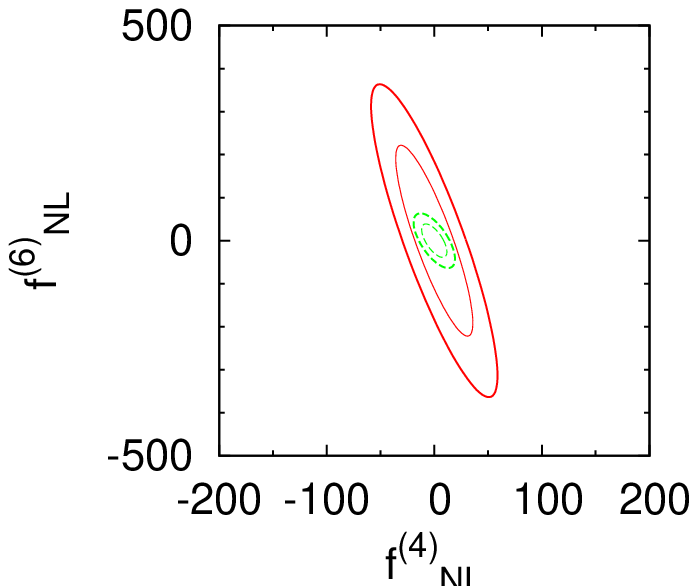}} &
      \hspace{-15mm}
      \resizebox{40mm}{!}{\includegraphics{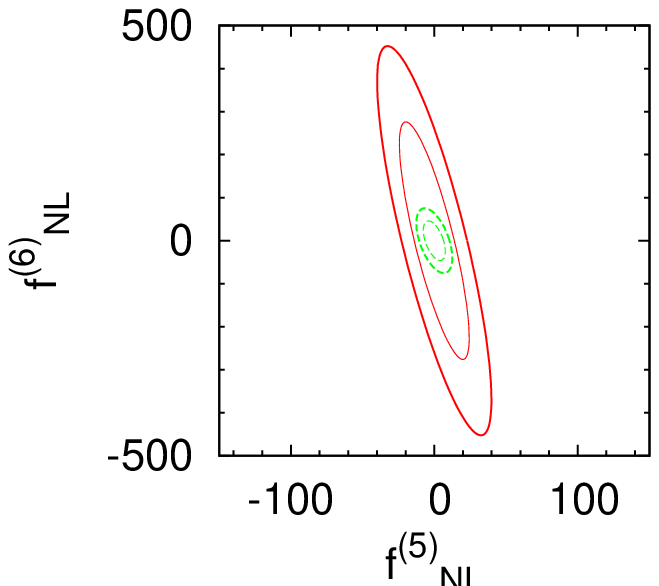}} 
    \end{tabular}
  \end{center}
  \caption{2d non-marginalized constraints on non-Gaussianity parameters for $N_\nu=3.04$.
  }
  \label{fig:Nnu3_Fij}
\end{figure}

We also performed the same analysis for the case of 
fixed $N_\nu=3.04$. Marginalized and non-marginalized 
constraints on $f^{(a)}_{\rm NL}$ are shown in Figs.~\ref{fig:Nnu3_margeFij} 
and \ref{fig:Nnu3_Fij}, respectively. 
Parameter uncertainties $\Delta f_{\rm NL}^{(a)}$
are listed in Table \ref{tbl:deltaf_304}.
In the context of isocurvature perturbations in extra radiation,
such the case is realized when, while the fraction of extra radiation 
in energy density of DR is quite small, the amplitude of 
isocurvature perturbations in extra radiation $S_\mathrm{X}$ 
is large enough for the total DR isocurvature perturbations 
$S_\mathrm{DR}$ to be yet non-negligible. 
On the other hand, this is also naturally realized without 
extra radiation; $S_\mathrm{DR}$ is non-zero
if there are isocurvature perturbations in the lepton number
and non-Gaussian isocurvature perturbations in the lepton number 
may be produced in the Affleck-Dine mechanism, as shown in Sec.~\ref{sec:Qball}.

Compared with the case of $N_\nu=4$, 
constraints on $f^{(a)}_{\rm NL}$ are less stringent 
for the case of $N_\nu=3.04$. 
This can be understood as follows.
As can be seen in Eqs. (3.23) and (3.24) 
of Ref. \cite{Kawasaki:2011rc}, given a fixed $S_{\rm DR}$, the initial perturbations
is roughly proportional to the $\hat R_{\rm DR}\sim N_{\rm eff}$, 
where the $\hat R_{\rm DR}$ is the energy fraction of DR
in the radiation component. Thus, 
apart from the effects of $N_{\rm eff}$ on the background evolution, 
the amplitude of CMB anisotropy from the dark radiation isocurvature perturbation
should be proportional $N_{\rm eff}$, given a fixed $S_{\rm DR}$.
Then the dependence of the bispectrum $b^{A_1P_1,A_2P_2A_3P_3}_{l_1l_2l_3}$ 
on $N_{\rm eff}$ can be determined by the number of $A_i$ equals to 
$S_{\rm DR}$. 
Therefore we can expect $b^{(1)P_1P_2P_3}_{l_1l_2l_3}\propto {N_{\rm eff}}^0$, 
$b^{(2)P_1P_2P_3}_{l_1l_2l_3}\propto b^{(3)P_1P_2P_3}_{l_1l_2l_3}\propto N_{\rm eff}$
$b^{(4)P_1P_2P_3}_{l_1l_2l_3}\propto b^{(5)P_1P_2P_3}_{l_1l_2l_3}\propto {N_{\rm eff}}^2$
and $b^{(6)P_1P_2P_3}_{l_1l_2l_3}\propto {N_{\rm eff}}^3$.
This can be converted into the dependence of $\Delta f_{\rm NL}$ on $N_{\rm eff}$.
We can expect $\Delta f_{\rm NL}^{(1)}\propto {N_{\rm eff}}^0$, 
$\Delta f_{\rm NL}^{(2)}\propto \Delta f_{\rm NL}^{(3)}\propto {N_{\rm eff}}^{-1}$
$\Delta f_{\rm NL}^{(4)}\propto \Delta f_{\rm NL}^{(5)}\propto {N_{\rm eff}}^{-2}$
and $\Delta f_{\rm NL}^{(6)}\propto {N_{\rm eff}}^{-3}$.
This rough estimate can be verified by comparing 
Tables \ref{tbl:deltaf_4} and \ref{tbl:deltaf_304}.

\begin{table}
\begin{center}\caption{Same as in Fig. \ref{tbl:deltaf_4}, 
but for the case with $N_{\rm eff}=3.04$.}
\label{tbl:deltaf_304}
\begin{tabular}{c|rrrrrr}
    \hline\hline
    survey & $f_{\rm NL}^{(1)}$ & $f_{\rm NL}^{(2)}$ & $f_{\rm NL}^{(3)}$
    & $f_{\rm NL}^{(4)}$ & $f_{\rm NL}^{(5)}$ & $f_{\rm NL}^{(6)}$ \\
    \hline\hline
    Planck & 21 & 126 & 27 & 187 & 257 & 339 \\
    \hline
    CVL & 3.5 & 18.3 & 5.0 & 27.2 & 26.4 & 39.3 \\
    \hline\hline
\end{tabular}
\end{center}
\end{table}

\section{Models for non-Gaussian isocurvature perturbations in dark radiation} \label{sec:model}

In this section, we refer to some of particle physics models
in which the DR isocurvature perturbations and their non-Gaussianities arise. 
We discuss two cases separately.
In one scenario, the DR isocurvature perturbation is carried by extra light species.
In the other scenario, ordinary neutrinos have large isocurvature perturbation.

\subsection{Extra light species} \label{sec:SUSYaxionmodel}

Let us consider the cosmological scenario considered in Ref.~\cite{Kawasaki:2011rc} where two scalars, the inflaton $\phi$ and the curvaton $\sigma$, which is light during inflation, contribute to both the adiabatic and isocurvature perturbations.
Using the $\delta N$ formalism~\cite{Sasaki:1995aw,Lyth:2004gb}, 
the expansion coefficients of $\zeta$, such as $N_{\phi},~N_{\sigma}$, are given as
\begin{equation}
\begin{split}
	&N_\phi = \frac{1}{M_{\rm P}^2}\frac{V}{V_\phi}, \\
	&N_\sigma = \frac{3+R}{6\sigma_i}\left( 
		\frac{\hat R_r R_r^{(\sigma)}}{R_r}+\frac{\hat R_X R_X^{(\sigma)}}{R_X} \right),\\
	&N_{\phi\phi} = \frac{1}{M_{\rm P}^2}\left( 1-\frac{VV_{\phi\phi}}{V_\phi^2}\right),\\
	&N_{\sigma\sigma}=\frac{2}{9\sigma_i^2}\frac{3+R}{4}
		\left(\frac{\hat R_r R_r^{(\sigma)}}{R_r}+\frac{\hat R_X R_X^{(\sigma)}}{R_X} \right) \\
		&~~~~~\times \left[ 
		3+4R-2R^2 -2(3+R)\left(\frac{\hat R_r R_r^{(\sigma)}}{R_r}+\frac{\hat R_X R_X^{(\sigma)}}{R_X} \right)
		\right].
\end{split}
\label{Ns}
\end{equation}
In a similar manner, those of $S_{\rm DR}$ are given as
\begin{equation}
\begin{split}
	&S_\sigma = -\frac{3+R}{2\sigma_i}\frac{\hat R_r \hat R_X}{\hat R_{\rm DR}}
	\left( \frac{R_r^{(\sigma)}}{R_r}-\frac{R_X^{(\sigma)}}{R_X} \right)(1-\hat c_\nu)
	,\\
	&S_{\sigma\sigma} = 
	\frac{3+R}{2\sigma_i^2}\frac{\hat R_r \hat R_X}{\hat R_{\rm DR}}
	\left( \frac{R_r^{(\sigma)}}{R_r}-\frac{R_X^{(\sigma)}}{R_X} \right)\frac{(1-\hat c_\nu)}{3} \\
	&~~~\times \left[ 2R^2-4R-3 + \frac{3+R}{\hat R_{\rm DR}}
	     \left(    \frac{\hat R_r R_r^{(\sigma)} (\hat c_\nu + \hat R_{\rm DR}) }{R_r}
			+\frac{\hat R_X R_X^{(\sigma)} (1+ \hat R_{\rm DR})  }{R_X}  \right)
	\right]
	,\\
	&S_\phi = S_{\phi\phi} = 0.
\end{split}
\label{Ss}
\end{equation}
The meanings of the symbols are as follows: 
$M_{P}$ is the reduced Planck mass.
$V$ is the potential of $\phi$ and 
$V_{\phi}$ and $V_{\phi\phi}$ are the first and second derivatives of $V$, respectively, evaluated when observable scales exit the horizon.
$R_{\sigma}$ is the ratio of the energy density of $\sigma$ to the total energy density at its decay
and $R \equiv 3R_\sigma/(4-R_\sigma)$.
$R_i$ is the ratio of the energy density of a fluid $i$ to the total energy density 
at the decay of $\sigma$, and $\hat{R}_i$ is that at the electron-positron annihilation.
The subscripts $r$, $X$ and DR mean the relativistic particles in the Standard Model, 
the extra radiation and the dark radiation, respectively.
$R^{(\sigma)}_i$ is the ratio of energy density of the fluid $i$ generated by $\sigma$ decay at that time.
$\sigma_i$ is the amplitude of the oscillation of $\sigma$ when it starts to oscillate.
$\hat{c}_{\nu} \simeq 0.405$ is the ratio of the energy density of neutrino to that of standard model relativistic particles
(photons and neutrinos) after the electron-positron annihilation.
We can get (\ref{Ns}) and (\ref{Ss}) by comparing the energy densities of various components before and after the events such as $\sigma$ decay, neutrino decoupling and electron-positron annihilation, when the energy ratio of the standard model radiation and that of the dark radiation change, on the uniform density slice.
Therefore (\ref{Ns}) and (\ref{Ss}) are written in terms of the energy ratio of each component at such events.
The effective number of neutrino species in this scenario is
\begin{equation}
	\Delta N_{\rm eff} = \frac{3 \hat R_X}{\hat c_\nu \hat R_r}.   \label{DNeff}
\end{equation}
We refer to Ref.~\cite{Kawasaki:2011rc} for details and derivations of these quantities.
Although these expression are rather lengthy, they are greatly simplified in some concrete situation,
as will be seen in the following examples.



Using these quantities, the non-Gaussianity parameters defined in Eq.~(\ref{fNLs}) are expressed as
\begin{eqnarray}
f_{\rm NL}^{\zeta,\zeta\zeta}(k_1,k_2,k_3) 
&\equiv&\frac{N_{\phi}^2N_{\phi\phi}+N_{\sigma}^2N_{\sigma\sigma}}{(N_\phi^2 + N_\sigma^2)^2}
+\frac{N_{\phi\phi}^3+N_{\sigma\sigma}^3}{(N_\phi^2 + N_\sigma^2)^3}
\ln(k_bL)\Delta_\zeta^2(k_1), 
\label{eq:f_zzz}\\
f_{\rm NL}^{S_{\rm DR},\zeta\zeta}(k_1,k_2,k_3) 
&\equiv&\frac{N_{\sigma}^2S_{\sigma\sigma}}{(N_\phi^2 + N_\sigma^2)^2}
+\frac{N_{\sigma\sigma}^2S_{\sigma\sigma}}{(N_\phi^2 + N_\sigma^2)^3}
\ln(k_bL)\Delta_\zeta^2(k_1), \notag
\\
f_{\rm NL}^{\zeta,S_{\rm DR}\zeta}(k_1,k_2,k_3)
&=&f^{\zeta,\zeta S_{\rm DR}}(k_1,k_2,k_3) \\
&\equiv&\frac{N_{\sigma}S_{\sigma}N_{\sigma\sigma}}{(N_\phi^2 + N_\sigma^2)^2}
+\frac{N_{\sigma\sigma}^2S_{\sigma\sigma}}{(N_\phi^2 + N_\sigma^2)^3}
\ln(k_bL)\Delta_\zeta^2(k_1), \notag
\\
f_{\rm NL}^{\zeta,S_{\rm DR}S_{\rm DR}}(k_1,k_2,k_3) 
&\equiv&\frac{S_{\sigma}^2N_{\sigma\sigma}}{(N_\phi^2 + N_\sigma^2)^2}
+\frac{N_{\sigma\sigma}S_{\sigma\sigma}^2}{(N_\phi^2 + N_\sigma^2)^3}
\ln(k_bL)\Delta_\zeta^2(k_1), \notag
\\
f_{\rm NL}^{S_{\rm DR},\zeta S_{\rm DR}}(k_1,k_2,k_3)
&=&f^{S_{\rm DR},S_{\rm DR}\zeta}(k_1,k_2,k_3) \\
&\equiv&\frac{N_{\sigma}S_{\sigma}S_{\sigma\sigma}}{(N_\phi^2 + N_\sigma^2)^2}
+\frac{N_{\sigma\sigma}S_{\sigma\sigma}^2}{(N_\phi^2 + N_\sigma^2)^3}
\ln(k_bL)\Delta_\zeta^2(k_1), \notag
\\
f_{\rm NL}^{S_{\rm DR},S_{\rm DR}S_{\rm DR}}(k_1,k_2,k_3) 
&\equiv&\frac{S_{\sigma}^2S_{\sigma\sigma}}{(N_\phi^2 + N_\sigma^2)^2}
+\frac{S_{\sigma\sigma}^3}{(N_\phi^2 + N_\sigma^2)^3}
\ln(k_bL)\Delta_\zeta^2(k_1). 
\label{eq:f_SSS}
\end{eqnarray}
Here, we include the ``quadratic'' type components~\cite{Suyama:2008nt,
Kawasaki:2008sn,Kawasaki:2008pa,Hikage:2008sk,Hikage:2009rt}, 
which consist of three quadratic terms of $\delta \phi_i$ in each $\zeta$ or $S_{\rm DR}$, in addition to the leading ``linear'' 
components, which consist of a quadratic term in one 
of three $\zeta$ or $S_{\rm DR}$ and two linear terms from others.
While $f_{\rm NL}^{\zeta,\zeta\zeta}$, etc. in Eqs.~\eqref{eq:f_zzz}-\eqref{eq:f_SSS} 
are in principle not constant, their scale-dependences 
due from the factor $\ln(k_bL)\Delta_\zeta^2(k_i)$ are quite moderate
since $P_\zeta(k)$ is nearly scale-invariant.
Therefore, so long as we consider observations sensitive to 
scales over only a few orders of magnitude, 
we can approximately ignore the scale-dependences.
When we in the next section consider CMB signatures of 
non-Gaussian isocurvature perturbations in dark radiation, 
we adopt this approximation and regard the quantities 
$f_{\rm NL}^{\zeta,\zeta\zeta}$, etc. as constants.
Then given constant $f_{\rm NL}^{\zeta,\zeta\zeta}$, etc., as shown in Eq.~(\ref{eq:localB}),
the bispectra can be written as 
\begin{eqnarray}
&&B^{\zeta\zeta\zeta}(k_1,k_2,k_3)
=f_{\rm NL}^{\zeta,\zeta\zeta}\left[P_\zeta(k_2)P_\zeta(k_3)
+\mbox{(2 cyclics of \{123\})}\right], 
\label{eq:B_zzz2}\\
&&B^{\zeta\zeta S_{\rm DR}}(k_1,k_2,k_3)
=f_{\rm NL}^{\zeta,\zeta S_{\rm DR}}\left[P_\zeta(k_2)P_\zeta(k_3)+P_\zeta(k_3)P_\zeta(k_1)\right]
+f_{\rm NL}^{S_{\rm DR},\zeta\zeta}P_\zeta(k_1)P_\zeta(k_2), \\
&&B^{\zeta S_{\rm DR}S_{\rm DR}}(k_1,k_2,k_3)
=f_{\rm NL}^{\zeta,S_{\rm DR}S_{\rm DR}}P_\zeta(k_2)P_\zeta(k_3) \\
&&\hspace{50mm}+f_{\rm NL}^{S_{\rm DR},S_{\rm DR}\zeta}
\left[P_\zeta(k_3)P_\zeta(k_1)+P_\zeta(k_1)P_\zeta(k_2)\right], \notag\\
&&B^{S_{\rm DR}S_{\rm DR}S_{\rm DR}}(k_1,k_2,k_3)
=f_{\rm NL}^{S_{\rm DR},S_{\rm DR}S_{\rm DR}}\left[P_\zeta(k_2)P_\zeta(k_3) 
+\mbox{(2 cyclics of \{123\})}\right].
\label{eq:B_SSS2} 
\end{eqnarray}

In the following subsections we consider two cases. 
One is the case where the $\sigma$ dominantly decays into
extra light species $X$, while the curvature perturbation is dominantly generated by the inflaton (Sec.~\ref{sec:KSVZ}).
The other is the case where the $\sigma$ decays into ordinary radiation and is the dominant source of the adiabatic perturbation, while extra light species are produced in thermal bath after the inflaton decay (Sec.~\ref{sec:DFSZ}).
Both cases are realized in the framework of supersymmetric (SUSY) axion model~\cite{Peccei:1977hh}
as mentioned in the previous work~\cite{Kawasaki:2011rc}, and originally in Ref.~\cite{Ichikawa:2007jv}.
We shall partly repeat discussions there.

\subsubsection{Dark radiation from particle decay} \label{sec:KSVZ}

Let us assume that the primordial curvature perturbation is dominantly produced by the 
inflaton : $N_{\phi} \gg N_{\sigma}$,\footnote{
As shown in (\ref{NSinKSVZ}), $S_{\sigma}$ is comparable to $N_{\sigma}$ in this model, then $N_{\phi} \gg N_{\sigma}$ is required in order to avoid the isocurvature mode comparable to the adiabatic mode.
}
and the inflaton decay only to the visible sector.
This makes the isocurvature mode uncorrelated with the adiabatic mode.
In this setup, we can approximate parameters as 
$R_r^{(\sigma)} \simeq 0$, and $R_X=R_X^{(\sigma)} \simeq R_\sigma \simeq 4R/3$.
We also assume $R_\sigma < 1$ since otherwise $\sigma$ dominates the Universe before it decays
and the Universe would be dominated by $X$.
Then Eqs.~(\ref{Ns}) and (\ref{Ss}) are simplified as
\begin{equation}
\begin{split}
	&N_\sigma \simeq \frac{1}{2\sigma_i}\hat{R}_X,\\
	&N_{\sigma\sigma} \simeq \frac{1}{2\sigma_i^2} \hat{R}_X, \\
	&S_{\sigma} \simeq \frac{3}{2\sigma_i}\frac{1-\hat{c}_{\nu}}{\hat{c}_{\nu}}\hat{R}_X\simeq 3\frac{1-\hat{c}_{\nu}}{\hat{c}_{\nu}}N_{\sigma}, \\
	&S_{\sigma\sigma}\simeq \frac{3}{2\sigma_i^2}\frac{1-\hat{c}_{\nu}}{\hat{c}_{\nu}}\hat{R}_X\simeq 3\frac{1-\hat{c}_{\nu}}{\hat{c}_{\nu}}N_{\sigma\sigma}.
\end{split}
\label{NSinKSVZ}
\end{equation}
Here and hereafter, we assume that the inflaton does not induce the non-Gaussianity, that is, $N_{\phi\phi}\simeq 0$.
The quantity $\hat{R}_X$ is related to $R_X$ as~\cite{Kawasaki:2011rc}
\begin{equation}
\hat{R}_X\simeq \frac{1}{1+\delta}\left(\frac{g_*(H=\Gamma_{\nu})}{g_*(H=\Gamma_{\sigma})}\right)^{1/3}R_X. \label{RXandhatRX}
\end{equation}
Here $\delta = ((11/4)^{1/3}-1)(1-c_{\nu})$, $c_{\nu}=\rho_{\nu}/\rho_r=21/43$ is the ratio of the energy of neutrinos to that of all visible matters at neutrino decoupling, $g_*(H=\Gamma_{\nu})$ is the relativistic degrees of freedom at that time and $g_*(H=\Gamma_{\sigma})$ is that at {\blue $\sigma$} decay.
From Eq.~(\ref{DNeff}), we can also express the effective number of neutrino species as
\begin{equation}
\Delta N_{\rm eff}\simeq \frac{4}{\hat c_{\nu} (1+\delta)}\left(\frac{g_*(H=\Gamma_\nu)}{g_*(H=\Gamma_\sigma)}\right)^{1/3}R.
\end{equation}
Thus $\Delta N_{\rm eff} \sim 1$ if $R$ is not much smaller than one.
Using (\ref{NSinKSVZ}), we get the relations among the non-Gaussianity parameters as
\begin{equation}
\begin{split}
	&f^{S_{\rm DR},\zeta\zeta}_{\rm NL}\simeq f^{\zeta,S_{\rm DR}\zeta}_{\rm NL} \simeq 3\frac{1-\hat{c}_{\nu}}{\hat{c}_{\nu}}f^{\zeta,\zeta\zeta}_{\rm NL},\\
	&f^{\zeta,S_{\rm DR}S_{\rm DR}}_{\rm NL}\simeq f^{S_{\rm DR},\zeta S_{\rm DR}}_{\rm NL} \simeq 9\left(\frac{1-\hat{c}_{\nu}}{\hat{c}_{\nu}}\right)^2f^{\zeta,\zeta\zeta}_{\rm NL}, \\
	&f^{S_{\rm DR},S_{\rm DR}S_{\rm DR}}_{\rm NL} \simeq 27\left(\frac{1-\hat{c}_{\nu}}{\hat{c}_{\nu}}\right)^3f^{\zeta,\zeta\zeta}_{\rm NL}.
\end{split}
\label{relationfKSVZ}
\end{equation}
 Thus these non-Gaussianity parameters are comparable.
 The magnitude of them is roughly given by
 \begin{align}
 f^{\zeta,\zeta\zeta}_{\rm NL}(k_1,k_2,k_3) & \simeq \frac{N_{\sigma}^2N_{\sigma\sigma}}{N_{\phi}^4}+
                                                        \frac{N_{\sigma\sigma}^3}{N_{\phi}^6}\ln (k_bL)\Delta_{\zeta}^2(k_1) \nonumber \\
                                                                     & \sim \epsilon^2\left(\frac{M_{\rm P}}{\sigma_i}\right)^4\hat{R}_X^3
                                                                     + \epsilon^3\left(\frac{M_{\rm P}}{\sigma_i}\right)^6\hat{R}_X^3\Delta_{\zeta}^2(k_1),
 \end{align}
where $\epsilon = \frac{1}{2}M_{\rm P}^2(V_{\phi}/V)^2$ is the slow-roll parameter.
It is easily found that the first term is of the order of $(P_{S_{\rm DR}}/P_\zeta)^2\times (1/\hat R_X)$
while the second term is of the order of $(P_{S_{\rm DR}}/P_\zeta)^3\times (\Delta_\zeta^2/\hat R_X^3)$.
Therefore, the non-linearity parameter can be large enough to be probed
for not so small $P_{S_{\rm DR}}$ compared to $P_\zeta$ and $\hat R_X \ll 1$.
Thus even if $R_X$ is very small and there is no significant deviation from $N_{\rm eff} = 3.046$, 
the DR isocurvature mode and its non-Gaussianity may be detected.


Now let us estimate $\hat R_X \sim R$ and $S_{\rm DR}$ in the SUSY KSVZ axion model~\cite{Kim:1979if}.
In a SUSY axion model~\cite{Rajagopal:1990yx}, the saxion $\sigma$, the scalar partner of the PQ axion, exists and has
a mass $m_{\sigma}$ which ranges from $\mathcal{O}({\rm keV})$ to $\mathcal{O}({\rm TeV})$, 
in accordance with the SUSY breaking scale. 
The saxion can have a large initial amplitude $\sigma_i$ during inflation, and may obtain quantum 
fluctuations $\delta\sigma \sim H_{\rm inf}/2\pi$ 
if it is much lighter than the Hubble parameter during inflation.
In this model, the dominant decay channel of the saxion is typically that into two axions.
Relativistic axions produced by the saxion decay behave as an extra radiation $X$, since they are decoupled from
ordinary matter almost completely.
Here the $R\simeq 3R_{\sigma}/4$ is given by~\cite{Kawasaki:2011rc}
\begin{equation}
	R \simeq 2\times 10^{-4}\left( \frac{T_{\rm R}}{10^6{\rm GeV}} \right)
	\left( \frac{1{\rm GeV}}{m_\sigma} \right)^{3/2}
	\left( \frac{f_a}{10^{12}{\rm GeV}} \right)^3
	\left( \frac{\sigma_i}{f_a} \right)^2,    \label{R_saxion}
\end{equation}
for $R\ll 1$ and $m_\sigma > \Gamma_\phi$ and
\begin{equation}
	R \simeq 2\times 10^{-3}
	\left( \frac{1{\rm GeV}}{m_\sigma} \right)
	\left( \frac{f_a}{10^{12}{\rm GeV}} \right)^3
	\left( \frac{\sigma_i}{f_a} \right)^2,    \label{R_saxion2}
\end{equation}
for $R\ll 1$ and $m_\sigma < \Gamma_\phi$, where $f_a$ is the PQ symmetry breaking scale,
$T_{\rm R}$ is the reheating temperature after the inflation
and $\Gamma_{\phi}$ is the decay rate of the inflaton.
Correspondingly, the magnitude of the DR isocurvature perturbation is given by
\begin{equation}
	S_{\rm DR} \simeq 4\times 10^{-7}\left( \frac{T_{\rm R}}{10^6{\rm GeV}} \right)
	\left( \frac{1{\rm GeV}}{m_\sigma} \right)^{3/2}
	\left( \frac{f_a}{10^{12}{\rm GeV}} \right)^2
	\left( \frac{H_{\rm inf}}{10^{10}{\rm GeV}} \right)
	\left( \frac{\sigma_i}{f_a} \right),    
\end{equation}
for $R\ll 1$ and $m_\sigma > \Gamma_\phi$ and
\begin{equation}
	S_{\rm DR} \simeq 4\times 10^{-6}
	\left( \frac{1{\rm GeV}}{m_\sigma} \right)
	\left( \frac{f_a}{10^{12}{\rm GeV}} \right)^2
	\left( \frac{H_{\rm inf}}{10^{10}{\rm GeV}} \right)
	\left( \frac{\sigma_i}{f_a} \right),    
\end{equation}
for $R\ll 1$ and $m_\sigma < \Gamma_\phi$.
Here $S_{\rm DR}$ is regarded as $S_{\rm DR} = \sqrt{ (k^3/2\pi^2) P_{S_{\rm DR}S_{\rm DR} } }$ which should be compared with $\sqrt{ (k^3/2\pi^2) P_{\zeta\zeta} }\simeq 5\times 10^{-5}$.
Thus the magnitude of the DR isocurvature perturbation can be sizable.

\subsubsection{Dark radiation from thermal bath} \label{sec:DFSZ}

Next, we consider the case $\sigma$ takes a role of the curvaton, and hence it dominantly sources the
adiabatic perturbation : $N_\sigma \gg N_\phi$.
Moreover, we assume that the inflaton decays into ordinary radiation with a branching ratio $r_\phi$,
and into $X$ with a branching ratio $1-r_\phi$.
The curvaton $\sigma$ is assumed to decay only into ordinary radiation.
In this model, we make use of following approximations :
$R^{(\sigma)}_r\simeq R_{\sigma}(\sim 4R/3)$, $R^{(\sigma)}_X\simeq 0$.
We also assume $R_\sigma<1$ since otherwise the $\sigma$ decay releases huge amount of entropy
and it dilutes the $X$ abundance significantly.
Under these assumptions, Eqs.~(\ref{Ns}) and (\ref{Ss}) are simplified as
\begin{equation}
\begin{split}
	&N_\sigma \simeq \frac{2R}{3\sigma_i},\\
	&N_{\sigma\sigma} \simeq \frac{2R}{3\sigma_i^2} , \\
	&S_{\sigma} \simeq -3\frac{1-\hat{c}_{\nu}}{\hat{c}_{\nu}}\hat{R}_X\frac{2R}{3\sigma_i} \simeq  -3\frac{1-\hat{c}_{\nu}}{\hat{c}_{\nu}}\hat{R}_X N_{\sigma}, \\
	&S_{\sigma\sigma}\simeq -3\frac{1-\hat{c}_{\nu}}{\hat{c}_{\nu}}\hat{R}_X \frac{2R}{3\sigma_i^2} \simeq -3\frac{1-\hat{c}_{\nu}}{\hat{c}_{\nu}}\hat{R}_X N_{\sigma\sigma}.
\end{split}
\label{NSinDFSZ}
\end{equation}
The relation between $\hat{R}_X$ and $R_X$ is given by Eq.~(\ref{RXandhatRX})
where $R_X$ is given by
\begin{equation}
R_X=(1-r_{\phi})(1-R_{\sigma}).   \label{RX_DFSZ}
\end{equation}
In this model, $\Delta N_{\rm eff}$ is given by
\begin{equation}
\Delta N_{\rm eff}\simeq \frac{3}{\hat c_{\nu} (1+\delta)}\left(\frac{g_*(H=\Gamma_\nu)}{g_*(H=\Gamma_\sigma)}\right)^{1/3}(1-r_{\phi}).
\end{equation}
The relationships among the non-Gaussianity parameters become
\begin{equation}
\begin{split}
	&f^{S_{\rm DR},\zeta\zeta}_{\rm NL}\simeq f^{\zeta,S_{\rm DR}\zeta}_{\rm NL} \simeq -3\frac{1-\hat{c}_{\nu}}{\hat{c}_{\nu}}\hat{R}_Xf^{\zeta,\zeta\zeta}_{\rm NL},\\
	&f^{\zeta,S_{\rm DR}S_{\rm DR}}_{\rm NL}\simeq f^{S_{\rm DR},\zeta S_{\rm DR}}_{\rm NL} \simeq 9\left(\frac{1-\hat{c}_{\nu}}{\hat{c}_{\nu}}\right)^2\hat{R}_X^2f^{\zeta,\zeta\zeta}_{\rm NL}, \\
	&f^{S_{\rm DR},S_{\rm DR}S_{\rm DR}}_{\rm NL} \simeq -27\left(\frac{1-\hat{c}_{\nu}}{\hat{c}_{\nu}}\right)^3\hat{R}_X^3f^{\zeta,\zeta\zeta}_{\rm NL}.
\end{split}
\label{relationfKSVZ}
\end{equation}
In this case, 
\begin{equation}
f^{S_{\rm DR},S_{\rm DR}S_{\rm DR}}_{\rm NL} \ll f^{\zeta,S_{\rm DR}S_{\rm DR}}_{\rm NL}\simeq f^{S_{\rm DR},\zeta S_{\rm DR}}_{\rm NL} \ll f^{S_{\rm DR},\zeta\zeta}\simeq f^{\zeta,S_{\rm DR}\zeta}_{\rm NL}  \ll f^{\zeta,\zeta\zeta}.
\end{equation}
This is because $\sigma$, which is the origin of the non-Gaussianity, dominantly decays to visible particles.
Since we assume that the primordial curvature perturbation is dominantly produced by $\sigma$.
the non-linearity parameter for the adiabatic perturbation is given by
\begin{equation}
f_{\rm NL}^{\zeta,\zeta\zeta}(k_1,k_2,k_3)\simeq \frac{3}{2R} + \frac{27}{8R^3}\ln (k_b L) \Delta^2_{\zeta}(k_1).
\end{equation}
We see that the non-Gaussianity in the adiabatic perturbation becomes large for $R\ll 1$
while that of the DR isocurvature mode is given by $f_{\rm NL}^{S_{\rm DR},\zeta\zeta}\sim f_{\rm NL}^{\zeta,S_{\rm DR}\zeta}
\sim (P_{\rm S_{\rm DR}}/P_\zeta)^{1/2}\times (1/R)$, 
can be order unity if $P_{S_{\rm DR}}$ is close to the observational upper bound, or $\hat R_X$ is close to unity.

The above situation is actually realized in the SUSY DFSZ axion model~\cite{Dine:1981rt}
once the saxion is identified as the curvaton $\sigma$ and the axion as the extra light species $X$.
In this model, the dominant decay channel of the saxion may be that into a Higgs boson pair.
Therefore, the energy of saxion is almost converted to visible particles.
On the other hand, there may be axions produced from the thermal bath during reheating.
Let us suppose that the reheating temperature is high so that it satisfies $T_{\rm R} \gtrsim T_{\rm D}$,
where $T_{\rm D}\simeq 10^7{\rm GeV}(f_a/10^{10}{\rm GeV})^{2.246}$ 
is the temperature at the axion decoupling from thermal bath~\cite{Graf:2010tv}.
Thus axions are thermalized after the inflaton decay.
The ratio of the axion energy density to the total energy density at the axion decoupling is given by $g_*(T=T_{\rm D})^{-1}$.
The inflaton decay branching ratio into $X$, $1-r_\phi$, is replaced by
the ratio of the energy density of axions to that of the whole radiation originating from the inflaton 
at the epoch of saxion decay. Thus it is estimated to be
\begin{equation}
	1-r_\phi = \frac{1}{g_*(T=T_{\rm D})} \left( \frac{g_*(H=\Gamma_\sigma)}{g_*(T=T_{\rm D})} \right)^{1/3},
\end{equation}
where $\Gamma_{\sigma}$ is the saxion decay rate.
From this, we see that $1-r_\phi$ is typically much smaller than $1$.
Eventually, the amount of axions is smaller than that of visible particles whether they are produced by the inflaton or the saxion. 
Then we have $R_X\ll 1$ from Eq.~(\ref{RX_DFSZ}).
The ratio of the saxion energy density to the total energy density at the epoch of saxion decay is given by
\begin{equation}
	R \simeq 7\times 10^{-2}\left( \frac{T_{\rm R}}{10^6{\rm GeV}} \right)
	\left( \frac{m_\sigma}{1{\rm GeV}} \right)^{1/2}
	\left( \frac{1{\rm TeV}}{\mu}\right)^2
	\left( \frac{f_a}{10^{15}{\rm GeV}} \right)^3
	\left( \frac{\sigma_i}{f_a} \right)^2,    \label{R_saxion3}
\end{equation}
for $m_\sigma > \Gamma_\phi$ and
\begin{equation}
	R \simeq 7\times 10^{-1}
	\left( \frac{m_\sigma}{1{\rm GeV}} \right)
	\left( \frac{1{\rm TeV}}{\mu}\right)^2
	\left( \frac{f_a}{10^{15}{\rm GeV}} \right)^3
	\left( \frac{\sigma_i}{f_a} \right)^2,    \label{R_saxion4}
\end{equation}
for $m_\sigma < \Gamma_\phi$, where $\mu$ denotes the higgsino mass.
The magnitude of the DR isocurvature perturbation is given by
\begin{equation}
	S_{\rm DR} \simeq 6\times 10^{-7}\left( \frac{T_{\rm R}}{10^6{\rm GeV}} \right)
	\left( \frac{m_\sigma}{1{\rm GeV}} \right)^{1/2}
	\left( \frac{1{\rm TeV}}{\mu}\right)^2
	\left( \frac{f_a}{10^{15}{\rm GeV}} \right)^2
	\left( \frac{H_{\rm inf}}{10^{13}{\rm GeV}} \right)
	\left( \frac{\sigma_i}{f_a} \right),  
\end{equation}
for $m_\sigma > \Gamma_\phi$ and
\begin{equation}
	S_{\rm DR} \simeq 6\times 10^{-6}
	\left( \frac{m_\sigma}{1{\rm GeV}} \right)
	\left( \frac{1{\rm TeV}}{\mu}\right)^2
	\left( \frac{f_a}{10^{15}{\rm GeV}} \right)^2
	\left( \frac{H_{\rm inf}}{10^{13}{\rm GeV}} \right)
	\left( \frac{\sigma_i}{f_a} \right),  
\end{equation}
for $m_\sigma < \Gamma_\phi$, where we have used $g_*(T=T_{\rm D}) = 228.75$.
Thus the DR isocurvature perturbation can be sizable for some parameter choices
even if $R_X$ is much smaller then unity.

\subsection{Large lepton asymmetry} \label{sec:Qball}

Next, let us consider the case where the neutrino number density has an isocurvature perturbation.
First note that neutrinos are in thermal equilibrium before the decoupling at $T\sim 1$\,MeV.
Thus neutrinos can only have an adiabatic perturbation unless they are produced after the decoupling
or they have a chemical potential, i.e., there is asymmetry in the neutrino sector.
We consider the latter possibility hereafter.
The lepton number is conserved well after the electroweak symmetry breaking (EWSB) since the 
sphaleron effect~\cite{Kuzmin:1985mm} is suppressed.
Therefore, if the asymmetry in the lepton number, $n_{L}-n_{\bar L}$, is created after the EWSB,
it survives thereafter.
Thus spatial fluctuations in the lepton asymmetry on large scales, if exist, are also conserved.
\footnote{
	If the lepton asymmetry is produced well before the EWSB, the sphaleron effect converts it into the 
	baryon number. In this case the lepton number must be same order as the baryon number
	and hence its effect is negligible.
	Note also that even in the case where the lepton asymmetry is produced well after the EWSB,
	the asymmetry in the charged lepton sector must be same as that in the baryon sector
	because of the electric charge conservation.
	We use the conventional terminology ``lepton asymmetry'' hereafter, but it actually means
	the asymmetry in the neutrino sector.
}
This opens up a possibility that an observable (non-Gaussian) isocurvature perturbation in the lepton asymmetry,
or the neutrino density isocurvature perturbation, is created after the EWSB~\cite{Lyth:2002my}.
The asymmetric part, $|n_\nu-n_{\bar\nu}|$, adds to the ordinary
neutrino number density and hence it may significantly contribute to the $N_{\rm eff}$ as well as $S_{\rm DR}$
if the lepton asymmetry and its isocurvature perturbation are large enough.
A concrete example was given in Ref.~\cite{Kawasaki:2002hq},
where it was shown that the late decay of Q-balls can create large lepton asymmetry.
It is interesting because we do not need an extra radiation particle $X$ to produce 
significant amount of DR isocurvature perturbation.
Let us follow the arguments of Ref.~\cite{Kawasaki:2002hq} and estimate $\Delta N_{\rm eff}$ and $S_{\rm DR}$ in this model.

A large lepton asymmetry is created through the Affleck-Dine (AD) mechanism~\cite{Affleck:1984fy,Dine:1995kz}.
Specifically, we make use of the $LL\bar e$ flat direction (also called as the AD field)~\cite{Dine:1995kz,Gherghetta:1995dv}.
It does not have the baryon number, and hence it can create large lepton asymmetry without producing 
too much baryon asymmetry.
If the AD field fragments into Q-balls~\cite{Coleman:1985ki} 
in which almost all the lepton number is confined~\cite{Kusenko:1997si,Enqvist:1997si,Kasuya:1999wu,Hiramatsu:2010dx}
and they evaporate after the EWSB, a large lepton asymmetry is released and it does not washed out.
Since the AD field may obtain quantum fluctuations in its angular component during inflation~\cite{Linde:1985gh,Kasuya:2008xp},
it results in the isocurvature fluctuation in the lepton asymmetry, i.e., 
the neutrino density (non-Gaussian) isocurvature perturbation.
Non-Gaussianity in the baryonic isocurvature perturbation generated through the AD mechanism
was studied in Ref.~\cite{Kawasaki:2008jy}.

We denote by $\psi$ the AD field along the $LL\bar e$ flat direction.
It is lifted by the dimension six operator in the superpotential, $W = \psi^6 / (6M^3)$ with $M$ being the cutoff scale.
We assume the gauge-mediated SUSY breaking model~\cite{Giudice:1998bp} in the following.
The scalar potential for the AD field is given by~\cite{de Gouvea:1997tn}
\begin{equation}
	V = (m_{3/2}^2-cH^2)|\psi|^2 + M_F^4\left( \log \frac{|\psi|^2}{M_{\rm mess}^2} \right)^2
	+\left( a_m m_{3/2} \frac{\psi^6}{6M^3}+{\rm h.c.} \right) + \frac{|\psi|^{10}}{M^{6}},
	\label{ADpot}
\end{equation}
for $|\psi|>M_{\rm mess}$, where $M_{\rm mess}$ is the messenger scale,
$m_{3/2}$ denotes the gravitino mass, $a_m$ is a constant of order unity
and $M_F^4 \simeq m_{\rm soft}^2 M_{\rm mess}^2$ with $m_{\rm soft}\sim 1$\,TeV.

The lepton number generated through the AD mechanism is estimated as
\begin{equation}
\begin{split}
	\frac{n_L}{s} &\simeq \frac{T_{\rm R} |\psi_{\rm os}|^2}{4m_{3/2}M_P^2}\sin (6\theta) \\
	&\sim 5\times 10^{-3} 
	\left( \frac{T_{\rm R}}{10^5\,{\rm GeV}} \right)
	\left( \frac{1\,{\rm GeV}}{m_{3/2}} \right)^{1/2}
	\left( \frac{M}{10^{20}\,{\rm GeV}} \right)^{3/2} \sin (6\theta),
\end{split}
\end{equation}
where $T_{\rm R}$ is the reheating temperature, $\theta$ is the initial angle of the AD field in the complex plane,
and $|\psi_{\rm os}| \sim (m_{3/2} M^3)^{1/4}$ is the AD field amplitude at the onset of its oscillation.
It can be checked that thermal effects on the AD field potential is neglected 
in this parameter choice~\cite{Allahverdi:2000zd,Anisimov:2000wx}.
The AD field fragments into Q-balls after it starts to oscillate, and 
they once dominate the Universe before they decay if $T_d < (T_{\rm R}/3)(|\psi_{\rm os}|/M_P)^2$,
where $T_d$ is the decay temperature of the Q-ball discussed later.
In this case, the expression becomes $n_L/s  \simeq (T_d/4m_{3/2}) \sin(6\theta)$.
Hereafter we regard the lepton asymmetry, denoted by the subscript $L$, as if it is an extra radiation
component, which has been denoted by $X$ in the previous sections.
The neutrino asymmetry in each flavor is expressed in terms of the chemical potential 
(or the degeneracy parameter) $\xi_{\nu_i}$ as
\begin{equation}
	\frac{n_L}{n_\gamma}=
	\sum_{i=e,\mu,\tau}\frac{n_{\nu_i}-n_{\bar\nu_i}}{n_\gamma} 
	= \sum_{i=e,\mu,\tau}\frac{1}{12\zeta(3)}
	\left( \frac{T_{\nu_i}}{T_\gamma} \right)^3
	(\pi^2 \xi_{\nu_i} +\xi_{\nu_i}^3 ).
\end{equation}
The neutrino chemical potentials contribute to the extra radiation energy density through the relation
\begin{equation}
	\Delta N_{\rm eff} = \frac{3\rho_L}{\rho_\nu} =
	\sum_{i=e,\mu,\tau}\left[ \frac{30}{7}\left( \frac{\xi_{\nu_i}}{\pi} \right)^2
	+   \frac{15}{7}\left( \frac{\xi_{\nu_i}}{\pi} \right)^4 \right].
\end{equation}
Note however that the chemical potential of the electron neutrino directly affects the helium abundance~\cite{Kang:1991xa}
and hence its contribution to the radiation energy density is constrained as
$\Delta N_{\rm eff} \lesssim \mathcal O(0.1)$ depending on 
the neutrino mixing angle $\theta_{13}$~\cite{Mangano:2010ei}.
Thus hereafter we neglect the contribution of the lepton asymmetry to the DR energy density,
although the effect of its isocurvature perturbation may not be neglected.\footnote{
In the limit of $\Delta N_{\rm eff}\ll 1$, the total curvature perturbation
is conserved for all scales of interest.
}

The angular component of the AD field may be light during inflation~\cite{Kasuya:2008xp}
and it leads to the isocurvature perturbation in the lepton asymmetry.
The isocurvature perturbation of the lepton asymmetry is calculated as
\begin{equation}
	S_{L} = n\cot (n\theta) \delta\theta - \frac{1}{2}n^2 (\delta \theta)^2,
\end{equation}
where $\sqrt{\langle\delta\theta^2\rangle} = H_{\rm inf}/(2\pi |\psi_i|)$, with $H_{\rm inf}$ being the 
Hubble scale during inflation and $n=6$ in the present model.
The DR isocurvature perturbation is then estimated as
\begin{equation}
\begin{split}
	S_{\rm DR}&=\frac{\rho_L}{\rho_{\rm DR}}\left( S_L + \frac{2}{3}\frac{\rho_\nu}{\rho_{\rm DR}}S_L^2\right)
	\simeq \frac{\rho_L}{\rho_{\nu}} \left( S_L + \frac{2}{3}S_L^2\right) \\
	& \sim 6\times 10^{-8} 
	 \left( \frac{T_{\rm R}}{10^5\,{\rm GeV}} \right)^2
	 \left( \frac{m_{3/2}}{1\,{\rm GeV}} \right)^{-1}
	 \left( \frac{M}{10^{20}\,{\rm GeV}} \right)^{9/4}
	 \left( \frac{H_{\rm inf}}{10^{14}\,{\rm GeV}} \right)^{3/4},
\end{split}
\end{equation}
where $\rho_{\rm DR} = \rho_L + \rho_\nu$ denotes the DR energy density.
In the second equality, we have neglected the contribution of the lepton asymmetry to the DR energy density for the reason discussed above.
In the last equality, we have considered only the leading term.
In the case of Q-ball domination, we obtain
\begin{equation}
	S_{\rm DR}\sim 2\times 10^{-8} 
	 \left( \frac{T_{d}}{10\,{\rm MeV}} \right)^2
	 \left( \frac{1\,{\rm GeV}}{m_{3/2}} \right)^2
	 \left( \frac{M}{10^{20}\,{\rm GeV}} \right)^{-3/4}
	 \left( \frac{H_{\rm inf}}{10^{14}\,{\rm GeV}} \right)^{3/4}.
\end{equation}
Expanding the DR isocurvature perturbation as $ S_{\rm DR}\simeq S_\theta\delta \theta+(1/2)S_{\theta\theta}(\delta\theta)^2$, 
we obtain
\begin{equation}
\begin{split}
	S_{\theta} &= \frac{\Delta N_{\rm eff}}{3} n \cot(n\theta),\\
	S_{\theta\theta} &= -\frac{\Delta N_{\rm eff}}{3} n^2\left[ 1-\frac{4}{3}\cot^2(n\theta) \right].
\end{split}
\end{equation}
Therefore, we obtain the non-linearity parameter for the DR isocurvature perturbation as
\begin{equation}
	f_{\rm NL}^{S_{\rm DR},S_{\rm DR}S_{\rm DR}}
	= \frac{S_\theta^2S_{\theta\theta}}{N_\phi^4 |\psi_i|^4}+
	 \frac{S_{\theta\theta}^3}{N_\phi^6 |\psi_i|^6}\ln(k_b L)\Delta_\zeta^2(k_1).
\end{equation}
It is easily checked that the first term is of the order of $(P_{S_{\rm DR}}/P_\zeta)^2\times (\tan^2(n\theta)/\Delta N_{\rm eff})$.
Thus the non-linearity parameter can be large enough to be probed
for not so small $P_{S_{\rm DR}}$ compared to $P_\zeta$ and $\Delta N_{\rm eff} \ll 1$ or $\tan^2(n\theta) \gg 1$.

Finally we comment on the Q-ball formation in the present model, which is essential for 
protecting the lepton number from the sphaleron process.
After the AD field starts to oscillate, the instability develops and Q-balls are formed.
We consider the ``delayed''-type Q-balls~\cite{Kasuya:1999wu}, which are formed when the AD field potential becomes
dominated by the logarithmic term in (\ref{ADpot}).
Then the charge of Q-ball is estimated as~\cite{Kasuya:1999wu}
\begin{equation}
	Q \sim \beta \left( \frac{M_F}{m_{3/2}} \right)^4 \sim 
	6\times 10^{20}\left( \frac{1\,{\rm GeV}}{m_{3/2}} \right)^4\left( \frac{M_F}{10^6\,{\rm GeV}} \right)^4,
\end{equation}
where $\beta=6\times 10^{-4}$, and the radius of the Q-ball is given by $R_Q \sim m_{3/2}^{-1}$.
Although almost all the lepton number created by the AD mechanism is absorbed into Q-balls,
they can decay into neutrinos from their surfaces.
The decay rate is given by $\Gamma_Q \sim A m_{3/2}^3/(192\pi^3 Q)$ where $A$ is a 
surface area of the Q-ball~\cite{Cohen:1986ct}. Then the Q-ball decay temperature, $T_d$, is calculated as
\begin{equation}
	T_d \simeq 3\times 10^{-2}\,{\rm GeV}\left( \frac{m_{3/2}}{1\,{\rm GeV}} \right)^{5/2}
	\left( \frac{10^6\,{\rm GeV}}{M_F} \right)^2.
\end{equation}
This is well below the electroweak scale. Thus the lepton number liberated by the Q-ball decay
is not converted into the baryon number.\footnote{
	The diffusion process from the Q-ball surfaces may transfer the lepton number in the Q-balls
	into surrounding plasma~\cite{Laine:1998rg,Banerjee:2000mb} even at the temperature above the electroweak scale.
	These leptonic charges are converted to the baryon number through the sphaleron process,
	but this amount can be smaller than (or comparable to) the observed baryon number~\cite{Kawasaki:2002hq}.
}
Notice that the Q-ball decay rate and hence its decay temperature depends on the charge $Q$.
Thus if the $Q$ depends on the initial angle of the AD field $\theta$, the decay rate fluctuates on large scales
and it causes the modulated reheating~\cite{Dvali:2003em} for the case of Q-ball domination. 
In this case, the magnitude of the DR isocurvature perturbation is modified up to an $\mathcal O(1)$ numerical factor.
In the GMSB, however, it is often the case that the ellipticity of the AD field orbit in the complex plane is small
and the $Q$ does not depend on $\theta$~\cite{Kasuya:1999wu}. Therefore, there is no such an effect.

\section{Summary} \label{sec:summary}

In this paper, we discussed non-Gaussianities in dark radiation
isocurvature perturbations. Extending our analysis in the previous 
work~\cite{Kawasaki:2011rc}, we first derived the primordial 
bispectrum originating from the non-Gaussian isocurvature 
perturbations in dark radiation. We presented
primordial bispectra of both the local and quadratic types. 
As far as primordial perturbations have nearly scale-invariant spectra, 
amplitude of primordial power spectra can be parameterized
with six non-Gaussian parameters, which consequently measure
the non-Gaussianities in the mixture of the
adiabatic and dark radiation isocurvature modes. 
We also presented CMB bispectrum from these non-Gaussianities, 
which allows us to forecast constraints on the non-Gaussian parameters 
from future CMB surveys including the Planck sattelite and a hypothetical CVL survey.
While these parameters can be more or less constrained from 
ongoing Planck satellite experiments, there can be still some room for future CMB surveys to
improve the constraint, and CMBpol~\cite{Bock:2009xw} and COrE~\cite{Collaboration:2011ck} missions are 
desirable to improve the constraints.
We referred to SUSY axion models as concrete models for non-Gaussian dark radiation isocurvature
perturbations and showed that they offer distinct signatures on 
amplitudes in the primordial bispectrum. 
We have also shown that non-vanishing $S_{\rm DR}$, imprinted in the lepton asymmetry, or the neutrino density isocurvature perturbation,
can be generated through the Affleck-Dine mechanism 
without producing sizable extra radiation energy density. Since observational 
signatures are the same as those in the isocurvature model of the extra radiation component, 
primordial non-Gaussianities in the neutrino density isocurvature perturbation can also be constrained
by CMB observations. 

Extra radiation with $\Delta N_{\rm eff}\simeq 1$ will be tested by the 
ongoing Planck survey with high significance and its origin may be
identified through the detection of extra radiation isocurvature perturbations.
Furthermore, isocurvature perturbations in dark radiation
can offer us unique information for consistent understanding of the 
early Universe and the particle physics theory.

\bigskip
\bigskip

\noindent 
\section*{Acknowledgment}

T.~S. and K.~M. would like to thank the Japan Society for the Promotion of Science for the financial report.
The authors acknowledge Kobayashi-Maskawa Institute for the Origin of
Particles and the Universe, Nagoya University for providing computing
resources useful in conducting the research reported in this paper.
This work is supported by Grant-in-Aid for Scientific research from
the Ministry of Education, Science, Sports, and Culture (MEXT), Japan,
No.\ 14102004 (M.K.), No.\ 21111006 (M.K. and K.N.), No.\ 22244030 (K.N.) and also 
by World Premier International Research Center
Initiative (WPI Initiative), MEXT, Japan.


\end{document}